\documentclass[notitlepage,prd,twocolumn,showpacs,preprintnumbers,nofootinbib,superscriptaddress,floatfix,tightenlines]{revtex4}
\usepackage{mathrsfs}
\usepackage{amssymb}
\usepackage{amsmath}
\usepackage{graphicx}

\usepackage[utf8]{inputenc}
\usepackage{amsfonts,amsbsy}
\usepackage{nicefrac}
\usepackage{xcolor}
\definecolor{lcolor}{rgb}{0.5,0,0}
\definecolor{citcolor}{rgb}{0,0.3,0.0}
\usepackage[breaklinks,colorlinks,urlcolor=blue,citecolor=citcolor,linkcolor=lcolor]{hyperref}
\usepackage{placeins}
\usepackage{comment}
\usepackage{bbm}

\usepackage[normalem]{ulem} 
\usepackage{enumitem}

\newcommand{\Tr}{\mathrm{Tr}}

\newcommand{\eq}{Eq.~}

\newcommand{\eqs}{Eqs.~}

\newcommand{\nr}[1]{(\ref{#1})}

\newcommand{\ud}{\mathrm{d}}

\newcommand{\xt}{\mathbf{x}}

\newcommand{\yt}{\mathbf{y}}

\newcommand{\zt}{\mathbf{z}}

\newcommand{\bt}{\mathbf{b}}
\newcommand{\rt}{\mathbf{r}}
\newcommand{\kt}{\mathbf{k}}

\newcommand{\Bt}{\mathbf{B}}
\newcommand{\vt}{\mathbf{v}}

\newcommand{\wt}{\mathbf{w}}
\newcommand{\Rt}{\mathbf{R}}
\newcommand{\rtbar}{\bar{\mathbf{r}}}

\newcommand{\btbar}{\bar{\mathbf{b}}}
\newcommand{\ktbar}{\bar{\mathbf{k}}}
\newcommand{\xtbar}{\bar{\mathbf{x}}}
\newcommand{\ytbar}{\bar{\mathbf{y}}}
\newcommand{\qt}{\mathbf{q}}
\newcommand{\qtbar}{\bar{\mathbf{q}}}

\DeclareMathOperator\arctanh{arctanh}
\DeclareMathOperator\arccoth{arccoth}

\begin{document}

\title{Proton hot spots and exclusive vector meson production}

\author{S.~Demirci} 
\affiliation{Department of Physics, P.O.~Box 35, 40014 University of Jyv\"{a}skyl\"{a}, Finland}
\affiliation{Helsinki Institute of Physics, P.O.~Box 64, 00014 University of Helsinki, Finland}

\author{T.~Lappi} 
\affiliation{Department of Physics, P.O.~Box 35, 40014 University of Jyv\"{a}skyl\"{a}, Finland}
\affiliation{Helsinki Institute of Physics, P.O.~Box 64, 00014 University of Helsinki, Finland}

\author{S.~Schlichting} 
\affiliation{Fakult\"at für Physik, Universit\"at Bielefeld, D-33615 Bielefeld, Germany}

\begin{abstract}
We explore consequences of the existence of gluonic hot spots inside the proton for coherent and incoherent exclusive vector meson production cross sections in deep inelastic scattering. By working in the dilute limit of the Color Glass Condensate framework to compute the cross sections for Gaussian hot spots of fluctuating color charges and employing a non-relativistic vector meson wave function, we are able to perform large parts of the calculation analytically. We find that the coherent cross section is sensitive to both the size of the target and the structure of the probe. The incoherent cross section is dominated by color fluctuations at small transverse momentum transfer ($t$), by proton and hot spot sizes as well as the structure of the probe at medium $t$ and again by color fluctuations at large $t$. While the $t$-dependence of the cross section is well reproduced in our model, the relative normalization between the coherent and the incoherent cross sections points to the need for additional fluctuations in the proton.
\end{abstract}
\maketitle

\section{Introduction}

Understanding the structure of the proton in terms of its fundamental quark and gluon constituents is an ever-important question in high energy physics. Since the proton is a composite bound state of QCD, its partonic structure is, in general, non-perturbative and can only be accessed within non-perturbative calculations, e.g. in lattice QCD, or extracted from experiments. One of the cleanest ways to study this structure experimentally is by deeply inelastic $e+p$ scattering (DIS) experiments. Perhaps the most widely used experimental data comes from the H1 and ZEUS experiments performed at the Hadron-Electron Ring Accelerator (HERA)~\cite{H1:2005dtp, H1:2003ksk, H1:2013okq, ZEUS:1999ptu, ZEUS:2002vvv}.  Currently, the active experimental program of ultraperipheral, photon mediated, interactions \cite{Baltz:2007kq,Klein:2020nvu} at the LHC and RHIC also gives clean access to proton structure via $\gamma+p$ collisions. In the future, a much more detailed picture of the structure of the proton will be reached with the Electron-Ion Collider EIC~\cite{Accardi:2012qut,AbdulKhalek:2021gbh,AbdulKhalek:2022erw}.

For a long time it has been standard to describe the proton in terms of collinear parton distribution functions (PDFs), which describe the (longitudinal) momentum distributions of partons inside the proton. However, in recent years it has become increasingly well understood that the (transverse) spatial distribution of partons in the proton can also have a significant impact on  hadron production in $p-p$ and $p-Pb$ collisions~\cite{Bzdak:2013zma,Schenke:2014zha,Albacete:2016pmp,Albacete:2016gxu,Weller:2017tsr,Mantysaari:2017cni,Moreland:2018gsh,Mantysaari:2020axf,Schlichting:2014ipa,Albacete:2016pmp,Mantysaari:2016ykx,Mantysaari:2016jaz,Mantysaari:2017dwh,Mantysaari:2018zdd}. In particular, experimental observations of so-called collective flow of soft hadrons in $p+p$ and $p+Pb$ collisions  provide a strong motivation to  study the (transverse) spatial distribution of partons in the proton. 

A common approach to the internal transverse coordinate structure that is frequently employed in phenomenological studies of $p-p$ and $p-Pb$ collisions is based on a picture of the proton consisting of a certain number of gluonic ``hot spots''~\cite{Albacete:2016pmp,Mantysaari:2016ykx,Mantysaari:2016jaz,Mantysaari:2017dwh, Traini:2018hxd,Mantysaari:2018zdd,Kumar:2021zbn,Demirci:2021kya,Mantysaari:2022ffw}.  Evidently, such a model is motivated by the valence quark picture of the proton, from which one can conjecture that perhaps most of the partons in the proton should be close to these valence quarks, which ultimately are the source of the smaller-$x$ ones (see Refs.~\cite{Dumitru:2020gla,Dumitru:2021tvw} for a recent explicit realization of this idea). For a review on proton and nuclear shape fluctuations see \cite{Mantysaari:2020axf}.

Our aim in this paper is to understand to what extent one can probe and constrain different aspects of such a hot spot model using $e+p \to e+p^{(*)}+V$ exclusive vector meson $(V)$ production cross sections in DIS. 
To this end we use the formulation of the  hot spot model from our previous work~\cite{Demirci:2021kya}, which shares the basic physics ingredients with several hot spot models recently used in the literature, see e.g. Refs.~\cite{Mantysaari:2016jaz, Mantysaari:2016ykx, Mantysaari:2017dwh, Kumar:2021zbn, Traini:2018hxd}. 
Our approach leads to a simple, analytically tractable model for the cross sections which includes different sources of fluctuations in the sub-nucleon degrees of freedom inside the proton. By exploiting this analytic approach, our aim is to understand how much one can actually learn about the impact parameter dependent structure of the proton independently of the properties of the probe in exclusive vector meson production. 

The starting point for our approach is  the Color Glass Condensate (CGC) formalism \cite{Gelis:2010nm} to describe the partonic structure of the high energy proton. In this framework the large-$x$ partons act as color charges which produce the small-$x$ gluons. These dominate the small-$x$ part of the proton and are taken to be sufficiently dense to be described by classical color fields. The correlations of the color charges are taken to be Gaussian as in the MV model \cite{McLerran:1993ni, McLerran:1993ka, McLerran:1994vd}. In this work we assume the proton to be dilute meaning that we expand the operators defining the observables into the lowest non-trivial order in the field strength, akin to the  leading order  in the twist expansion.  

In our model~\cite{Demirci:2021kya} a non-trivial spatial structure of the proton emerges, as the large-$x$ color charges of the proton are taken to be distributed into hot spots. These hot spots have a Gaussian shape and are statistically distributed according to a Gaussian distribution, in such a way that the center of mass is at the center of the proton. Nevertheless, the locations of individual hot spots fluctuate from event to event and have to be averaged over to calculate the cross section. 

We combine our hot spot model with  other commonly used physics ingredients to calculate exclusive vector meson production cross sections in $\gamma^{(*)}p$ collisions. The scattering process is modelled using the dipole picture \cite{Kowalski:2006hc}. In this picture the incoming photon fluctuates into a quark-antiquark dipole which interacts with the classical color field of the proton, after which the dipole forms a vector meson. This calculation requires the use of a photon and a vector meson wave function, or more specifically the overlap of the two \cite{Kowalski:2006hc}. We take the vector meson wave function  to be non-relativistic to make the model as simple and the calculation as analytically tractable as possible. We use the  non-relativistic wave function  from Ref.~\cite{Lappi:2020ufv}.

The dipole picture naturally has an impact parameter dependence. This makes it a good way to understand the transverse structure of the target proton. The impact parameter is the Fourier conjugate of the Mandelstam variable $t$, which describes the squared momentum change of the proton, approximated to be purely transverse due to the high energy nature of the collision. This means that larger transverse momentum transfers probe smaller size-structures inside the proton.

When the virtuality of the photon or the mass of the quark in the dipole is large, the photon wave function greatly favors small dipole sizes. This is an approximation one might use when computing $J/\Psi$ exclusive vector meson production cross sections. We will study how good this approximation actually is in our model. For this reason we study processes where the dipole consists of a $c\bar{c}$, $b\bar{b}$ or $t\bar{t}$ pairs. The first two are phenomenologically relevant and the last acts as the true non-relativistic limit and is studied as a way to give a reference point for the other two cases.

We use the Good-Walker~\cite{Good:1960ba,Miettinen:1978jb} framework to compute coherent (target proton remains intact) and incoherent (target proton dissociates) exclusive vector meson production cross sections, both of which have been measured experimentally at HERA. The coherent cross section is proportional to the square of the average of the scattering amplitude \cite{Kowalski:2006hc}. This means that it is sensitive to the averages of the hot spot and color field fluctuations. The sum of the coherent and incoherent cross sections is proportional to the average of the square of the scattering amplitude \cite{Miettinen:1978jb, Lappi:2010dd, Mantysaari:2020axf}. Now because the incoherent cross section is the difference of the two, it is sensitive to the fluctuations in the scattering amplitude. In our work the sources for fluctuations are the event-by-event fluctuations of the color field of the proton and the fluctuating positions of the hot spots. Both of these two sources of fluctuations are averaged over analytically in the dilute limit for the proton. We study different contributions to the incoherent cross section separately to see how different features of the cross section are sensitive to different properties of the target, and to what extent they are independent of the probe. 

The paper is organized as follows. First we briefly review the dipole picture for exclusive vector meson production in Sec.~\ref{ssec:DipolePicture} and the photon and vector meson wave function overlap needed for the scattering amplitude in Sec.~\ref{ssec:WF}. Next we recall in Sec.~\ref{ssec:HSModel} the basic properties of the hot spot model that we developed in Ref.~\cite{Demirci:2021kya} and discuss how the color and hot spot averages of the dipole cross section and its square are computed in Sec.~\ref{sec:CSAverages}. Subsequently we compute the coherent and incoherent exclusive vector meson production cross sections and separate the color and hot spot fluctuation parts of the incoherent cross section in Sec.~\ref{sec:CrossSections}. Next we discuss our results, present numerical evaluations of the cross sections and explain how our model depends on different parameters in Sec.~\ref{sec:Analysis}. We finish with a brief summary of our most important findings and conclusions in Sec.~\ref{sec:Conclusions}. In the appendices we discuss the computation of the scattering amplitude averages in more detail, and also study the small dipole size approximation and the asymptotic behavior of the cross-section in different limits.

\section{Theoretical framework} \label{sec:Theory}

\subsection{Exclusive vector meson production in the dipole picture} \label{ssec:DipolePicture}

\begin{figure} 
\centerline{\includegraphics[scale=0.65]{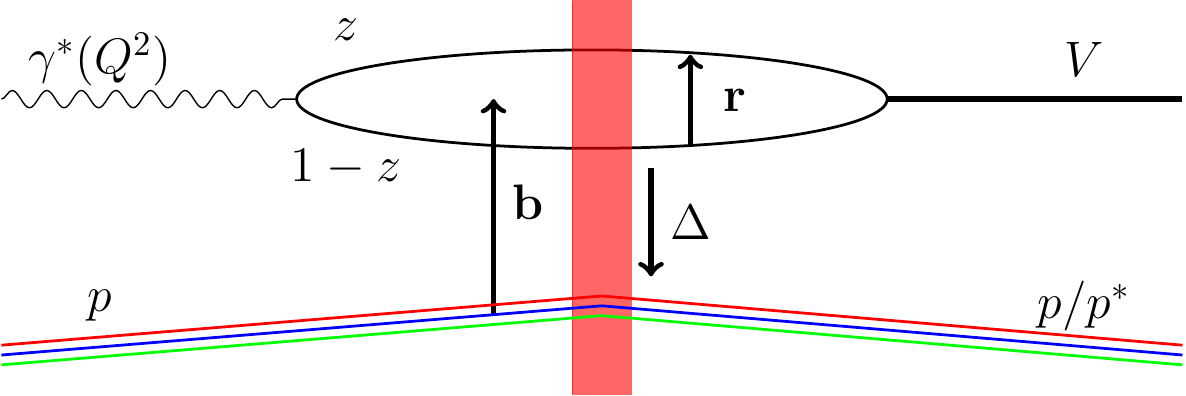}}

\caption{Illustration of the scattering process with an incoming photon, target proton and its color field and the outgoing vector meson. The relevant kinematic variables have been shown in the figure. The virtuality of the incoming photon $\gamma^*$ is  $Q^2$, $p/p^*$ is the proton/dissociated proton, $z$ is the fraction of the photon momentum carried by the quark (or antiquark), $V$ is the outgoing vector meson, $\rt$ is the size of the dipole, $\bt$ is the impact parameter of the dipole, $\Delta$ is the momentum transfer between the dipole and the proton system and the red rectangle represents the classical color field of the proton.}
\label{fig:ProtonKinematics}
\end{figure}

We describe exclusive vector meson production in the dipole picture using the Color Glass Condensate (CGC) formalism. In this picture, at the lowest order, the incoming virtual photon fluctuates into a quark-antiquark dipole which interacts with the color field of the proton. After the scattering the dipole forms the vector meson in the final state of the interaction, which can still decay before it is measured by a detector.

The scattering amplitude for the exclusive production of a transversely (T) or longitudinally (L) polarized  vector meson can be expressed as \cite{Kowalski:2006hc, Mantysaari:2020axf} 
\begin{multline}
\label{eq:VMamplitude}
\mathcal{A}^{\gamma^{*}p \rightarrow Vp}_{T,L}(Q^2,\Delta) = i \int d^2 \rt \int d^2 \bt \int \frac{dz}{4 \pi} 
\\ \times
(\Psi ^{*} \Psi _V)_{T,L}(Q^2,\rt,z)
\\  \times
\exp \left\{ -i\left[\bt+\left(\frac{1}{2}-z\right)\rt\right]\cdot \Delta \right\}
\frac{d \sigma ^{\text{p}}_{\text{dip}}}{d^2 \bt}(\bt, \rt).
\end{multline}
Here $\bt$ refers to the impact parameter of the dipole with respect to the target proton, $\rt$ is the transverse size vector of the dipole, $Q^2$ is the virtuality of the photon and $z$ is the fraction of the plus component of the light cone momentum of the photon carried by the quark in the dipole. The transverse momentum transfer to the target is denoted by $\Delta$, which according to Eq.~(\ref{eq:VMamplitude}) is the Fourier conjugate to $\bt + (\frac{1}{2}-z) \rt$. \footnote{Note Eq.~(\ref{eq:VMamplitude}) contains the corrected phase factor~\cite{Hatta:2017cte} $(\frac{1}{2}-z)\rt\cdot\Delta$  (instead of the $(1-z)\rt\cdot\Delta$ that has been used in e.g. \cite{Kowalski:2006hc} and  several other previous works). However,  the $\rt$ dependent phase factor does not influence the final results much \cite{Mantysaari:2020axf} and will in fact completely vanish in the non-relativistic limit, where the light cone momentum fraction is fixed to $z=1/2$, which we use in the following.}  The kinematic variables are visualized in the Fig.~ \ref{fig:ProtonKinematics}. 

By using the Good-Walker~\cite{Good:1960ba,Miettinen:1978jb,Kovner:2001vi,Caldwell:2010zza,Lappi:2010dd} picture for fluctuations in the target,
the coherent cross section can be written as a square of the average of the amplitude \cite{Kowalski:2006hc}
\begin{equation} \label{eq:CohCS}
\frac{d \sigma_{T,L}^{\gamma^*p\rightarrow Vp}}{dt} = \frac{1}{16\pi}|\langle \langle \mathcal{A}^{\gamma^{*}p \rightarrow Vp}_{T,L}(Q^2,\Delta) \rangle \rangle |^2
\end{equation} 
whereas the incoherent cross section can be written as the variance of the scattering amplitude as \cite{Miettinen:1978jb,Caldwell:2010zza,Lappi:2010dd,Mantysaari:2020axf}
\begin{multline} \label{eq:IncohCS}
\frac{d \sigma_{T,L}^{\gamma^*p\rightarrow Vp^*}}{dt} = \frac{1}{16\pi}\Big(\langle \langle | \mathcal{A}^{\gamma^{*}p \rightarrow Vp}_{T,L}(Q^2,\Delta) |^2 \rangle \rangle 
\\ 
- |\langle \langle \mathcal{A}^{\gamma^{*}p \rightarrow Vp}_{T,L}(Q^2,\Delta) \rangle \rangle |^2 \Big).
\end{multline} 
While the wave function overlap $(\Psi ^{*} \Psi _V)_{T,L}(Q^2,\rt,z)$ characterizes the properties of the probe, the information on the target color fields is contained in what is called the dipole cross section, $\frac{d \sigma ^{\text{p}}_{\text{dip}}}{d^2 \bt}(\bt, \rt)$. Hence, the computation of exclusive vector meson cross sections  can be separated into computing the expectation value and variance of the dipole cross section in the proton hot spot model on one hand,  and computing the wave function overlap $(\Psi ^{*} \Psi _V)_{T,L}(Q^2,\rt,z)$ describing the overlap of the dipole with the (virtual) photon and vector meson states on the other hand. We will first discuss the latter one, before moving to the calculation of the dipole cross section in the hot spot model of Ref.~\cite{Demirci:2021kya}.

\subsection{Wave function overlap in the non-relativistic limit} \label{ssec:WF}
Within this study we will employ the leading order photon wave functions along with the strict non-relativistic limit for the vector meson wave function to compute their overlap. Below we provide the relevant expressions, separately for the transverse and longitudinal polarizations. We first recall that the definition of the overlap is \cite{Kowalski:2006hc}
\begin{equation}
\begin{split}
&
(\Psi_V^*\Psi_{\gamma})_T = \frac{1}{2} \sum_{\substack{\lambda=+1,-1 \\ h,h'=+,-}} (\Psi^{\lambda}_{V,hh'})^* \Psi^{\lambda}_{\gamma,hh'}
\\ &
(\Psi_V^*\Psi_{\gamma})_L = \sum_{h,h'=+,-} (\Psi^{\lambda=0}_{V,hh'})^* \Psi^{\lambda=0}_{\gamma,hh'},
\end{split}
\end{equation}
where $h,h'$ refer to the helicities of the two quarks in the dipole and where $\lambda$ refers to the polarization of the photon. The longitudinally and transversely polarized photon wave functions read \cite{Lappi:2020ufv}
\begin{equation}
\label{eq:asd1}
\Psi^{\lambda=0}_{\gamma,hh'}(z,|\rt|)=-e_{Q} e\sqrt{N_c}\delta_{h,-h'}2Qz(1-z)\frac{K_0(\varepsilon |\rt|)}{2\pi}
\end{equation}
and
\begin{equation}
\label{eq:asd2}
\begin{split}
&
\Psi^{\lambda=\pm 1}_{\gamma,hh'}(z,|\rt|)=-e_{Q} e\sqrt{2N_c}
\Big[ m_Q\frac{K_0(\varepsilon |\rt|)}{2\pi} \delta_{h\pm} \delta_{h'\pm} 
\\ &
\pm i e^{\pm i\theta_{\rt}} \frac{\varepsilon K_1(\varepsilon |\rt|)}{2\pi}(z\delta_{h\pm}\delta_{h'\mp}-(1-z)\delta_{h\mp}\delta_{h'\pm}) 
\Big],
\end{split}
\end{equation}
where
\begin{equation}
\varepsilon = \sqrt{Q^2z(1-z)+m_{Q}^2}.
\end{equation}
Here $e=\sqrt{4\pi \alpha_{\text{em}}}$ denotes the electric charge, and $e_{Q}$ refers to the fractional charge of the quark , i.e. $e_{c}=e_{t}=\frac{2}{3}$ for charm (c) and top (t) quarks and $e_{b}=-\frac{1}{3}$ for bottom (b) quarks.

In the non-relativistic limit, the non-vanishing components of the vector meson wave function are given by \cite{Lappi:2020ufv}
\begin{equation}
\label{eq:asd3}
\Psi^{\lambda=0}_{V,+-}(z,|\rt|)=\Psi^{\lambda=0}_{V,-+}(z,|\rt|)=A_Q \frac{\sqrt{2}\pi}{\sqrt{m_{Q}}}\delta \left(z-\frac{1}{2}\right)
\end{equation}
and
\begin{equation}
\label{eq:asd4}
\Psi^{\lambda=1}_{V,++}(z,|\rt|)=\Psi^{\lambda=-1}_{V,--}(z,|\rt|)=A_Q \frac{2\pi}{\sqrt{m_{Q}}}\delta \left(z-\frac{1}{2} \right),
\end{equation}
where $m_{Q}$ refers to the mass of the non-relativistic heavy quark. The constant $A_Q$ can be found from the leptonic decay width of the vector meson, which can be calculated from the same light cone wave function. Specifically for the $c\bar{c}$ case, the numerical value can be obtained from the decay width of the $J/\Psi$ as~\cite{Lappi:2020ufv}
\begin{equation}\label{eq:decaywidth}
\Gamma(J/\Psi \rightarrow e^-e^+) = A_c^2 \frac{4\pi e_c^2 \alpha_{\text{em}}}{m_c^2}.
\end{equation}
Using the  experimental leptonic decay width the value of $A_c$  is found  in Ref.~\cite{Lappi:2020ufv} to be
\begin{equation}
A_c = 0.211 \text{GeV}^{\frac{3}{2}}.
\end{equation}
When studying the cases with different quark masses, we do not want to introduce uncertainties related to the decay constants into our calculation. Thus for heavier mesons we always compare cross sections divided by the leptonic decay widths, which cancels the constant $A_Q$.

By combining Eqs.~\nr{eq:asd1}, \nr{eq:asd2}, \nr{eq:asd3} and~\nr{eq:asd4}, one can then compute the overlaps of the wave functions as
\begin{equation}
(\Psi_V^*\Psi_{\gamma})_T = -A_Q \sqrt{2m_{Q}N_c} e_{Q} e K_0(\varepsilon |\rt|) \delta \left(z - \frac{1}{2} \right)
\end{equation}
and
\begin{equation}
(\Psi_V^*\Psi_{\gamma})_L = -A_Q \sqrt{\frac{2N_c}{m_{Q}}} 2 e_{Q} e Q z(1-z) K_0(\varepsilon |\rt|) \delta \left(z - \frac{1}{2} \right)
\end{equation}
which can be employed directly in the calculation of the exclusive vector meson production cross sections. 

\subsection{The hot spot model} \label{ssec:HSModel}

\begin{figure} 
\centering
\includegraphics[scale=0.65]{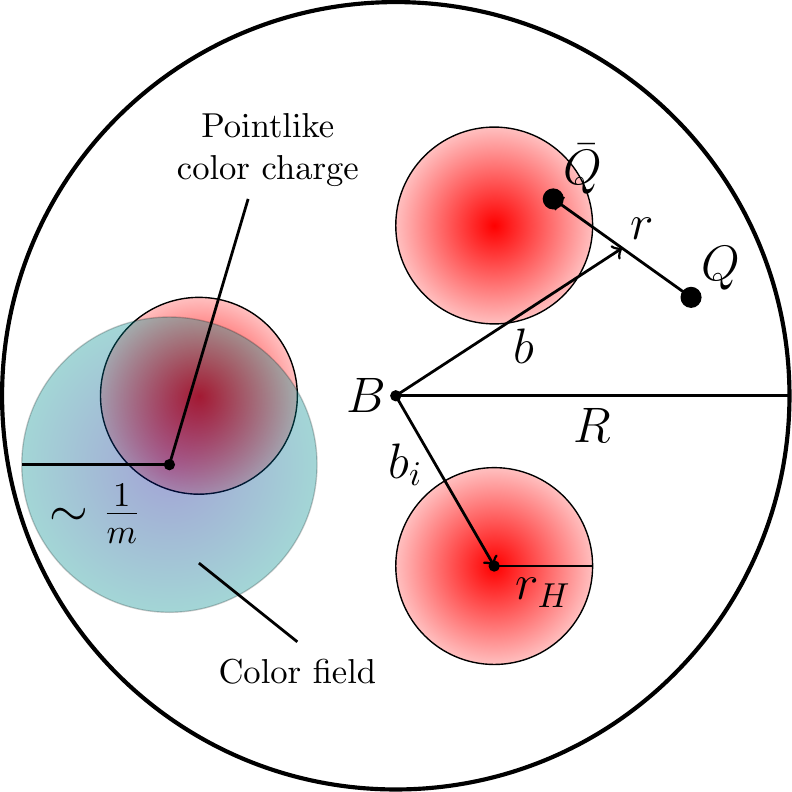}
\caption{The transverse profile of a hot spot model proton with a dipole probe. The red circles represent the hot spots containing the large-$x$ color charges of the proton. The blue circle represents the classical color field of small-$x$ gluons generated by a pointlike color charge within the hot spot. The other quantities are as follows: $\Bt$ is  the center of mass of the proton, $\bt_i$ is the position of the center of the $i$:th hot spot, $r_H$ is the radius of a hot spot, $R$ is the radius parameter of the proton. The quark and antiquark forming the dipole are denoted by $Q(\bar{Q})$.   The impact parameter of the dipole (with respect to the center of mass of the target) is denoted by $\bt$, and  $\rt$ is the size of the dipole.}
\label{fig:ProtonDipole}
\end{figure}

We use the hot spot model formulated in Ref.~\cite{Demirci:2021kya} to
quantify the event-by-event fluctuations of the structure of the proton. 
Within this model, an expectation value averaged over target fluctuations 
can be computed as a double average over the positions $\bt_{i}$ of $i=1,\cdots,N_q$ hot spots
inside the proton on the one hand, and individual realizations of color charges
$\rho_{a}(\xt)$ inside the hot spots on the other hand. 
Event-by-event fluctuations of the proton structure thus consist of both
color charge fluctuations as well as hot spot location fluctuations, and we will
separately investigate the effects of both sources of fluctuations on the 
incoherent cross section.

Within the CGC description, the partons in the proton are divided into large-$x$ color
charges acting as sources for small-$x$ classical gluon fields. Hence, it is
natural to express the hot spot structure of the proton in terms of the
impact parameter dependence of the color charge density $\rho_{a}(\xt)$, which vanishes on average 
$\langle \rho_{a}(\xt) \rangle_{CGC}=0$ to satisfy color neutrality, but fluctuates 
on an event by event basis. Specifically, we 
consider the color charge density $\rho_{a}(\xt)$ to be concentrated around the location $\bt_i$ of
the hot spots, such that fluctuations of the color charge density
are determined by 
\begin{equation} \label{eq:rhorho} \langle
\rho^a(\xt) \rho^b(\yt) \rangle_{CGC} = \sum_{i=1}^{N_q} \mu^2 \left(
\frac{\xt+\yt}{2} - \bt_i \right) \delta^{(2)}(\xt-\yt) \delta^{ab},
\end{equation}
where $\mu^2(\xt)$ denotes the transverse profile of the hot spot, and following the McLerran-Venugopalan (MV) model~\cite{McLerran:1993ni, McLerran:1993ka, McLerran:1994vd} we take the correlations of the color charges at different transverse points to be independent. 
We  take a Gaussian profile for the hot spots in transverse coordinate space
\begin{equation}
\label{eq:hotspotprofile} \mu^2(\xt)=\frac{\mu^2_0}{2\pi r_H^2} \exp
\left[ -\frac{\xt^2}{2r_H^2} \right], \end{equation}
where $r_H$ is the radius of the hot spot and $\mu^2_0$ is
a parameter describing the amount of color charge in the hot spot. We note that the normalization in Eq.~(\ref{eq:hotspotprofile}) is such that
that the dimensions of $\mu^2(\xt)$ and $\mu_0^2$ are different.

By performing the CGC average of the correlation functions of the color charge density $\rho_{a}(\xt)$, 
the model thus accounts for color charge fluctuations
inside the proton. When performing these averages, we will assume Gaussian statistics
for the color charges as in the McLerran-Venugopalan (MV) model~\cite{McLerran:1993ni, McLerran:1993ka, McLerran:1994vd}. This means that the CGC averages of higher order correlators such as $\langle \rho^a(\xt) \rho^b(\yt) \rho^c(\xt')  \rho^d(\yt')  \rangle_{CGC}$, can
be expressed in terms of the two-point function in Eq.~\nr{eq:rhorho} using Wick's theorem.

By taking into consideration the number of valence quarks, which
generally carry a large fraction of the proton momentum and are thus
part of the CGC color charge, it would be natural to assume that the number
of hot spots would be $N_q=3$. However, this is not a necessity as the
valence quarks can also emit large-$x$ gluons, which can proceed and emit
other large-$x$ partons. Such kind of a processes could increase the
number of hot spots in the proton, and it is also conceivable that the
number of hot spots fluctuates on event-by-event basis. Within our work we
allow the number of hot spots ($N_q$) to be a free parameter in our
calculations until the numerical analysis, but we do not take into account event-by-event
fluctuations of the number of hot spots ($N_q$).

The spatial positions of the hot spots fluctuate on an event-by-event basis, and
the distribution of the hot spot positions in the transverse plane is
taken to be Gaussian
 \begin{equation} T(\bt)=\frac{1}{2\pi
R^2}\exp \left[ -\frac{\bt^2}{2R^2} \right]. 
\end{equation} 
where $R$ characterizes the size of the proton. By following our previous paper \cite{Demirci:2021kya} the averages over the positions of the hot spots and
the color charge fluctuations around the hot spots can be combined into a \emph{double average}, such that the expectation value of 
an observable $\mathcal{O}$ is determined by
\begin{multline} \label{eq:defavg} \langle \langle \mathcal{O} \rangle
\rangle = \left( \frac{2 \pi R^2}{N_q} \right) \int \prod^{N_q}_{i=1}
\Big[ \ud^2\bt_i T(\bt_i-\mathbf{B})\Big] \\  \times \delta^{(2)}\left(
\frac{1}{N_q}\sum _{i=1}^{N_q}\bt_i - \mathbf{B}\right)  \langle
\mathcal{O} \rangle _{CGC}, \end{multline} 
where, as illustrated in Fig.~\ref{fig:ProtonDipole}, $N_q$ is the number of hot spots, 
$\bt_i$ is the center of the $i$:th hot spot and $\Bt$ is the center of
mass of the proton. The prefactor is
chosen in such a way that the double average is normalized, i.e.
$\langle \langle 1 \rangle \rangle = 1$. We note that the $\delta$ constraint in Eq.~(\ref{eq:defavg}) fixes the center of mass of the hot spot system $\frac{1}{N_q}\sum _{i=1}^{N_q}\bt_i$ to the center of mass of the proton $\mathbf{B}$. This procedure does not take into account the color charge fluctuations, but we believe
that they give a much smaller contribution to it than the hot spot
fluctuations. Since in this work the proton acts as a target for a dipole
probe, we can further set $\Bt=0$ without loss of generality and instead let the impact parameter of the dipole
change.

We finally note that very similar hot spot models have been used in a variety of studies. In
particular our model has most of the same physics ingredients and
parameters as  the so called IP-Glasma model \cite{Schlichting:2014ipa,
Mantysaari:2016ykx}. The averaging over hot spot
configurations and color charges in the IP-Glasma model is performed
numerically by a Monte Carlo procedure. In our work we take the dilute
limit in the color charges of the proton, which allows us to perform the
averages analytically.
In addition to taking the dilute limit, we also do not include the so
called $Q_s$ fluctuations in our model which were used in
Ref.~\cite{Mantysaari:2016ykx} by letting the saturation scale of each hot spot
fluctuate independently. We also do not have repulsion between
individual hot spots as in Refs.~\cite{Albacete:2016pmp, Mantysaari:2022ffw}.

\subsection{Average of the dipole cross section and its square} \label{sec:CSAverages}
Next we proceed to calculate the two quantities
\begin{equation}
\left\langle \left\langle
\frac{d \sigma ^{\text{p}}_{\text{dip}}}{d^2 \bt}(\bt, \rt)
\right\rangle \right\rangle
\quad \text{and} \quad
\left\langle \left\langle
\frac{d \sigma ^{\text{p}}_{\text{dip}}}{d^2 \bt}(\bt, \rt)
\frac{d \sigma ^{\text{p}}_{\text{dip}}}{d^2 \btbar}(\btbar, \rtbar)
\right\rangle \right\rangle
\end{equation}
as well as their Fourier transforms, where the double average now refers to averaging over both the color fields and the hot spot positions of the proton as discussed above.

The starting point of the calculation is the operator expression for the dipole cross-section, which is given by twice the forward elastic scattering amplitude for a color neutral dipole to scatter off the target color field
\begin{equation}
\frac{d \sigma ^{\text{p}}_{\text{dip}}}{d^2 \bt}(\bt, \rt) =
2 N(\bt, \rt).
\end{equation}
The dipole scattering amplitude $N(\bt,\rt)$  is obtained from the target color fields as a two-point function of Wilson lines
\begin{equation}
N(\bt, \rt) = 1 - \frac{1}{N_c} \Tr [V(\xt)V^{\dagger}(\yt)],
\end{equation}
where the (anti)quark transverse positions $(\xt,\yt)$ are connected to the dipole impact parameter $(\bt)$ and dipole radius ($\rt$) coordinates $(\bt,\rt)$ by the transformations
\begin{equation}
\bt = \frac{\xt + \yt}{2}, \qquad \rt = \xt - \yt 
\end{equation}
or equivalently
\begin{equation}
\xt = \bt + \frac{\rt}{2}, \quad \yt = \bt - \frac{\rt}{2},
\end{equation}
as illustrated in Fig.~\ref{fig:ProtonDipole}. By $N_c=3$ we denote the number of colors and $V(\xt)$ is a light-like fundamental Wilson line at the transverse coordinate $\xt$, defined as a path-ordered exponential
\begin{multline} \label{eq:WilsonLine}
V(\xt) =
\mathcal{P}_+ \exp 
\left[ 
ig \int_{-\infty}^{\infty} dz^+ A^-_{a}(z^+,\xt) t^{a}
\right]
\\  =
\mathcal{P}_+ \exp 
\left[ 
ig \int_{-\infty}^{\infty} dz^+ \int d^2 \zt G(\xt-\zt) \rho_a(z^+,\zt)t^a
\right].
\end{multline}
It describes the eikonal multiple scattering of a color charge off  a classical color field. The connection between the classical color field and the color charge densities is found by solving the classical Yang-Mills equation. These Wilson lines are the basic building blocks of observables in the CGC. To get the final result one needs to average over the color charge densities $\rho$ that the Wilson lines depend on. The Green's function relating the color field in the Wilson line to the color charge is defined as
\begin{equation}
G(\xt - \yt) = \int \frac{d^2 \kt}{(2 \pi)^2} \frac{\exp{(i \kt \cdot (\xt - \yt))}}{\kt ^2 + m^2},
\end{equation}
where $m$ is an infrared regulator. The IR regulator suppresses the Coulomb-like long range tails of the color fields, so that the field of an individual point charge extends to a distance  of order $\frac{1}{m}$.

Now we invoke the dilute expansion for the proton and expand the dipole cross section to the first order in the proton color charge density, or equivalently, the field strength. Expressing the Wilson lines in terms of color charges, expanding in the dilute limit
and performing the double average over the color and hot spot fluctuations then yields
\begin{equation}
\begin{split}
&  \left\langle \left\langle
\frac{d \sigma ^{\text{p}}_{\text{dip}}}{d^2 \bt}(\bt, \rt)
\right\rangle \right\rangle
= 
\frac{g^2 (N_c^2 - 1) N_q}{2 N_c} 
\\ & \qquad \qquad \times
\int d^2 \zt
\Omega \left( \bt + \frac{\rt}{2}, \bt - \frac{\rt}{2}, \zt, \zt \right) F_1(\zt , \Bt),
\end{split}
\end{equation}
where the function $F_1(\zt, \Bt)\equiv \langle \mu^2(\zt-\bt_i) \rangle_{\mathrm{HS}}$ describes the average color charge density. We have  also defined the shorthand notation
\begin{equation}
\label{eq:defOmega}
\begin{split}
\Omega(\xt, \yt, \zt, \vt) \equiv & G(\xt - \zt)G(\xt - \vt) + G(\yt - \zt)G(\yt - \vt) 
\\ &
- 2G(\xt - \zt)G(\yt - \vt),
\end{split}
\end{equation}
which is convenient as the dilute expansion of the Wilson lines always has this structure of disconnected and connected transverse coordinate parts. Specifically, this structure ensures that the dipole cross section vanishes identically in the limit $\xt=\yt$, where the quark and antiquark hit the target at the same transverse position. In this configuration the $Q\bar{Q}$-dipole appears color neutral and thus does not interact with the color field of the target. More details on this derivation are given in Appendix~\ref{app:dip1point}.

Within our hot spot model the average color charge density of the proton in the transverse plane $F_1(\zt, \Bt)$ has already been computed in  \cite{Demirci:2021kya} and is given by
\begin{multline} \label{eq:F1}
F_1(\zt, \Bt) \equiv \langle \mu^2(\zt-\bt_i) \rangle_{\text{Hot spot}}
\\  
=
\left( \frac{ \frac{\mu^2_0}{2 \pi r_H^2}}{1+\left(\frac{N_q-1}{N_q}\right)\frac{R^2}{r_H^2}} \right)
\exp \left\{ -\frac{1}{2}  \frac{(\zt-\Bt)^2}{r_H^2+\left(\frac{N_q-1}{N_q}\right)R^2} \right\}.
\end{multline}
Note that while to a first approximation (in the limit $R\gg r_H, \ N_q\gg 1$) this is just a Gaussian whose width is given by the proton radius parameter $R$, the constraint of fixing the center of mass of the hot spot system gives a nontrivial dependence on the number of hot spots $N_q$ and the hot spot radius $r_H$.
We will call the quantity setting the scale of the impact parameter dependence in the exponential \emph{the coherent radius}:
\begin{equation} \label{eq:CohRadius}
R_C^2 = r_H^2 +\left( \frac{N_q-1}{N_q} \right) R^2. 
\end{equation}
This quantity represents the effective size of the hot spot system of color charges, taking into account both the locations of the hot spots and the size of an individual hot spot.

Next, as the $\Psi_V^*\Psi_{\gamma}$ wave function overlap does not depend on the impact parameter ($\bt$) of the dipole, we can compute the Fourier transform of the dipole cross sections with respect to the impact parameter
\begin{equation} \label{eq:1PointDipFourier}
\begin{split}
&
\int d^2 \bt e^{-i \bt \cdot \Delta}
\left\langle \left\langle
\frac{d \sigma ^{\text{p}}_{\text{dip}}}{d^2 \bt}(\bt, \rt)
\right\rangle \right\rangle
= \frac{g^2 \mu_0^2 (N_c^2-1) N_q}{2 \pi N_c} 
\\ & \times
\exp \left \{ -\frac{1}{2} \left( r^2_H + \left( \frac{N_q-1}{N_q} \right) R^2 \right) \Delta^2 \right \}
\\ & \times
\Big \{ \cos \left( \frac{1}{2}\Delta \cdot \rt \right) \Psi \left( 0,\Delta,m^2 \right) - \Psi \left( \rt,\Delta,m^2 \right) \Big \}.
\end{split}
\end{equation}
which then directly enters the scattering amplitude in Eq.~(\ref{eq:VMamplitude}). In order to obtain Eq.~(\ref{eq:1PointDipFourier}) we have expressed  the products of Green functions in \eq\nr{eq:defOmega} in momentum space, and evaluated the integrals by a Feynman parametrization. This provides the following integral representation of the function $ \Psi \left( \rt,\Delta,m^2 \right)$
\begin{multline} \label{eq:PsiFunction}
\Psi \left( \rt,\Delta,m^2 \right) = \int_0^{\frac{1}{2}} d\lambda \cos \left( \lambda\Delta\cdot\rt \right) 
\\  \times
\frac{|\rt|K_1 \left( |\rt|\sqrt{-\Delta^2\lambda^2 + \frac{1}{4}\Delta^2 + m^2} \right) }{\sqrt{-\Delta^2\lambda^2 + \frac{1}{4}\Delta^2 + m^2}},
\end{multline}
where $\lambda$ is a Feynman parameter, $\Delta$ is the momentum transfer between the dipole and the proton system and $m$ is the IR regulator. We note that the Feynman parameter integral can be evaluated exactly in the limit $\rt\to0$ of a zero size dipole as
\begin{equation} \label{eq:PsiFunctionR0}
\begin{split}
\Psi \left( 0,\Delta,m^2 \right) 
&
= \int_0^{\frac{1}{2}} d\lambda  \frac{1}{-\Delta^2\lambda^2 + \frac{1}{4}\Delta^2 + m^2}
\\ &
=
\frac{2 \arctanh \left( \frac{|\Delta|}{\sqrt{\Delta^2 + 4m^2}} \right) }{|\Delta|\sqrt{\Delta^2 + 4m^2}},
\end{split}
\end{equation}
and it can be expanded to in powers of $\rt$, to obtain an analytical expression of the coherent amplitude in the small dipole size limit $(m|\rt|,|\Delta \rt| \ll 1)$. This limit is  discussed in more detail in Appendix~\ref{app:smallR}. In Appendix~\ref{app:smallRCrossSections} we compute the cross sections at the limit of small $\rt$, in \ref{app:DeltaCohApproximations0AndInfty} and \ref{app:LargeT} we discuss some analytical limits attainable at some regions of $t$ in the small-$\rt$ limit and in \ref{app:FullVsLimits} we compare the limits to the full results.

Next we will calculate the Fourier transform of the average of the square of the dipole cross section, $\left\langle \left\langle
\frac{d \sigma ^{\text{p}}_{\text{dip}}}{d^2 \bt}(\bt, \rt)
\frac{d \sigma ^{\text{p}}_{\text{dip}}}{d^2 \btbar}(\btbar, \rtbar)
\right\rangle \right\rangle$ which is needed for the calculation of the incoherent cross section. We will separate this contribution into four parts, based on the criteria  a) whether the color structure in the amplitude and conjugate amplitude is connected (C) or disconnected (DC) and b) whether the gluon fields in the amplitude and conjugate amplitude originate from the same single (1) hot spot or two different (2) hot spots.  Based on this separation the four individual parts  can be evaluated as
\begin{equation}
\begin{split}
&
\left[
\int d^2 \bt d^2 \bar{\bt} e^{-i \bt \cdot \Delta + i \bar{\bt} \cdot \Delta}
\left\langle \left\langle
\frac{d \sigma ^{\text{p}}_{\text{dip}}}{d^2 \bt}(\bt, \rt)
\frac{d \sigma ^{\text{p}}_{\text{dip}}}{d^2 \bar{\bt}}(\bar{\bt}, \bar{\rt})
\right\rangle \right\rangle \right]_{1,DC}
\\ &
=
\frac{g^4\mu_0^4(N_c^2-1)^2N_q}{(2\pi)^2N_c^2}\exp \left( -r^2_H\Delta^2 \right)
\\ & \times
\Big\{ \Psi(\rt)\Psi(\bar{\rt}) +\cos \left(\frac{1}{2}\Delta\cdot\rt \right) \cos \left( \frac{1}{2}\Delta\cdot\bar{\rt} \right) \Psi(0)\Psi(0)
\\ &
- \cos \left( \frac{1}{2}\Delta\cdot\bar{\rt} \right)\Psi(\rt)\Psi(0) - \cos \left(\frac{1}{2}\Delta\cdot\rt \right) \Psi(\bar{\rt})\Psi(0)
 \Big\}
\end{split}
\end{equation}
\begin{equation}
\begin{split}
&
\left[
\int d^2 \bt d^2 \bar{\bt} e^{-i \bt \cdot \Delta + i \bar{\bt} \cdot \Delta}
\left\langle \left\langle
\frac{d \sigma ^{\text{p}}_{\text{dip}}}{d^2 \bt}(\bt, \rt)
\frac{d \sigma ^{\text{p}}_{\text{dip}}}{d^2 \bar{\bt}}(\bar{\bt}, \bar{\rt})
\right\rangle \right\rangle \right]_{2,DC}
\\ &
=
\frac{g^4\mu_0^4(N_c^2-1)^2N_q(N_q-1)}{N_c^2(2\pi)^2}\exp \left( -(R^2+r^2_H)\Delta^2 \right)
\\ & \times
\Big\{ \Psi(\rt)\Psi(\bar{\rt}) +\cos \left( \frac{1}{2}\Delta\cdot\rt \right) \cos \left( \frac{1}{2}\Delta\cdot\bar{\rt} \right) \Psi(0)\Psi(0)
\\ &
- \cos \left( \frac{1}{2}\Delta\cdot\bar{\rt} \right)\Psi(\rt)\Psi(0) - \cos \left( \frac{1}{2}\Delta\cdot\rt \right) \Psi(\bar{\rt})\Psi(0)
 \Big\}
\end{split}
\end{equation}
\begin{equation}
\begin{split}
&
\left[
\int d^2 \bt d^2 \bar{\bt} e^{-i \bt \cdot \Delta + i \bar{\bt} \cdot \Delta}
\left\langle \left\langle
\frac{d \sigma ^{\text{p}}_{\text{dip}}}{d^2 \bt}(\bt, \rt)
\frac{d \sigma ^{\text{p}}_{\text{dip}}}{d^2 \bar{\bt}}(\bar{\bt}, \bar{\rt})
\right\rangle \right\rangle \right]_{1,C}
\\ &
=
\frac{g^4\mu_0^4(N_c^2-1)}{64\pi^4N_c^2}N_q
\\ & \times
\int d^2\kt d^2 \bar{\kt} 
\frac{\exp \left( -r^2_H(\kt+\bar{\kt} \right)^2)}{( \left( \kt+\frac{\Delta}{2} \right)^2+m^2)( \left( \kt-\frac{\Delta}{2} \right)^2+m^2)}
\\ & \times
\frac{1}{( \left( \bar{\kt}+\frac{\Delta}{2} \right)^2+m^2)( \left(\bar{\kt}-\frac{\Delta}{2} \right)^2+m^2)}
\\ & \times 
\Bigg \{
8\cos \left( \frac{\Delta}{2}\cdot\rt \right)\cos \left( \frac{\Delta}{2}\cdot\bar{\rt} \right)
\\ &
-8\cos \left( \frac{\Delta}{2} \cdot \rt \right)\exp \left( i\bar{\kt} \cdot \bar{\rt} \right)
-8\cos \left( \frac{\Delta}{2} \cdot \bar{\rt} \right)\exp \left( i\kt \cdot \rt \right)
\\ &
+4\exp \left( i\kt \cdot \rt +i \bar{\kt} \cdot \bar{\rt} \right)
+4\exp \left( i\kt\cdot\rt - i\bar{\kt}\cdot\bar{\rt}  \right)
\Bigg \}
\end{split}
\end{equation}
and
\begin{equation}
\begin{split}
&
\left[
\int d^2 \bt d^2 \bar{\bt} e^{-i \bt \cdot \Delta + i \bar{\bt} \cdot \Delta}
\left\langle \left\langle
\frac{d \sigma ^{\text{p}}_{\text{dip}}}{d^2 \bt}(\bt, \rt)
\frac{d \sigma ^{\text{p}}_{\text{dip}}}{d^2 \bar{\bt}}(\bar{\bt}, \bar{\rt})
\right\rangle \right\rangle \right]_{2,C}
\\ &
=
\frac{g^4\mu_0^4(N_c^2-1)}{64\pi^4N_c^2}N_q(N_q-1)
\\ & \times
\int d^2\kt d^2 \bar{\kt} 
\frac{\exp \left( -(R^2+r^2_H)(\kt+\bar{\kt})^2 \right) }{( \left( \kt+\frac{\Delta}{2} \right)^2+m^2)( \left( \kt-\frac{\Delta}{2} \right)^2+m^2)}
\\ & \times
\frac{1}{( \left(\bar{\kt}+\frac{\Delta}{2} \right)^2+m^2)( \left(\bar{\kt}-\frac{\Delta}{2} \right)^2+m^2)}
\\ & \times 
\Bigg\{
8\cos \left( \frac{\Delta}{2}\cdot\rt \right)\cos \left( \frac{\Delta}{2}\cdot\bar{\rt} \right)
\\ &
-8\cos \left( \frac{\Delta}{2} \cdot \rt \right)\exp \left( i\bar{\kt} \cdot \bar{\rt} \right)
-8\cos \left( \frac{\Delta}{2} \cdot \bar{\rt} \right) \exp \left( i\kt \cdot \rt \right)
\\ &
+4\exp \left( i\kt \cdot \rt +i \bar{\kt} \cdot \bar{\rt} \right)
+4\exp \left( i\kt\cdot\rt - i\bar{\kt}\cdot\bar{\rt} \right)
\Bigg\}.
\end{split}
\end{equation}
While the incoherent cross section is of course only sensitive to the sum of all of these contributions, the above distinction is useful to organize the calculation and separate the effects of different sources of fluctuations. Specifically, for the contributions labeled 1 (as in ``1 hot spot'') the gluon fields in both the amplitude and the conjugate amplitude originate from the same hot spot, making  the contribution correlated in the hot spot sense. These contributions are explicitly proportional to the number of hot spots $N_q$. On the other hand for the contributions labeled 2 (as in ``2 hot spot''), the gluon fields originate from two different hot spots and these terms are proportional to the number of hot spot pairs $N_q(N_q-1)$. Somewhat counterintuitively these two hot spot contributions are also correlated in terms of hot spot location fluctuations, due to the fact that the center of mass of the hot spot system is fixed. A more detailed derivation of these expressions in presented in Appendix~\ref{app:dip2point}.

Since for the color-disconnected (DC) diagrams, the CGC averages can be performed separately in the amplitude and its conjugate, the structure of the resulting expressions is similar to the coherent amplitude in Eq.~(\ref{eq:1PointDipFourier}) and again involves the same function $\Psi \left( \rt,\Delta,m^2 \right)$ defined in \eq\nr{eq:PsiFunction}. Conversely, the color-connected (C) contributions contain non-trivial color correlations between the amplitude and conjugate amplitude, which are suppressed by a relative factor $1/(N_c^2-1)$ and involve a different structure of the Green's functions. We note that in our calculation, the clear separation into color connected and disconnected parts is easily possible due to the dilute expansion and the fact that correlations of color charges are assumed to be Gaussian.

\subsection{Cross sections} \label{sec:CrossSections}

Now we will use the results obtained above for the target averages of the dipole amplitude and its square to compute the average and the square average of the vector meson scattering amplitude in Eq.~(\ref{eq:VMamplitude}). This will allow us  to evaluate the coherent and incoherent cross-sections in the hot spot model.
Evaluating the square of the average, one has
\begin{equation}
\begin{split}
&
\!\!\!\!\!\!\!\!\!\!\!\!\!\! \left| \left\langle \left\langle \mathcal{A}^{\gamma^{*}p \rightarrow Vp}_{T,L}(Q^2,\Delta) \right\rangle \right\rangle\right|^2
\\ 
= & \int d^2 \rt d^2 \rtbar \int d^2 \bt d^2 \btbar \int \frac{dz}{4 \pi} \frac{d\bar{z}}{4 \pi} 
\\ & \times
(\Psi ^{*} \Psi _V)_{T,L}(Q^2,\rt,z) (\Psi ^{*} \Psi _V)^{*}_{T,L}(Q^2,\rtbar,\bar{z})
\\ & \times
\exp \left\{ -i \left[ \bt+ \left( \frac{1}{2}-z \right) \rt \right] \cdot \Delta \right\}
\\ & \times
\exp \left\{ i \left[\btbar+ \left(\frac{1}{2}-\bar{z} \right)\rtbar \right]\cdot \Delta \right\}
\\ & \times
\left\langle \left\langle
\frac{d \sigma ^{\text{p}}_{\text{dip}}}{d^2 \bt}(\bt, \rt)
\right\rangle \right\rangle \left\langle \left\langle
\frac{d \sigma ^{\text{p}}_{\text{dip}}}{d^2 \btbar}(\btbar, \rtbar)
\right\rangle \right\rangle
\end{split}
\end{equation}
whereas the average of the square reads
\begin{equation}
\begin{split}
&
\!\!\!\!\!\!\!\!\!\!\!\!\!\!\left\langle \left\langle \left|\mathcal{A}^{\gamma^{*}p \rightarrow Vp}_{T,L}(Q^2,\Delta)\right|^2 \right\rangle \right\rangle
\\ 
= & \int d^2 \rt d^2 \rtbar \int d^2 \bt d^2 \btbar \int \frac{dz}{4 \pi} \frac{d\bar{z}}{4 \pi} 
\\ & \times
(\Psi ^{*} \Psi _V)_{T,L}(Q^2,\rt,z) (\Psi ^{*} \Psi _V)^{*}_{T,L}(Q^2,\rtbar,\bar{z})
\\ & \times
\exp \left\{ -i \left[\bt+ \left(\frac{1}{2}-z \right)\rt \right]\cdot \Delta \right\}
\\ & \times
\exp \left\{ i \left[\btbar+ \left(\frac{1}{2}-\bar{z} \right)\rtbar \right]\cdot \Delta \right\}
\\ & \times
\left\langle \left\langle
\frac{d \sigma ^{\text{p}}_{\text{dip}}}{d^2 \bt}(\bt, \rt)
\frac{d \sigma ^{\text{p}}_{\text{dip}}}{d^2 \btbar}(\btbar, \rtbar)
\right\rangle \right\rangle.
\end{split}
\end{equation}

We first note, that in the non-relativistic approximation, the $z$-integral can trivially be performed. The $z$-dependent piece of the scattering amplitude reads
\begin{equation}
\int \frac{dz}{4\pi} (\Psi_V^*\Psi_{\gamma})_{T,L} e^{-i \left( \frac{1}{2}-z \right) \rt \cdot \Delta},
\end{equation}
which, for the transverse and longitudinal amplitude respectively, yields
\begin{multline} \label{eq:WFTOverlapIntegrated}
\int \frac{dz}{4\pi} (\Psi_V^*\Psi_{\gamma})_T e^{-i \left( \frac{1}{2}-z \right) \rt \cdot \Delta} 
\\ 
= -A_Q \frac{\sqrt{2m_{Q}N_c}}{4\pi}e_{Q} e K_0(\varepsilon'|\rt|) \equiv C_T K_0(\varepsilon'|\rt|)
\end{multline}
and
\begin{multline} \label{eq:WFLOverlapIntegrated}
\int \frac{dz}{4\pi} (\Psi_V^*\Psi_{\gamma})_L e^{-i \left( \frac{1}{2}-z \right) \rt \cdot \Delta} 
\\ 
= -A_Q \sqrt{\frac{2N_c}{m_{Q}}} \frac{e_{Q} e Q}{8\pi} K_0(\varepsilon' |\rt|) \equiv C_L K_0(\varepsilon'|\rt|),
\end{multline}
where
\begin{equation}
\varepsilon' = \sqrt{\frac{Q^2}{4}+m_{Q}^2}\;.
\end{equation}
We have also defined the notations
\begin{eqnarray}
C_T &\equiv& -A_Q \frac{\sqrt{2m_{Q}N_c}}{4\pi}e_{Q} e\;,  \label{eq:CT} \\
C_L &\equiv& -A_Q \sqrt{\frac{2N_c}{m_{Q}}} \frac{e_{Q} e Q}{8\pi}, \label{eq:CL}
\end{eqnarray}
for  the different coefficients for the transverse and longitudinal scattering amplitudes.

Now by plugging the wave function overlap \eqref{eq:WFTOverlapIntegrated},\eqref{eq:WFLOverlapIntegrated} and the dipole cross section average \eqref{eq:1PointDipFourier} into the definition of the amplitude \eqref{eq:VMamplitude}, we get
\begin{equation}
\begin{split}
&
\left| \left\langle \left\langle \mathcal{A}^{\gamma^{*}p \rightarrow Vp}_{T,L}(Q^2,\Delta) \right\rangle \right\rangle \right|^2
\\ &
=
C^2_{T,L}\frac{g^4 \mu_0^4 (N_c^2-1)^2 N_q^2}{(2\pi N_c)^2} 
\\ & \times
\exp \left( - \left[r^2_H + \left( \frac{N_q-1}{N_q} \right)R^2 \right] \Delta^2 \right)
\\ & \times
\Bigg[ 
\int d^2\rt
K_0( \varepsilon' |\rt|)
\\ & \times
\left\{ \cos \left( \frac{1}{2}\Delta \cdot \rt \right) \Psi \left( 0,\Delta,m^2 \right) - \Psi \left( \rt,\Delta,m^2 \right) \right\}
\Bigg]^2 .
\end{split}
\end{equation}
Since the coherent cross-section, as well as all color disconnected contributions to the incoherent cross-section give rise to the same contribution for the dipole size dependent part of the expressions, it is convenient to define the notation 
\begin{equation} \label{eq:CohIntegral}
\begin{split}
&
Z(\Delta,m^2,\varepsilon')
\equiv
\int d^2\rt
K_0( \varepsilon' |\rt|)
\\ & \times
\left\{ \cos \left( \frac{1}{2}\Delta \cdot \rt \right) \Psi \left( 0,\Delta,m^2 \right) - \Psi(\rt,\Delta,m^2) \right\}.
\end{split}
\end{equation}
In terms of this function  the coherent vector meson production cross-section in Eq.~(\ref{eq:CohCS}) is then given by
\begin{equation} \label{eq:CohAmplitudeSq}
\begin{split}
&
 \frac{d \sigma_{T,L}^{\gamma^*p\rightarrow Vp}}{dt} 
=
\frac{C^2_{T,L}}{16\pi}\frac{g^4 \mu_0^4 (N_c^2-1)^2 N_q^2}{(2\pi N_c)^2} 
\\ & \times
\exp \left( - \left[ r^2_H + \left( \frac{N_q-1}{N_q} \right) R^2 \right] \Delta^2 \right)
Z(\Delta,m^2,\varepsilon')^2 .
\end{split}
\end{equation}

Now we can clearly see the structure of this expression. It  is factorized into  a constant times a Gaussian in $\Delta$ times some function $Z$. The width of the Gaussian depends on the parameters $R$ and $r_H$ 
characterizing the sizes of the proton and the hot spot in the combination that we call the coherent radius \nr{eq:CohRadius}. In a naive   picture these would be the only parameters determining the $t$-dependence of the cross section. Here, however, this $t$-dependence is modified by the $Z$-function, which depends on  Coulomb tails of the color charges (through the IR regulator $m$), and the photon and vector meson wave functions. In  the fully non-relativistic limit the photon wave function forces the dipole size $\rt$ to be small. In this limit one can further evaluate the Feynman parameter and dipole size ($\rt$) integrals analytically. This limit is discussed in more detail in  Appendix \ref{app:smallR}.

Let us now turn to the incoherent cross section. 
We separate it into four distinct parts
\begin{multline} \label{eq:IncohCS2}
\frac{d \sigma_{T,L}^{\gamma^*p\rightarrow Vp^*}}{dt} = \frac{1}{16\pi}\Big(\left\langle \left\langle \left| \mathcal{A}^{\gamma^{*}p \rightarrow Vp}_{T,L}(Q^2,\Delta) \right|^2 \right\rangle \right\rangle 
\\ 
- \left|\left\langle \left\langle \mathcal{A}^{\gamma^{*}p \rightarrow Vp}_{T,L}(Q^2,\Delta) \right\rangle \right\rangle \right|^2 \Big)
\\
=
\left. \frac{d \sigma_{T,L}^{\gamma^*p\rightarrow Vp^*}}{dt} \right|_{1,DC}
+
\left. \frac{d \sigma_{T,L}^{\gamma^*p\rightarrow Vp^*}}{dt} \right|_{2,DC}
\\
+
\left. \frac{d \sigma_{T,L}^{\gamma^*p\rightarrow Vp^*}}{dt} \right|_{1,C}
+
\left. \frac{d \sigma_{T,L}^{\gamma^*p\rightarrow Vp^*}}{dt} \right|_{2,C}
,
\end{multline}
which correspond to the contributions from hot spot fluctuations with gluons coming from one (1,DC) or two (2,DC) hot spots, and the contributions due to color charge fluctuations with gluons from one (1,C) or two (2,C) hot spots. Due to its disconnected color structure, the square of the coherent amplitude is subtracted from the color disconnected contribution, and distributed proportionally in $N_q$ between the one and two hot spot contributions. Using the function $Z$ introduced for the coherent case in \eq\nr{eq:CohIntegral}, 
the contributions due to hot spot fluctuations can then be expressed as   
\begin{multline} \label{eq:OneHSDC}
 \left. \frac{d \sigma_{T,L}^{\gamma^*p\rightarrow Vp^*}}{dt} \right|_{1,DC}
= 
\frac{C^2_{T,L}}{16 \pi} \frac{g^4\mu_0^4(N_c^2-1)^2}{(2\pi N_c)^2}
N_q 
\\  \times
\left[
\exp \left( -\Delta^2r_H^2 \right)
- \exp\left( - \left[ r^2_H + \left( \frac{N_q-1}{N_q} \right) R^2 \right] \Delta^2 \right)
\right]
\\  \times
Z(\Delta,m^2,\varepsilon')^2
\end{multline}
and
\begin{multline} \label{eq:TwoHSDC}
 \left. \frac{d \sigma_{T,L}^{\gamma^*p\rightarrow Vp^*}}{dt} \right|_{2,DC}
=
\frac{C^2_{T,L}}{16 \pi} \frac{g^4\mu_0^4(N_c^2-1)^2}{(2\pi N_c)^2}
N_q(N_q-1) 
\\  \times
\bigg[
\exp \left( -\Delta^2 \left[ R^2+r_H^2 \right] \right)
\\  - \exp\left( - \left[ r^2_H + \left( \frac{N_q-1}{N_q} \right) R^2 \right] \Delta^2 \right)
\bigg]
Z(\Delta,m^2,\varepsilon')^2 .
\end{multline}

The contributions due to color charge fluctuations have a more complicated structure due to the two non-factorizing color charge two-point correlators, and we were not able to find an equally simple form for them as we have for the function $Z$. For these contributions we are forced to leave the two transverse momentum integrals for numerical evaluation. Let use define another auxiliary notation for these integrals as:
\begin{equation} \label{eq:IncohIntegral}
\begin{split}
&
K(A,\Delta,m^2,\varepsilon'^2) \equiv 
\\ &
\int d^2\kt d^2 \bar{\kt} \frac{\exp \left(-A \left( \kt+\bar{\kt} \right)^2 \right)}{
( \left(\kt+\frac{\Delta}{2} \right)^2+m^2)
( \left(\kt-\frac{\Delta}{2} \right)^2+m^2)}
\\ & \times
\frac{1}{
( \left(\bar{\kt}+\frac{\Delta}{2} \right)^2+m^2)
( \left(\bar{\kt}-\frac{\Delta}{2} \right)^2+m^2)
}
\\ & \times
\left( 
\frac{1}{\varepsilon'^2 + \frac{\Delta^2}{4}} - \frac{1}{\varepsilon'^2 + \kt^2}
\right)
\left( 
\frac{1}{\varepsilon'^2 + \frac{\Delta^2}{4}} - \frac{1}{\varepsilon'^2 + \ktbar^2}
\right) .
\end{split}
\end{equation}
Using this notation the contributions due to color charges fluctuations can be expressed in a similarly compact form as
\begin{equation} \label{eq:OneHSConn}
\left.  \frac{d \sigma_{T,L}^{\gamma^*p\rightarrow Vp^*}}{dt} \right|_{1,C}
=
\frac{C^2_{T,L}}{16 \pi}\frac{g^4 \mu_0^4 (N_c^2-1)}{2\pi^2 N_c^2} N_q
K(r_H^2,\Delta,m^2,\varepsilon'^2)
\end{equation}
and
\begin{multline} \label{eq:TwoHSConn}
\left.  \frac{d \sigma_{T,L}^{\gamma^*p\rightarrow Vp^*}}{dt} \right|_{2,C}
=
\frac{C^2_{T,L}}{16 \pi}\frac{g^4 \mu_0^4 (N_c^2-1)}{2\pi^2 N_c^2} N_q(N_q-1)
\\ \times K(R^2+r_H^2,\Delta,m^2,\varepsilon'^2).
\end{multline}

Before we proceed to evaluate the cross section numerically, some important comments are in order. We first note that in the limit of a single hot spot $N_q=1$, the color disconnected (DC) contributions vanish as they should. This is due to the fact that for $N_q=1$ the only hot spot is always at the center of mass of the proton, and thus there are no hot spot fluctuations. In this case all contributions to the incoherent cross section come from the color charge fluctuations associated with the color connected (C) contributions. We further note that at first sight the two hot spot disconnected (``2,DC'') contribution might  seem like an uncorrelated contribution that should not influence the incoherent cross section since it is disconnected in both color and the hot spot averaging. However in reality it is correlated through the fixed center of mass of the hot spot system and thus contributes to the incoherent cross-section. 

By closer inspection of Eqns.~\eqref{eq:OneHSDC} and \eqref{eq:TwoHSDC} one finds that, similarly to the coherent cross section  \eq\nr{eq:CohAmplitudeSq}, the  color disconnected contributions to the incoherent cross section factorize into two parts. There are firstly parts that are Gaussian in $\Delta$, depending on the parameters of the hot spot system. The $t$-dependence is then modified by the $\Delta$-dependence of the residual part of the cross section $Z(\Delta,m^2,\varepsilon')^2$,  which also depends on the IR regulator $(m)$ and the properties of the probe $(\varepsilon')$. In contrast, the color connected contributions to the incoherent cross section in Eqns.~\eqref{eq:OneHSConn} and \eqref{eq:TwoHSConn} do not factorize like this except for the $-t \rightarrow \infty$ limit with the small dipole size approximation that is true in the non-relativistic quark limit. We again discuss the small dipole size approximation of these results in Appendix \ref{app:smallR}.

\begin{figure*}[tbh]
\centering
\includegraphics[width=0.49\textwidth]{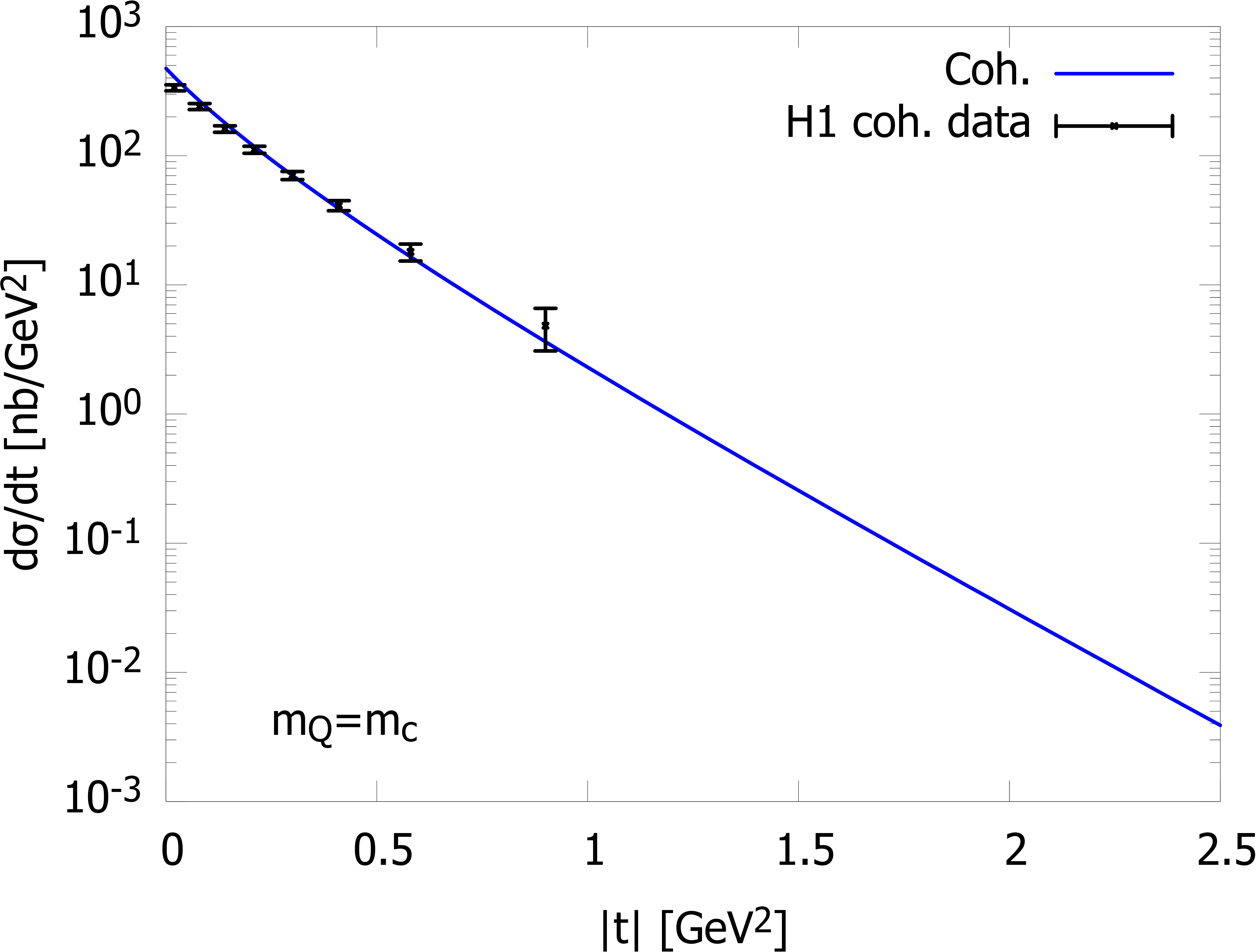}
\includegraphics[width=0.49\textwidth]{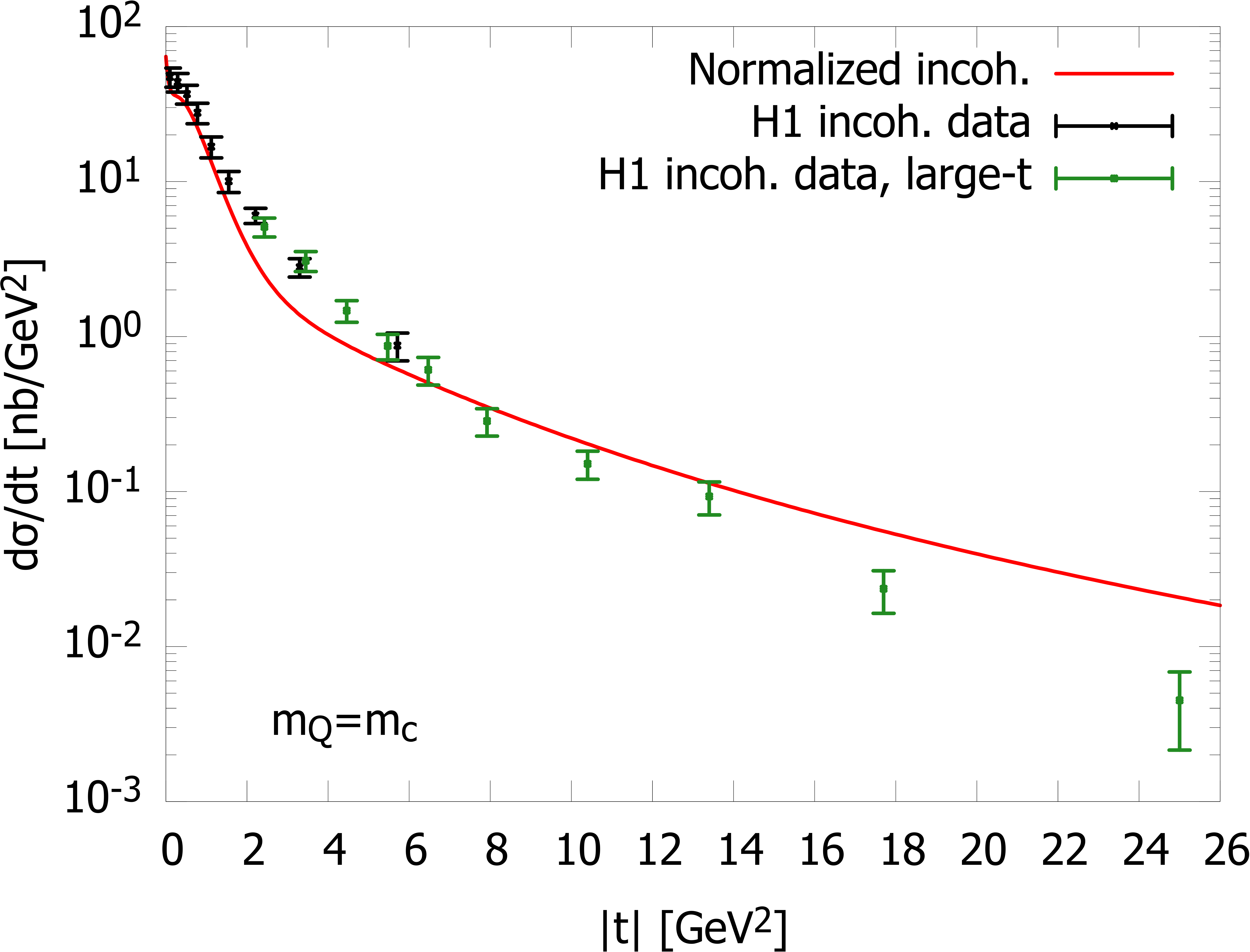}
\caption{Coherent and incoherent exclusive $J/\Psi$ production cross sections compared to HERA data. The normalization of the incoherent cross section has been increased by a factor of 2.5 as explained in the text. The H1 coherent data is from \cite{H1:2013okq}, the lower-$|t|$ incoherent data is also from \cite{H1:2013okq} and the higher-$|t|$ data is from \cite{H1:2003ksk}. Note that the different incoherent data sets have a slightly different $Q^2$ and a different center of mass energy range. Note  also that the $t$ range for the incoherent cross section data goes much further than in most works except for at least \cite{Kumar:2021zbn}.}
\label{fig:CSNormalizedFits}
\end{figure*}

\section{Analysis of results} \label{sec:Analysis}

In this section we will study the coherent and incoherent cross sections and their dependence on the model parameters.

If not stated otherwise, we will be using the following set of default parameters, where the proton size parameter $R=\sqrt{3.3}\text{GeV}^{-1}$, the hot spot radius $r_H=\sqrt{0.7}\text{GeV}^{-1}$, the IR regulator $m=0.22\text{GeV}$, the number of hot spots $N_q=3$ and the photon virtuality $Q^2=0.1\text{GeV}^{2}$. The radii are the same as the ones used in \cite{Mantysaari:2016jaz}, the IR regulator was chosen to be of the order of the QCD scale and the number of hot spots was motivated by the number of valence quarks in a proton. For the charm quark we use the mass of $m_c=1.275\text{GeV}$, for the bottom quark $m_b=4.18\text{GeV}$ and for the top quark we use $m_t=173\text{GeV}$. 

The full cross section measured experimentally in electron-proton scattering can be approximately expressed as the sum of the transverse and longitudinal cross sections as \cite{Kowalski:2006hc}
\begin{equation}
\sigma \approx \sigma_T + \sigma_L.
\end{equation}
In Fig.~\ref{fig:CSNormalizedFits} we compare  the predictions of our model for the coherent and incoherent exclusive $J/\Psi$ cross sections  to HERA data. We note that our model has a free normalization parameter $g^4\mu_0^4$ which can be adjusted e.g. to the coherent cross section at low momentum transfer $|t|$. However, we find that the model underpredicts the ratio of the incoherent to the coherent cross section. This clearly points to the absence of a necessary additional physical source of fluctuations, operative over a broad range in $t$, from our model. In order to better compare the $|t|$-dependence in spite of this mismatch, we have separately adjusted the normalization of the incoherent cross section in Fig.~\ref{fig:CSNormalizedFits}, increasing it by a factor of 2.5 relative to the coherent cross section, to better match the experimental data. While the relative normalization does not work as well as one could have hoped, by normalizing the two cross sections separately, the model seems to reproduce the general features of $|t|$-dependence of the data rather well. The normalization mismatch prevents us from performing a fit of the parameters of our model to experimental data. Instead, we will now focus on investigating further how the different features of the $t$-dependence emerge and how they depend on the physical parameters.

\begin{figure*} 
\centering
\includegraphics[scale=0.135]{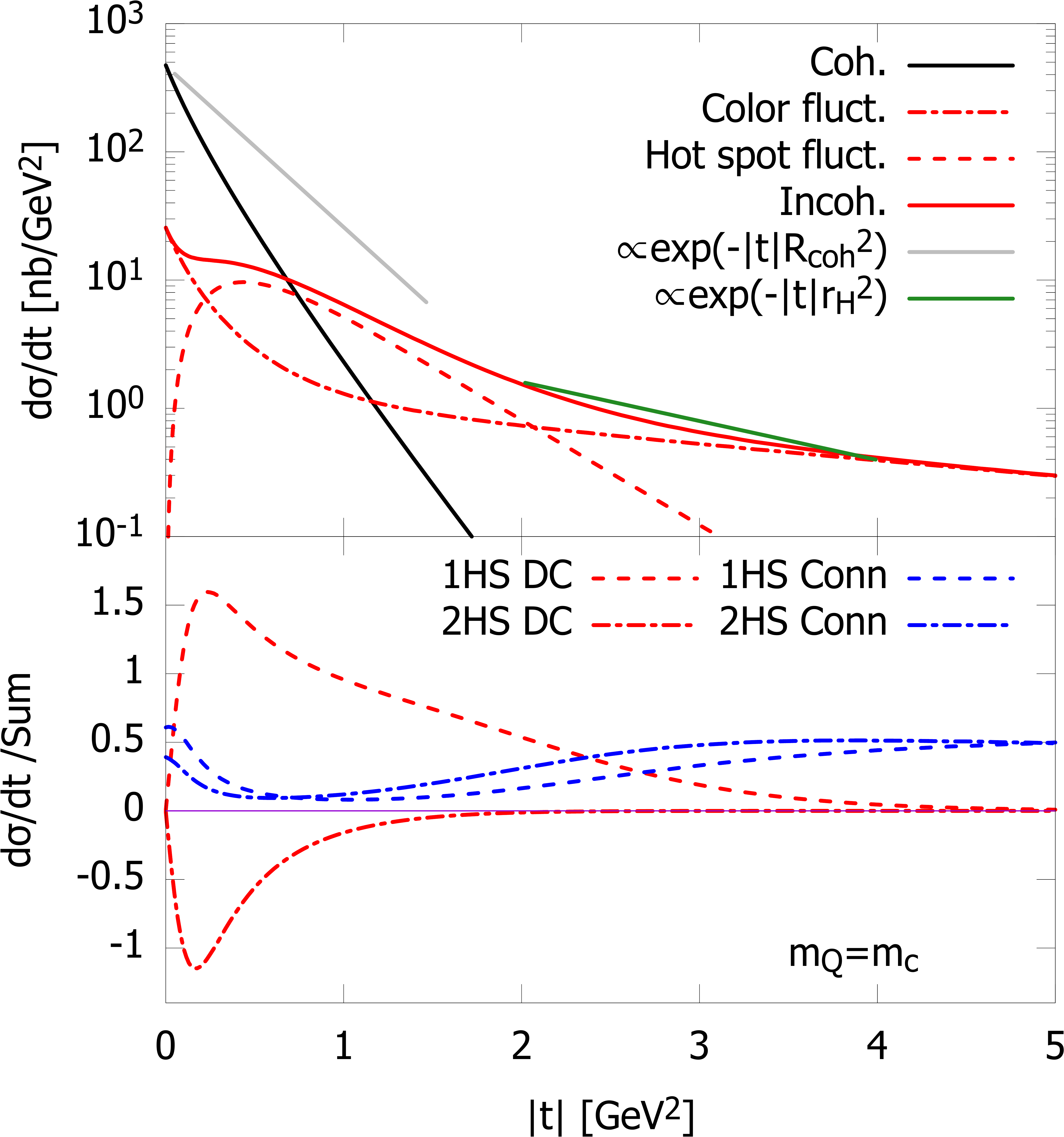}
\includegraphics[scale=0.135]{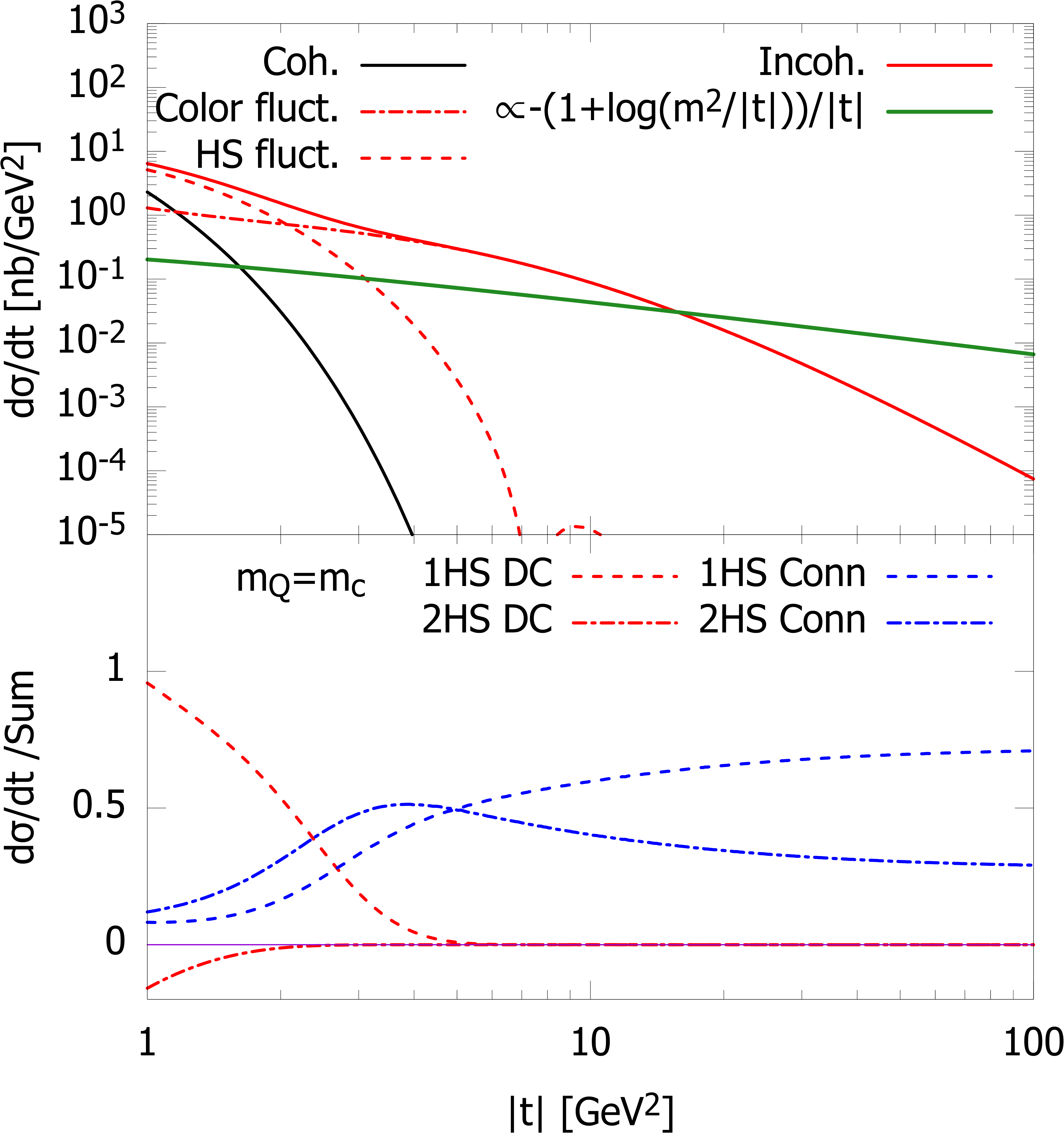} \\ 
\includegraphics[scale=0.135]{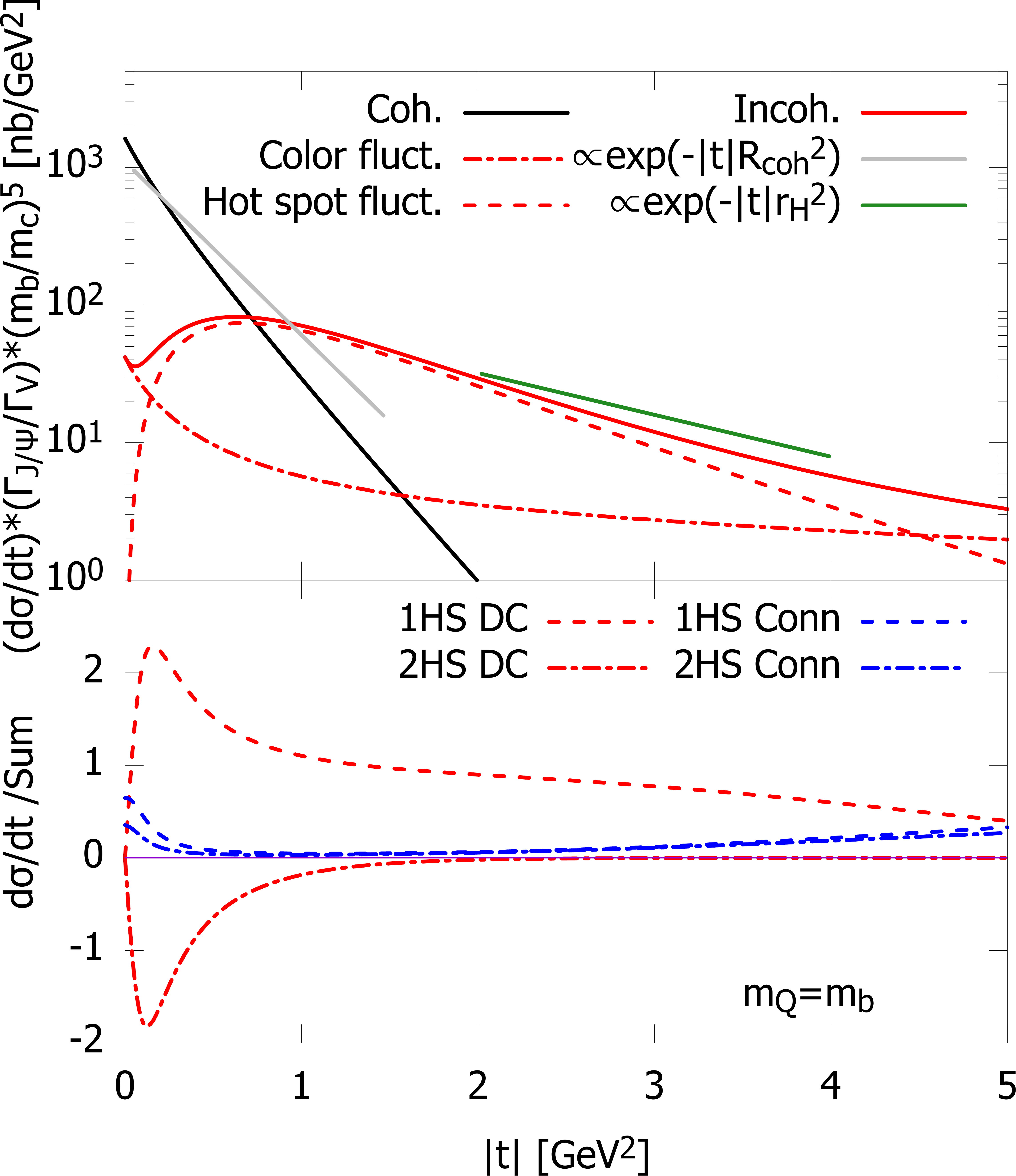}
\includegraphics[scale=0.135]{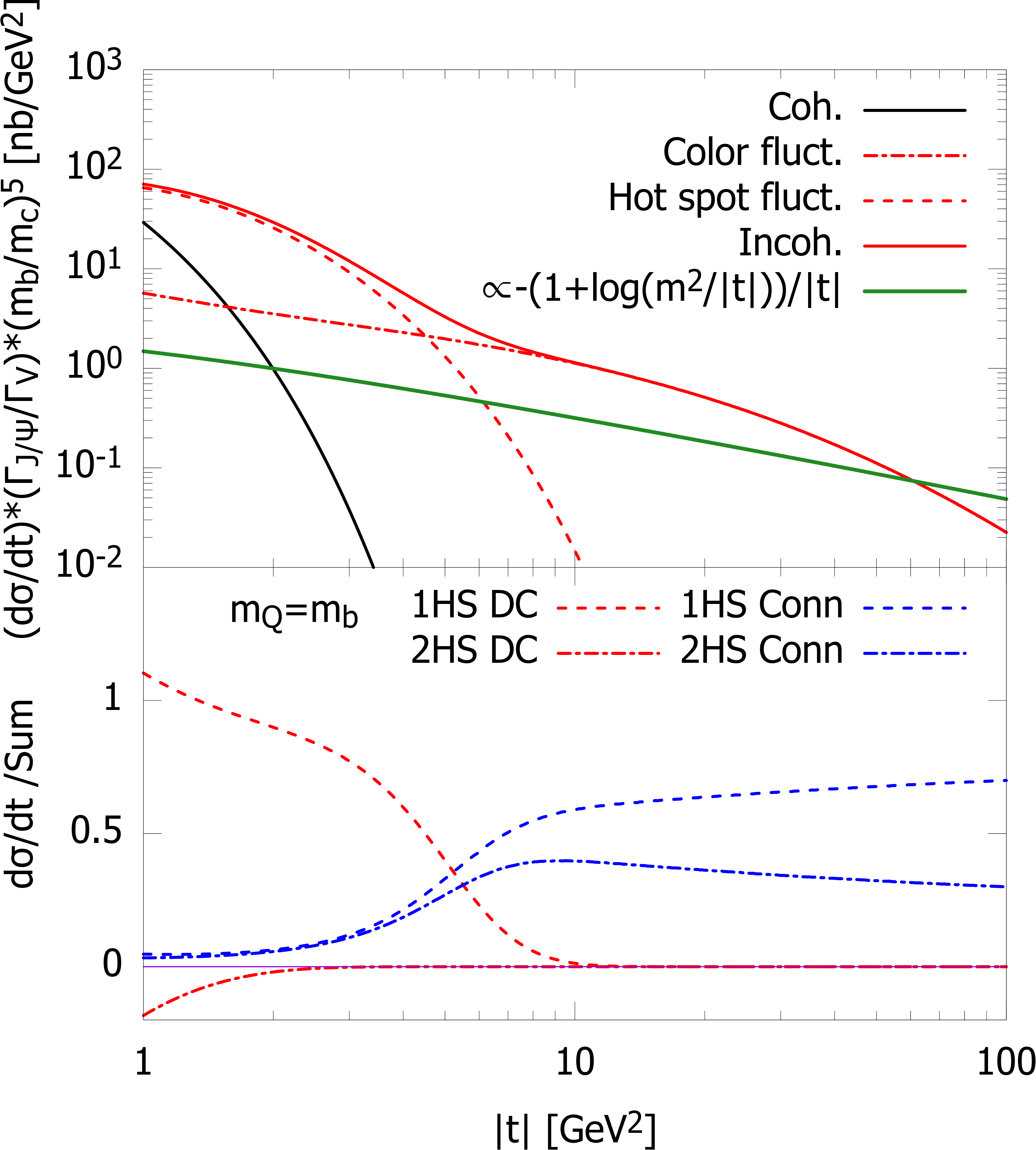} \\ 
\includegraphics[scale=0.135]{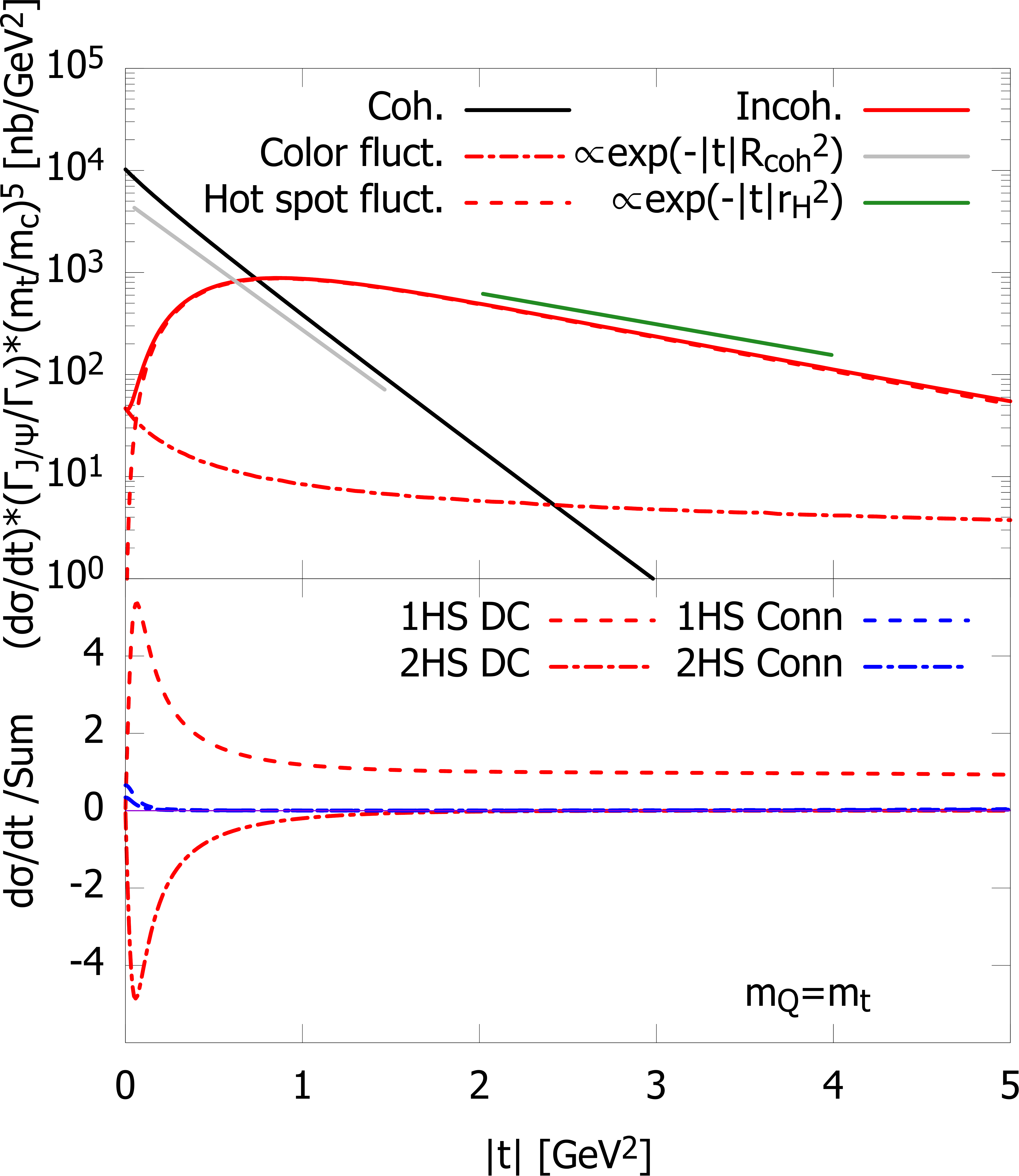}
\includegraphics[scale=0.135]{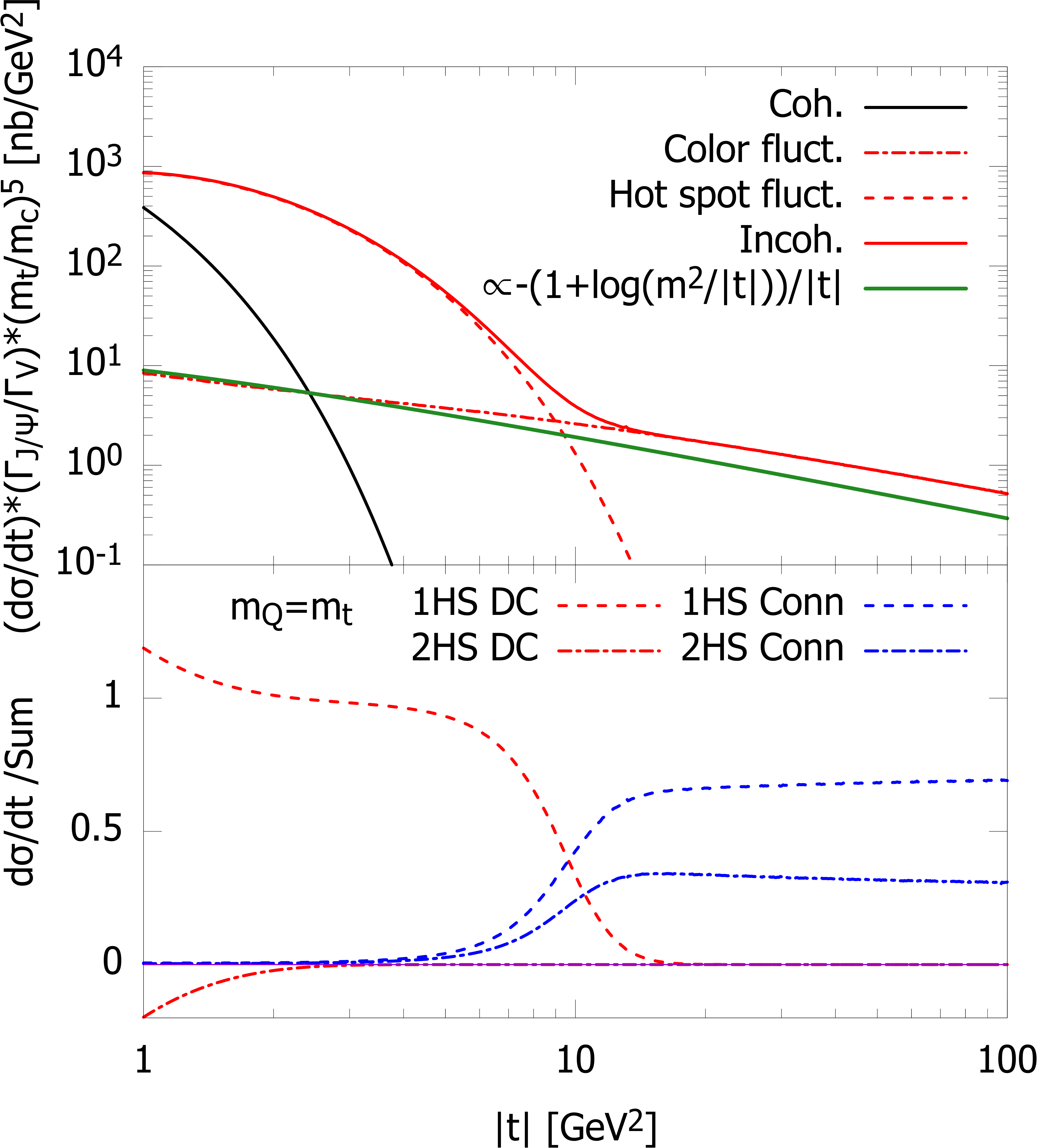}
\caption{Coherent and incoherent cross sections as a function of $t$ for $c\bar{c}$ (top panel), $b\bar{b}$ (middle panel) and $t\bar{t}$ (bottom panel). Cross-sections for the $b\bar{b}$ and $t\bar{t}$ mesons $V$ are scaled as $(d\sigma/dt)~(\Gamma_{J/\Psi}/\Gamma_{V})~(m_{Q}/m_c)^5$
to absorb the leading quark mass dependence. 
We also show a breakup of the incoherent cross section into different contributions.
The top plots in each subfigure show the coherent cross section and the incoherent cross section with an additional separation of the incoherent cross section into hot spot and color fluctuation contributions. The bottom plots of the subfigures show the relative contribution of each part of the incoherent cross section to the total incoherent cross section. Also shown is the  expected behavior of the cross sections in specific  ranges of $|t|$, as discussed in the text.
}
\label{fig:CSParts}
\end{figure*}

In Fig.~\ref{fig:CSParts} one can see plots of the cross sections  with our default parameters for the three quarks masses. The figures in the top row show the $c$-quark, the middle row the $b$-quark and the bottom row the $t$-quark.  The left-hand side plots have a linear scale in $t$ and the right-hand side plots   a logarithmic one.  We show both in order to highlight different aspects of the cross sections. The top panels of the subfigures show the total coherent and the incoherent cross sections, where  -- unlike in Fig.~\ref{fig:CSNormalizedFits} --  the relative normalization of the coherent and incoherent cross sections is the same. However we have scaled the cross sections with the ratio of the decay widths, and by the quark masses to the fifth power. This scaling cancels out the leading quark mass dependence in the fully non-relativistic limit, which corresponds to the small dipole size approximation discussed in App.~\ref{app:smallR}. This limit can also be seen, up to a logarithmic dependence on the mass, from the coherent cross  section at $t=0$ without an expansion in $\rt$, as shown in App.~\ref{app:Delta0Coh}.
Scaling with the decay width also fully cancels out the dependence on the wave function normalization constant $A_q$ (see \eq\nr{eq:decaywidth}) so that we do not need to pick any specific value for $A_q$.

The total incoherent cross section is further separated into its color and hot spot fluctuation contributions. These can be seen in the top panels as well. The bottom panels show the relative contributions of all four different parts of the incoherent cross section to the full incoherent cross section. The two hot spot color disconnected contribution becomes negative due to the way we subtract the coherent cross section proportionally in $N_q$. However the sum of all contributions is always equal to unity, which is why the one hot spot color disconnected contribution can be larger than the total incoherent cross section. 

We have also added lines to Fig.~\ref{fig:CSParts}  representing the  behavior of the cross sections that could be expected from a non-relativistic heavy quark limit in specific ranges in $t$. The normalization of these lines is adjusted by hand for better visibility. Plots using the actual normalization constants of the limits are shown in Fig.~\ref{fig:CSAsymptotics} in Appendix \ref{app:smallR}. 
The non-relativistic expectations shown in the figure are the following ones:
\begin{itemize}
\item By looking at the coherent cross section \eq\nr{eq:CohAmplitudeSq}, one would be tempted to neglect the dependence of the  function $Z$ on $\Delta$. Thus, to a first approximation the coherent cross section at small $t$ is expected to behave as
\begin{equation}
 \frac{d \sigma_{T,L}^{\gamma^*p\rightarrow Vp}}{dt} \propto \exp(R_C^2 t),
\end{equation}
with the coherent radius given by \eq\nr{eq:CohRadius}. This behavior, however, receives corrections from the long range Coulomb tails regulated by $m$. These corrections to the slope, as discussed in Appendix~\ref{app:DeltaCohApproximations0AndInfty},  are only extremely slowly suppressed in the non-relativistic limit, as $\sim 1/\ln m_Q$. 
\item Similarly, the hot spot fluctuation part is expected to dominate the cross section in the range $ -t \sim 1/r_H^2 \gg 1/R_C^2$ where one is probing distance scales of the hot spot radius, but not yet the very smallest scale color charge fluctuations. Here the first term of \eq\nr{eq:OneHSDC} dominates and one may expect 
\begin{equation}
 \frac{d \sigma_{T,L}^{\gamma^*p\rightarrow Vp^* }}{dt} \propto \exp(r_H^2 t).
\end{equation}
\item Ultimately, at very large $t$ color charge fluctuations dominate the incoherent  cross section. In this region the expectation in the non-relativistic limit takes the form 
\begin{equation}
 \frac{d \sigma_{T,L}^{\gamma^*p\rightarrow Vp^* }}{dt} \propto \frac{1+\ln\frac{m^2}{-t}}{t},
\end{equation}
resulting from the power-law Coulomb tails of the color fields around the color charges, regulated by $m$  as discussed in more detail in  Appendix \ref{app:LargeT} (see \eq\nr{eq:larget}).
\end{itemize}

Now let us discuss what we see in the plots of Fig.~\ref{fig:CSParts}. 
The coherent cross section is much larger than the incoherent cross section at small values of $t$. This is due to coherent cross section measuring averages of the system which  are smoother as functions of the transverse position $\bt$ that is the Fourier conjugate to $\Delta=\sqrt{-t}$.  The incoherent cross section on the other hand measures fluctuations of the system and is therefore a mix of exponentially decaying hot spot fluctuation parts and color fluctuation parts that decay approximately like a power law in $t$. This is also why the incoherent cross section is dominated by color fluctuations at both very small and large $t$. The influence of the different contributions to the incoherent cross section as a function of $t$ can best be seen in the bottom panels. Here we clearly see that, at intermediate values of $t$ the hot spot fluctuations dominate. Then the color fluctuation contributions start to take over at large $t$, where one starts to resolve the color charge fluctuations inside the target. 

Finally let us discuss what changes as we change the mass of the heavy quark. Let us start from our non-relativistic limit of the top quark. Here our dipole is truly small due the large $\varepsilon' \sim m_{Q}$ in the photon wave function. We see that in this case the analytic expectations in the non-relativistic limit are well satisfied, apart from the change in slope at small $t$ caused by the Coulomb tails (see discussion in Appendix~\ref{app:DeltaCohApproximations0AndInfty}, \eq\nr{eq:Coht0}). This is due to the fact that a large quark mass renders the $Z$-function almost independent of $t$.
However, when we study the phenomenologically relevant cases of the bottom and the charm quark, we see that this is not the case. Especially the large $t$ tail of the incoherent cross section $(-[1+\log(m^2/-t)]/-t)$  does not describe the cross section for the charm and the bottom at all, which means that the large-$t$ and small-$\rt$ approximation is not valid for realistic quark masses. Similarly, also the exponential behavior of the coherent and  incoherent cross section at small to moderate $t$ is not entirely determined by the properties of the target. Instead, the $Z$-function that describes the interaction of the target with a color dipole dominates the $t$-dependence. Due to this discrepancy, the slope of the coherent cross section does not have a clear interpretation as a measure of the overall size of the proton as one might expect. 
This is one of the main results of this paper.  The cross sections for realistic quark masses do not only measure the target but some more involved convolution of the target and probe structures.

\begin{figure*} 
\centering
\includegraphics[height=0.25\textheight]{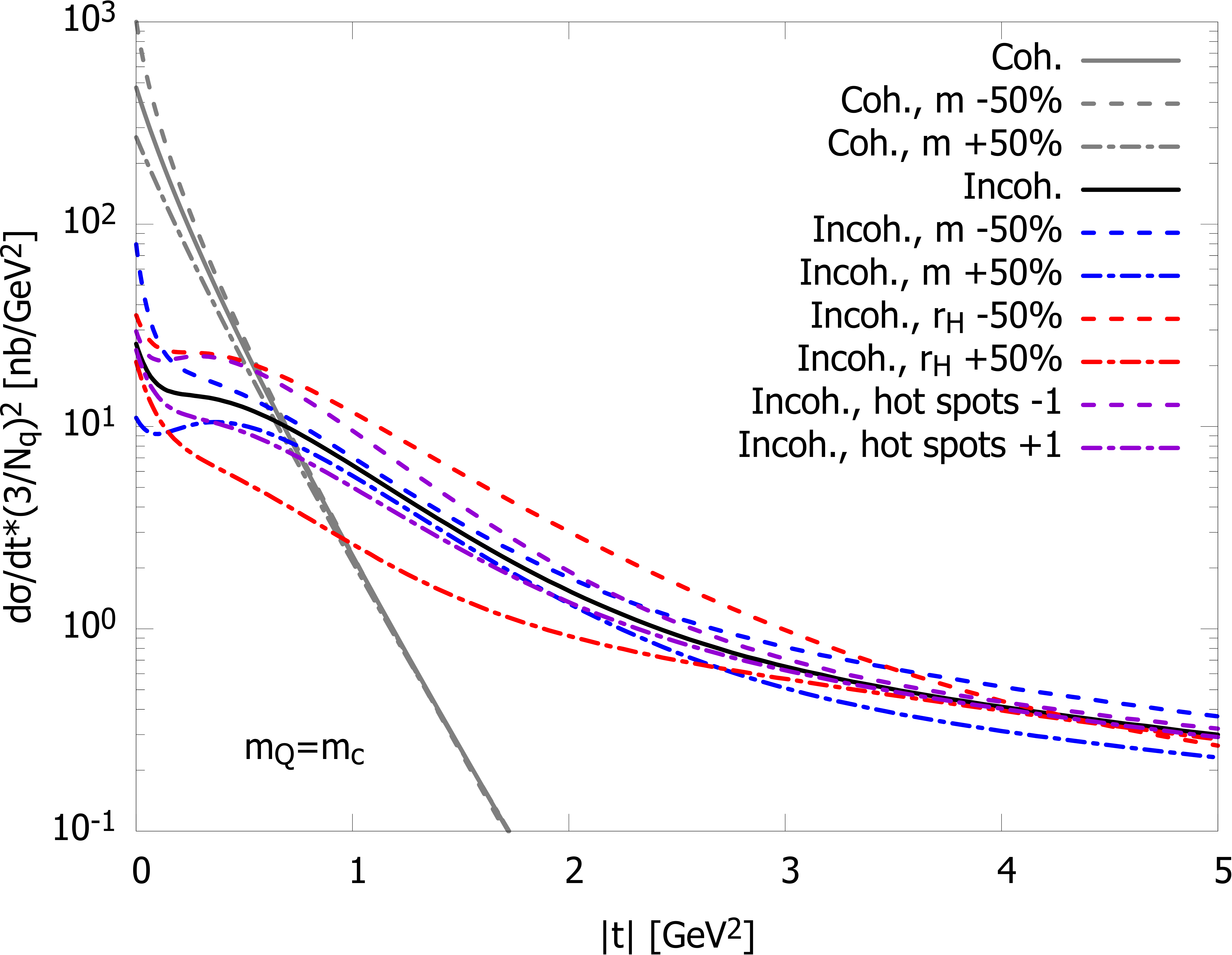}
\includegraphics[height=0.25\textheight]{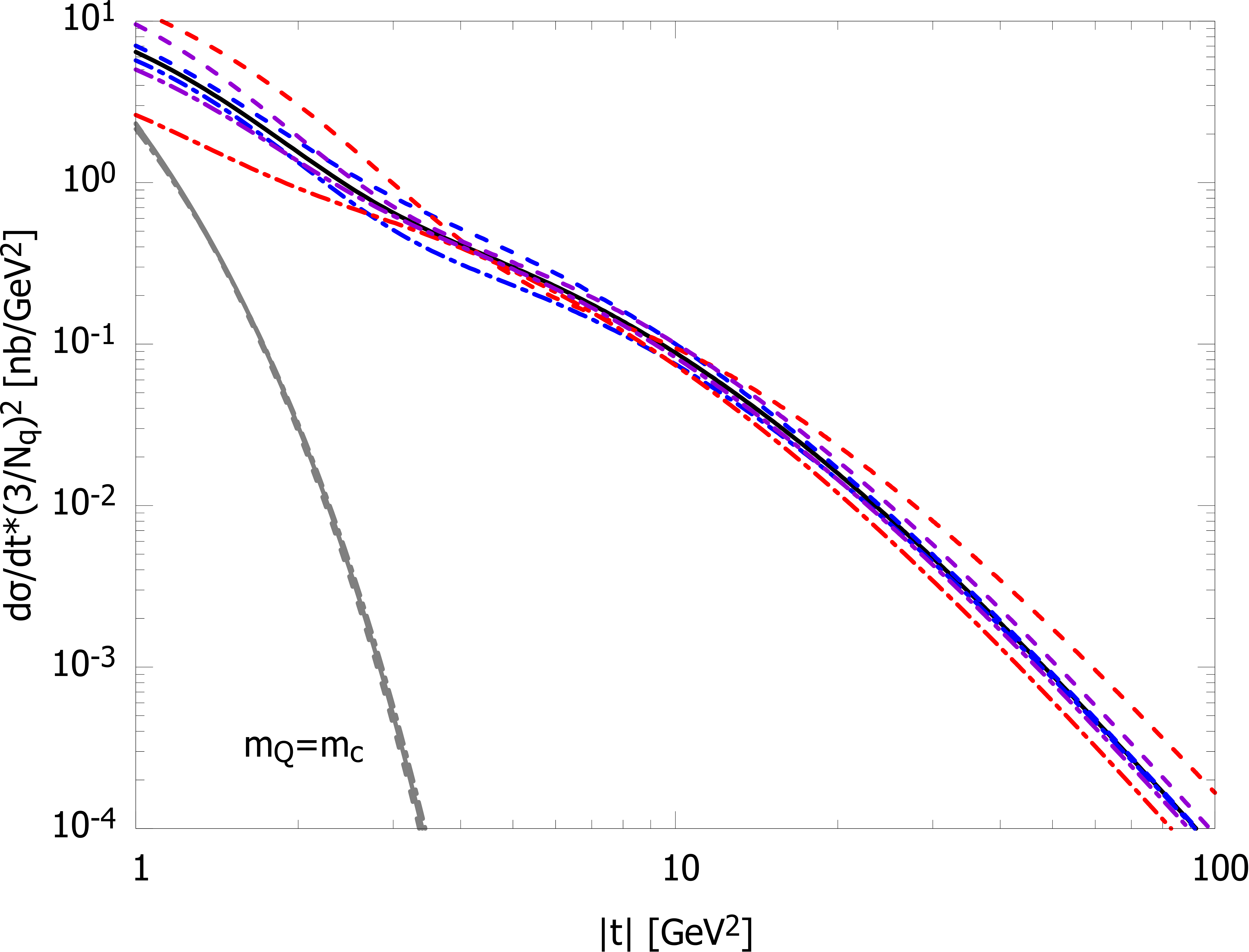} \\
\includegraphics[height=0.25\textheight]{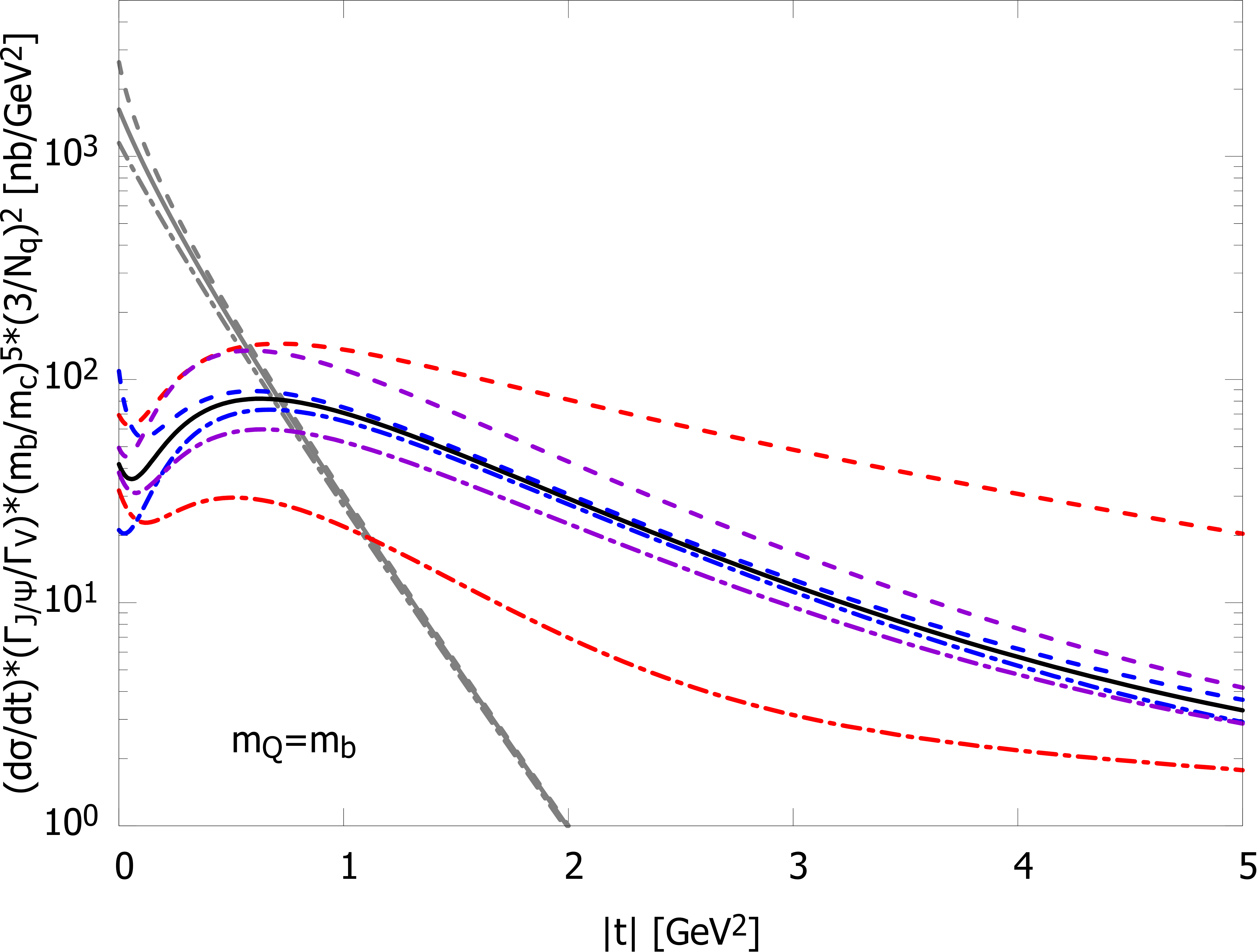}
\includegraphics[height=0.25\textheight]{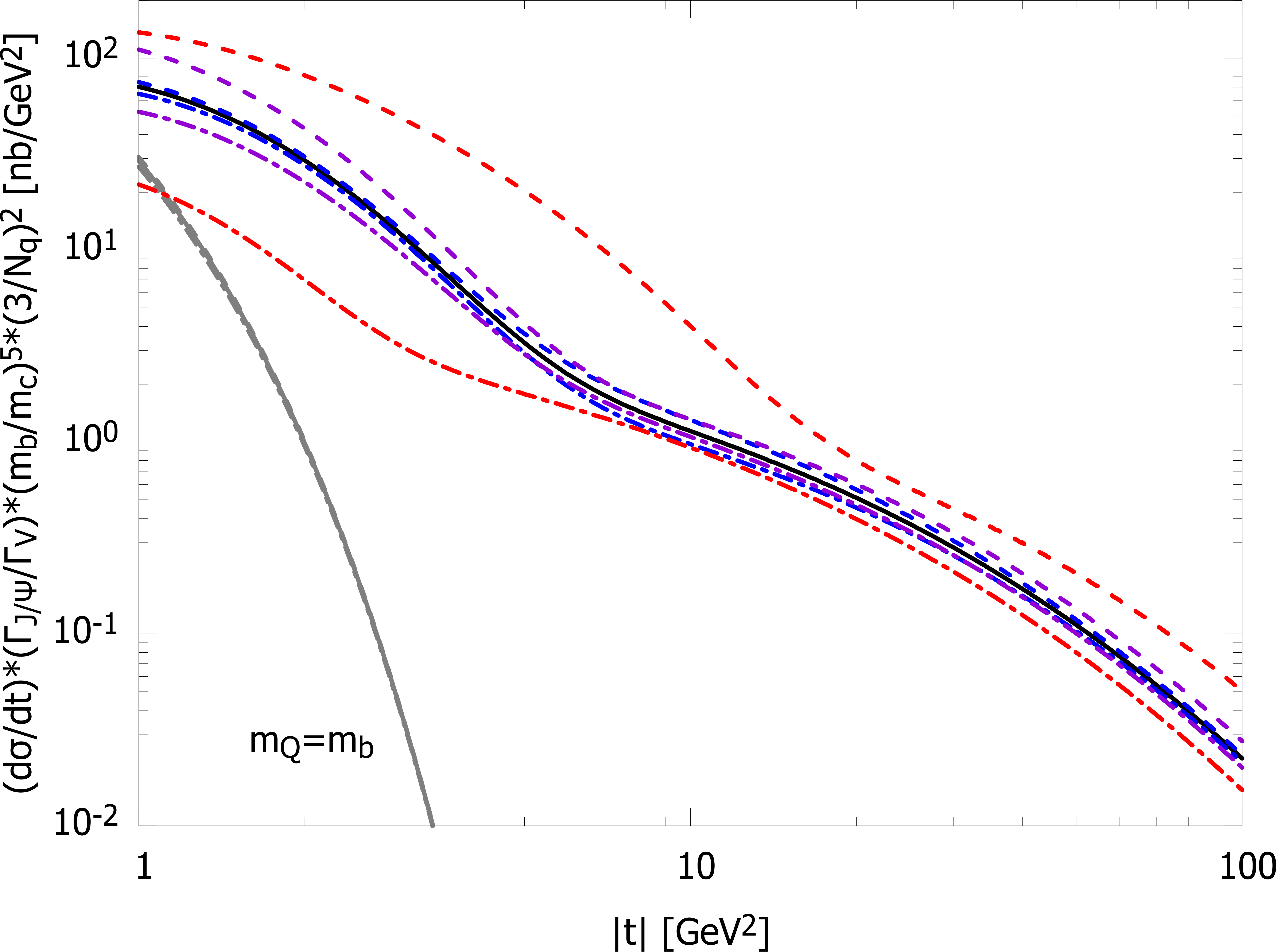} \\
\includegraphics[height=0.25\textheight]{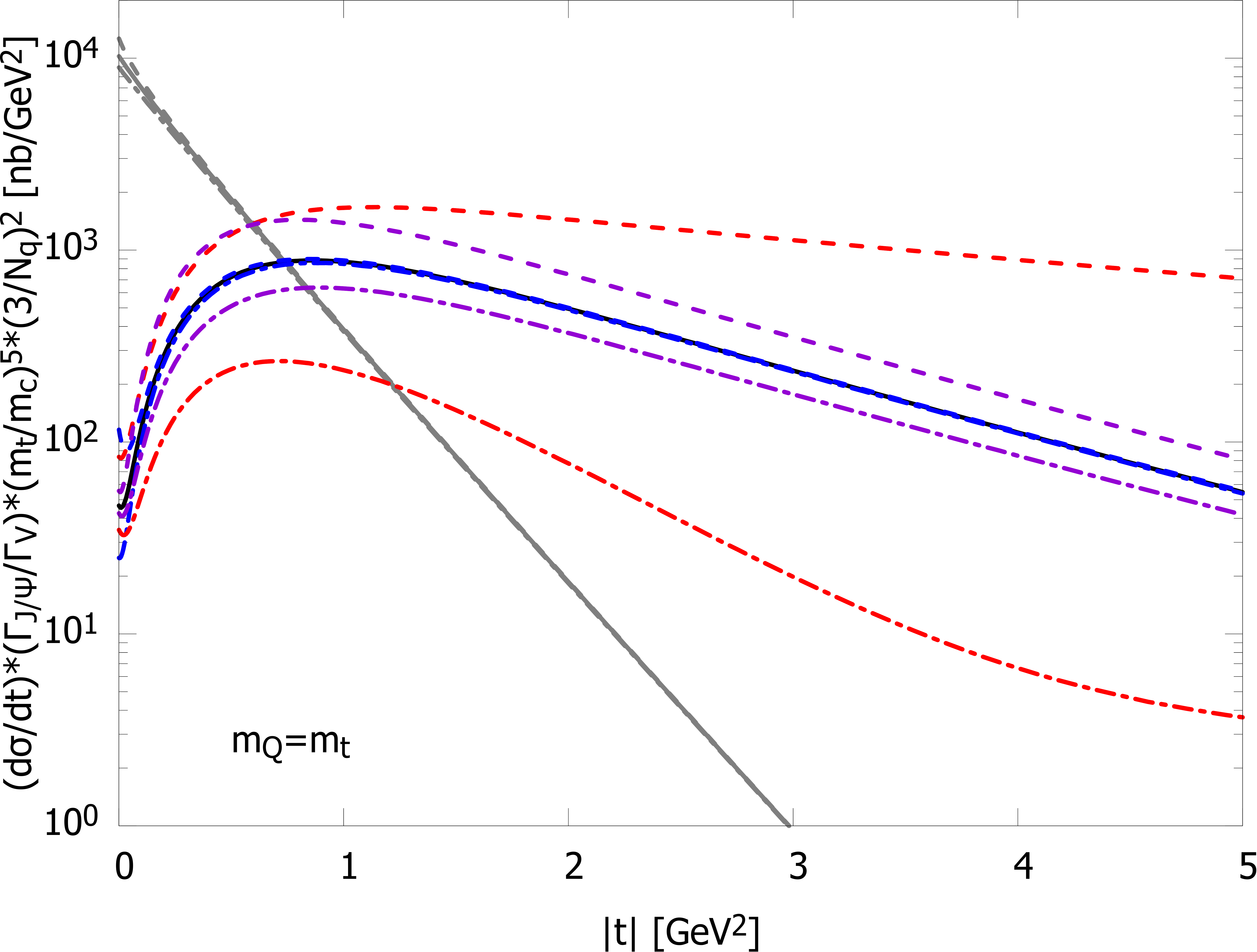}
\includegraphics[height=0.25\textheight]{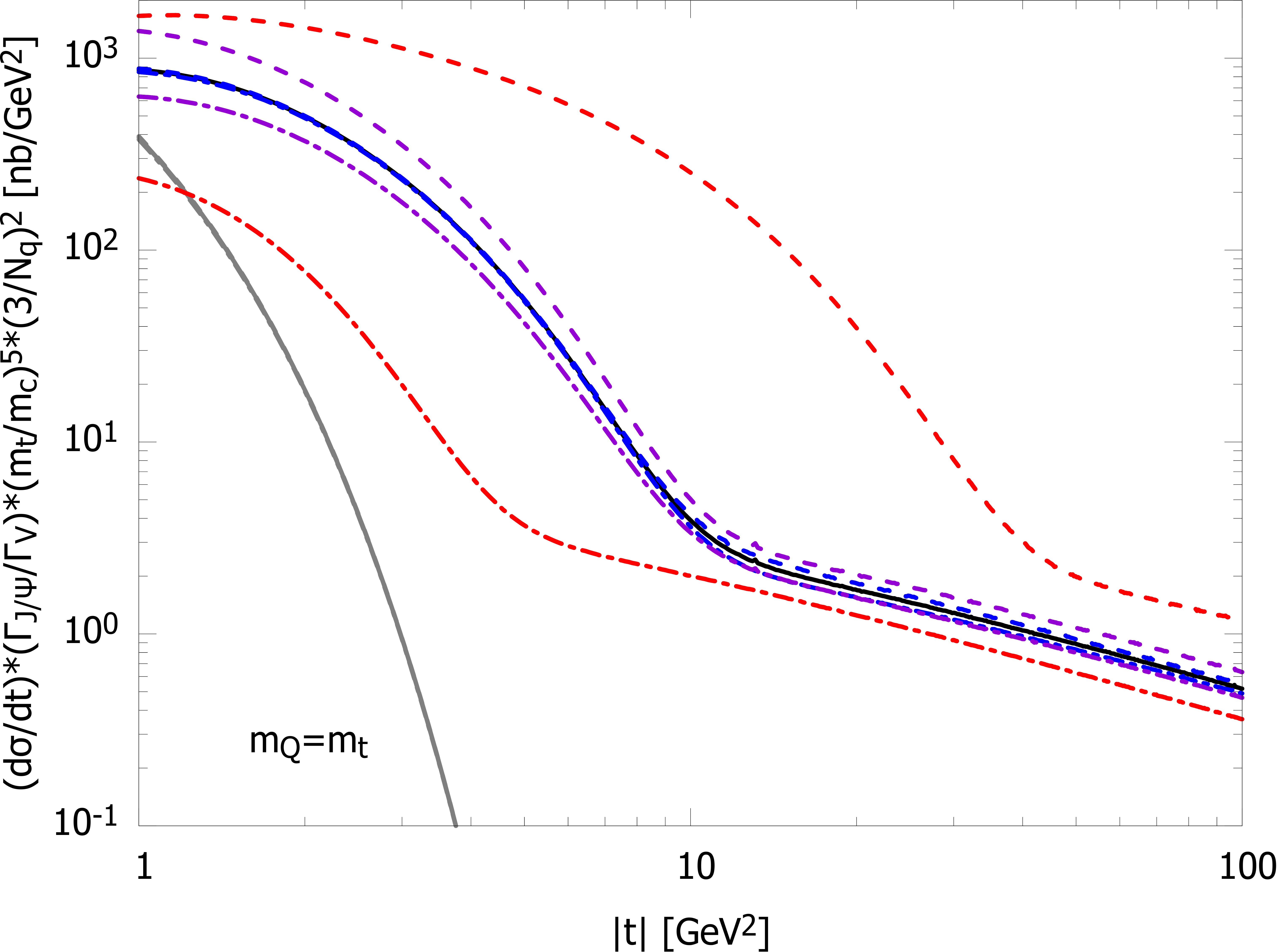}
\caption{Effect of variations of the model parameters $m,r_{H}$ and $N_q$ on the $t$-dependence of the coherent and incoherent cross sections for $c\bar{c}$ (top panel), $b\bar{b}$ (middle panel) and $t\bar{t}$ (bottom panel). Cross-sections are scaled as $(d\sigma/dt)~(\Gamma_{J/\Psi}/\Gamma_{V})~(m_{Q}/m_c)^5~(3/N_q)^2$ to absorb the leading $m_{Q}$ and $N_q$ dependence. 
The legend in the top-left plot is valid for all the other plots. 
The parameter variations are made in such a way that the coherent radius $R_C$ is kept fixed, as discussed in the text. 
}
\label{fig:CSParameters}
\end{figure*}

Now let us move to the plots in Fig.~\ref{fig:CSParameters}. Here we vary our model parameters $m,r_H$ and $N_q$ that describe the hot spot structure of the proton to see how the cross sections depend on them. In all cases, the plots are made so that the coherent radius $R_C$  defined in \eq\nr{eq:CohRadius} is kept constant. 
Thus, when we vary  the hot spot radius $r_H$, the  proton size parameter $R$ is changed to compensate.
Similarly, when we vary the number of hot spots $N_q$, we keep the hot spot radius $r_H$ fixed and vary the proton size parameter $R$ to keep $R_C$ fixed. 
With a constant coherent radius $R_C$  the only parameter affecting the $t$-slope of the coherent cross section is the IR regulator $m$, which affects the $t$-distribution via the $Z$-function.  We also normalize all cases to have the same amount of color charge, which in practice means that we add a normalizing factor of $\frac{3}{N_q}$ in front of the sum over the hot spots in the color charge density two point correlator in Eq.~\eqref{eq:rhorho}.  This ensures that the proton always has the same overall amount of color charges as in the case of $N_q=3$. We thus remove the trivial effect of the $N_q$ variation on the normalization of the cross section.
We do not plot separately a variation of the proton size parameter $R$, since this case would either result in a variation of the coherent radius $R_C$, or when keeping $R_C$ fixed is already included in the other cases.

We see that the larger the quark mass is, the smaller the effect of varying the IR regulator $m$ has on the cross sections. This can be understood by noting that a large quark mass forces the dipoles to be small. In terms of equations, we  can see that taking the small-$r$ limit and expand at large $-t/m^2$ (i.e. small $m$), the $Z$-function goes as $\log(\varepsilon' / t)$ with corrections of $O(m^2)$ (see Eq.~\eqref{eq:CohSmallRLargeTSmallM} in Appendix \ref{app:DeltaCohApproximations0AndInfty}). Thus, for small dipoles, the cross section becomes independent of $m$ in the limit $-t\gg m^2$.
For the $c$ and $b$ quarks, however, the IR regulator $m$  has a significant effect on the coherent cross section at small $t$. Overall in the hot spot picture it is not very straightforward to interpret the $t$-dependence of the coherent cross section in terms of the spatial distribution of color charges, because one is very sensitive to the Coulomb tail of the field around them. On the other hand, the $t$-dependence of the incoherent cross section is, perhaps surprisingly, not very sensitive to the IR cutoff $m$ apart from the behavior at very small values of $t$. 

Varying the hot spot radius, in the charm quark case, does not really change the slope of the medium-$|t|$ incoherent cross section. This is due to the dominance of the $Z$-function  in this region as discussed before. The hot spot radius does, however, have an effect on the normalization. But when looking at the heavier quarks, we can see that eventually one recovers the non-relativistic expectation, where the  the slope changes with a changing $r_H$. This effect is especially noticeable for the top quark case.

Now let us discuss the variation of $N_q$. By normalizing the amount of color charge in the proton with the normalization factor $(3/N_q)^2$ and by changing the proton size parameter $R$ to keep the coherent radius $R_C$ fixed, we keep the coherent cross section unchanged when varying $N_q$. For the incoherent cross section a variation in $N_q$ most importantly means a change in the $N_q$ weighting between the one and two hot spot contributions. After the normalization by $(3/N_q)^2$, the coefficient of the one hot spot contributions \eqref{eq:OneHSDC}, \eqref{eq:OneHSConn} goes as $\propto 1/N_q$ and the coefficient of the two hot spot contributions \eqref{eq:TwoHSDC}, \eqref{eq:TwoHSConn}  as $\propto (N_q-1)/N_q$. Thus as $N_q$ increases, the one hot spot contributions go down and the two hot spot contributions go up. However, as we see in Fig.~\ref{fig:CSParts} for $N_q=3$, the one hot spot contribution is dominant at all values of $t$. Correspondingly, the incoherent cross section always increases with decreasing $N_q$ and vice versa. The increase is not fully uniform in $t$, however, because of the varying secondary effect from the two hot spot contributions, which have a different $N_q$ scaling.

\begin{figure*} 
\centering
\includegraphics[scale=0.15]{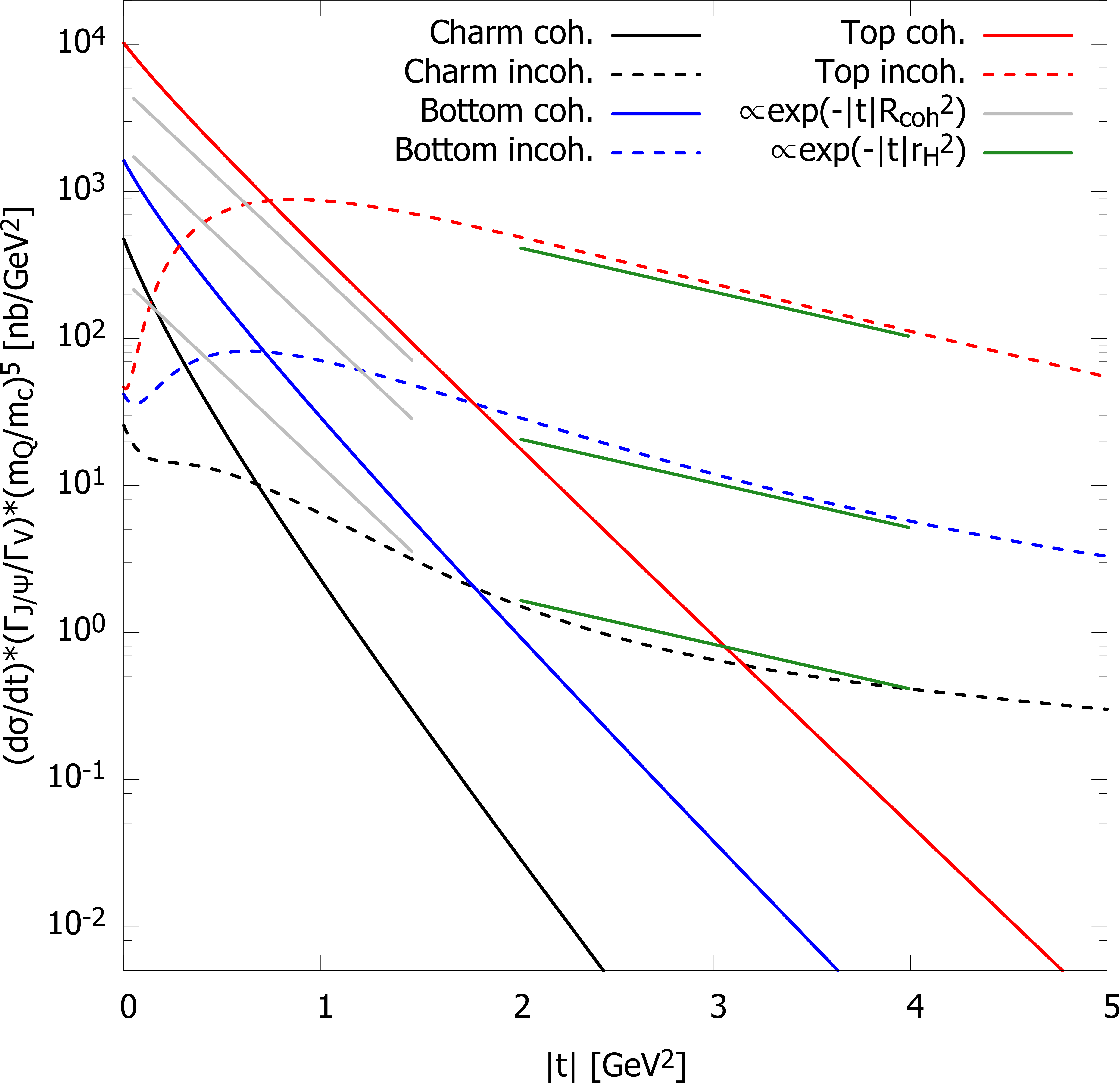}
\includegraphics[scale=0.15]{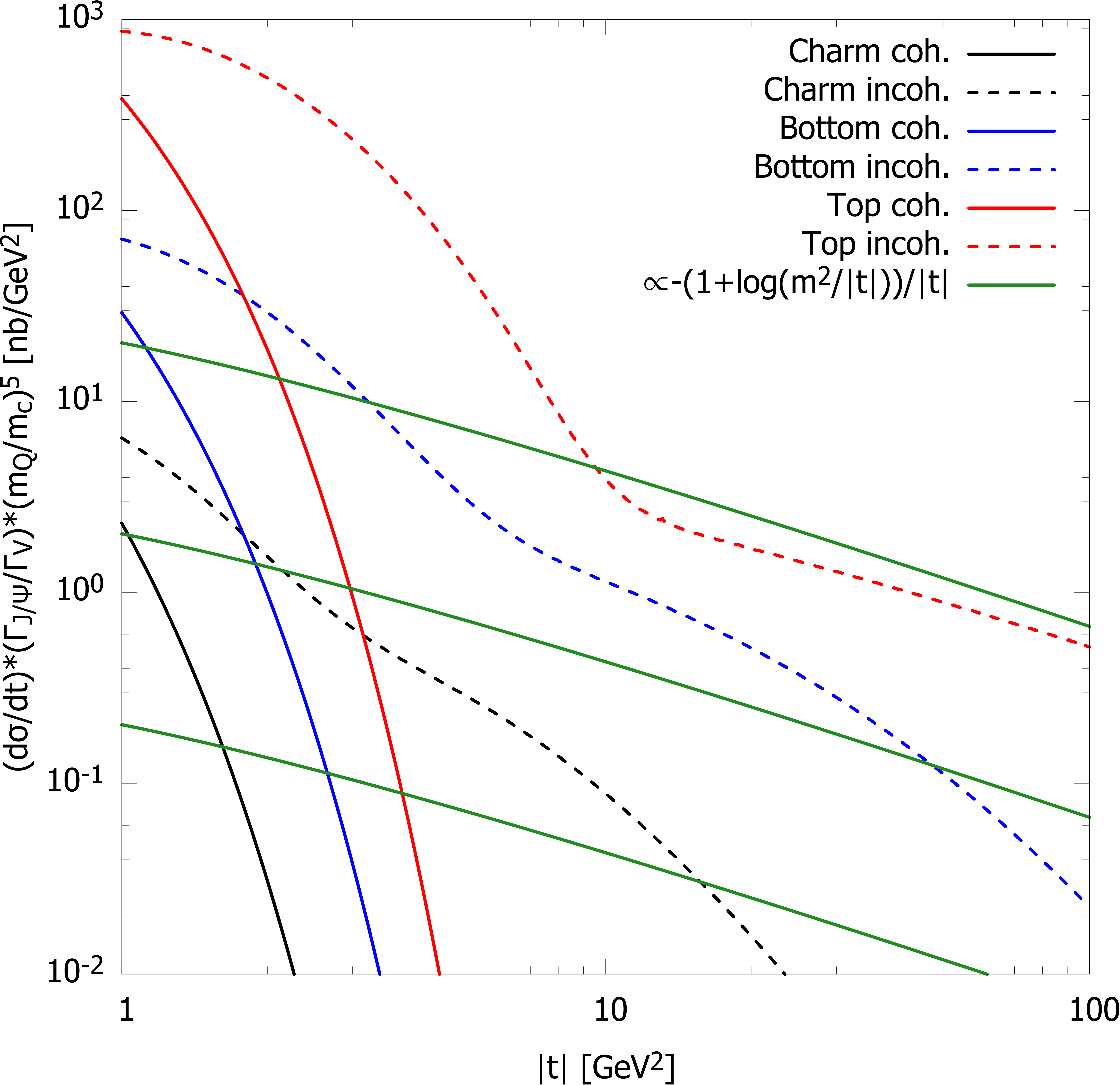}
\caption{Quark mass dependence of the cross-section $(d\sigma/dt)~(\Gamma_{J/\Psi}/\Gamma_{V})~(m_{Q}/m_c)^5$. Numerical results for $c\bar{c}$, $b\bar{b}$ and $t\bar{t}$ vector meson production are compared to the expected behavior of the cross sections in the relevant $t$-ranges. 
}
\label{fig:CSCharmBottomTop}
\end{figure*}

In Fig.~\ref{fig:CSCharmBottomTop} we combine cross sections with different quark masses in the same plot, in order to better visualize the quark mass dependence. As we know, the larger the quark mass, the smaller the dipole tends to be due to the larger $\varepsilon' \sim m_{Q}$ in the photon wave function. 
We again see that the larger the mass of the quark, the better the expected behaviors of the cross sections work. The cross sections have been scaled by the meson leptonic decay width and $m_Q^5$. This cancels the expected power law quark mass dependence, as discussed in Appendices \ref{app:Delta0Coh}, \ref{app:smallR}, but leaves a logarithmic one in the coherent cross section (see \eq\nr{eq:coherentzerot}) and the hot spot fluctuation part  of the incoherent cross section (see \eq\eqref{eq:CohIntegralSmallr}). In the numerical evaluation  we see that e.g. the normalization of the charm and bottom cross sections differ from the power law by roughly an order of magnitude, so the deviations from the expected power law dependence because of these logarithms are quite substantial. 
 
\section{Conclusions} \label{sec:Conclusions}

In our previous work  \cite{Demirci:2021kya} we took commonly used physics ingredients used to describe collisions involving hadrons and built an analytical model for the hot spot structure of the color field of a high energy nucleon. In this paper we have computed  coherent and incoherent exclusive vector meson cross sections in this model, in order to study how such measurements can be used to understand the subnucleon scale geometry. Our emphasis has been on keeping the model simple and analytically tractable, for instance by using the dilute expansion of the proton. This has  allowed us to perform averages over both color and hot spot degrees of freedom in an analytical fashion. 

Using parameter values which we think are in an acceptable, physically relevant range, we have found a clear separation  into contributions sensitive to different types of fluctuations for the incoherent cross section. The hot spot fluctuations parts vanish at zero momentum transfer ($t=0$) but become the dominant contribution quickly as $t$ increases. After this, as $t$ increases and the probe starts resolving the color fluctuations in the target, the color fluctuation contributions become dominant. We found that the hot spot fluctuation parts, as well as the coherent cross section, depend on the geometrical parameters of the hot spot system through an exponential in $t$, modified by logarithms, whereas the color fluctuation contributions decrease as power law in $t$, up to logarithmic corrections.

Our analytically tractable approach allowed us to gain several important insights into the relation of the hot spot structure of the nucleon and the exclusive vector meson production cross section. Firstly, in our model the normalization between the coherent and incoherent cross sections did not match experimental observations at HERA; however the form of the $t$-distribution does look mostly correct as seen in Fig.~\ref{fig:CSNormalizedFits}, pointing to the presence of additional fluctuations not included in the model.  Secondly we found that  any realistic quark mass ($c,b$) results in a $t$-dependence of the cross section, which is sensitive to both the structure of the target and the probe, and not only the target. Thirdly we found that we can not use the small dipole size approximation in this model when studying $J/\Psi$ production, and not even in the case of $\Upsilon$ when studying the incoherent cross section.

Generally, the sensitivity of the $t$-dependence of the cross section  on the meson wave function, or the wave function overlap, dependence seems to be larger than anticipated. Usually one assumes that already for the $J/\Psi$ the charm quarks form a dipole probe that can be treated as  small.  However in our model it seems that with charm, or even bottom quarks, we are still sensitive to the structure of the probe. Thus the usual intuition that the slope of the coherent cross section measures the overall size of the proton breaks down. This can also be seen when we  expand the cross sections in the small dipole size approximation, which yields a  drastically different asymptotic behavior of the cross sections for  realistically heavy quark masses.
By studying different quark masses we have seen that for heavier quarks the $t$-distribution does eventually become less dependent on the probe. We therefore conclude, that in order to disentangle projectile and target properties from exclusive vector meson production, it will be interesting and important to simultaneously study charmonium and bottomonium production in future experimental measurements.

Due to the discrepancy in the relative normalization of the coherent and incoherent cross section, between the experimental data and our microscopic model, we did not perform an actual fit to experimental data. We instead considered values of the model parameters that are, in our estimation, physically relevant and that have been used in previous applications to both exclusive vector meson and proton-nucleus collisions. 

There are studies, using similar versions of the hot spot model, where the normalization of the cross sections works significantly better \cite{Mantysaari:2016jaz, Mantysaari:2022ffw}, but at this stage we can only speculate on the reasons for this discrepancy. 
In a Monte Carlo implementation  one can define the center of mass for the gluon field,
whereas we fix the center of mass of the proton to be the center of the hot spot system of color charges, but this we expect to have a negligible effect. One thing that is clearly  different is the dilute expansion, which actually removes the gluon saturation aspect of our model. This might, and probably does, reduce the magnitude of gluon field fluctuations in the proton and could contribute to the normalization discrepancy, although not obviously in the correct direction.
In addition to the sources of fluctuation included here, many models include additional
 saturation scale $(Q_s)$ fluctuations, letting  the amount of color charge change in individual hot spots fluctuate on an event-by-event basis. Such additional fluctuations clearly do increase the incoherent cross section across the $t$-spectrum \cite{Mantysaari:2016ykx}. One simplification used here is also the treatment of the wave function overlap, for which we used the non-relativistic wave functions to keep our model simple and analytically tractable. In this limit the quark and the anti-quark in the dipole always carry half of the photon momentum $(z=1/2)$, which decouples  the dipole size $\rt$ from the transverse momentum transfer $\Delta$. This can actually make the $t$-distribution at large $t$ behave differently, as such a coupling would effectively make the typical dipoles smaller at large momentum transfer.

Clearly, as a possible future direction, it would be interesting to study how exactly the aforementioned differences to Monte Carlo implementations affect the cross sections we found here. One could e.g. try use a more general wave function with an actual dipole momentum fraction ($z$) dependence and see how, in general, the change of the wave function affects the results. This would be interesting because we found a large wave function dependence on the cross sections. One could also try to incorporate the $Q_s$ fluctuations  or saturation to our simple model, but it is not obvious how to do this maintaining the analytical treatment of the fluctuations. One could also include $N_q$ fluctuations which should not be very difficult to do in our model. Finally, one additional possibility for further work could be to try and fit the microscopic hot spot model to experimental data, with or without separate normalizations for the coherent and incoherent cross sections.

\section*{Acknowledgements}
S.D. acknowledges the support of the Vilho, Yrjö and Kalle Väisälä Foundation.
S.S acknowledges support by the Deutsche Forschungsgemeinschaft (DFG, German Research Foundation) through the CRC-TR 211 'Strong-interaction matter under extreme conditions'– project number 315477589 – TRR 211.
S.D. and T.L have  been supported by the Academy of Finland, by the Centre of Excellence in Quark Matter (project 346324) and project 321840. This work has also been supported under the European Union’s Horizon 2020 research and innovation programme by the STRONG-2020 project (grant agreement No 824093) and by the European Research Council, grant agreements ERC-2015-CoG-681707 and ERC-2018-AdG-835105. The content of this article does not reflect the official opinion of the European Union and responsibility for the information and views expressed therein lies entirely with the authors. 

\appendix

\section{Average of the dipole cross section} \label{app:dip1point}

We start from the definition of the dipole scattering amplitude written with respect to the Wilson line dipole as
\begin{equation}
\frac{d\sigma  ^{\text{p}}_{\text{dip}}}{d^2\bt}(\bt,\rt) = 2\left( 1-\frac{1}{N_c}\Tr\left[ V(\xt)V^{\dagger}(\yt) \right] \right),
\end{equation}
where the Wilson lines  $V$ are the path ordered exponentials of the color field \eqref{eq:WilsonLine}.
We work in the dilute limit of the proton and we proceed to expand the Wilson lines to the lowest order in the proton color charge density or equivalently the proton saturation scale. This yields
\begin{equation}
\begin{split}
&
\frac{d\sigma  ^{\text{p}}_{\text{dip}}}{d^2\bt}(\bt,\rt) \approx
\\ &
-\frac{2g^2}{N_c}\Bigg(
\Tr\left[ \int d^2 \zt d^2 \wt G(\xt-\zt)G(\yt-\wt) \rho_a(\zt)\rho_b(\wt)t^at^b \right]
\\ &
- \frac{1}{2}\Tr\left[ \int d^2 \zt d^2 \wt G(\xt-\zt)G(\xt-\wt) \rho_a(\zt)\rho_b(\wt)t^at^b \right]
\\ &
- \frac{1}{2}\Tr\left[ \int d^2 \zt d^2 \wt G(\yt-\zt)G(\yt-\wt) \rho_a(\zt)\rho_b(\wt)t^at^b \right]
\Bigg),
\end{split}
\end{equation}
where $G$ is the Green's function
\begin{equation}
G(\xt-\yt)=\int\frac{d^2\kt}{(2\pi)^2}\frac{\exp(i\kt\cdot(\xt-\yt))}{\kt^2+m^2}.
\end{equation}
Next we take the trace over the $SU(N_c)$ generators and take the Gaussian averages over the color charge configurations with an MV-model color charge two point function with the added hot spot model distribution of color charges as defined in 
\begin{equation}
\left\langle \rho_a(\xt) \rho_b(\yt) \right\rangle_{CGC} = \sum_{i=1}^{N_q} \mu^2 \left( \frac{\xt-\yt}{2}-\bt_i \right) \delta^{(2)}(\xt-\yt)\delta^{ab},
\end{equation} 
where $b_i$ are the coordinates for the centers of the hot spots in the proton. 

Now by taking the trace and  the color charge average, and integrating over the transverse coordinate delta function, the dipole cross section takes the form
\begin{equation}
\begin{split}
&
\left\langle \frac{d\sigma  ^{\text{p}}_{\text{dip}}}{d^2\bt}(\bt,\rt) \right\rangle _{CGC} \approx
\frac{g^2(N_c^2-1)}{2N_c}
\int d^2 \zt
\Big[
G(\xt-\zt)G(\xt-\zt)
\\ &
+ G(\yt-\zt)G(\yt-\zt) -2G(\xt-\zt)G(\yt-\zt)
\Big] 
\\ &
\times \sum_{i=1}^{N_q} \mu^2(\zt-\bt_i).
\end{split}
\end{equation}
By taking the hot spot average, we get
\begin{multline}
\left\langle \left\langle \frac{d\sigma  ^{\text{p}}_{\text{dip}}}{d^2\bt}(\bt,\rt) \right\rangle \right\rangle \approx
\frac{g^2(N^2-1)N_q}{2N}
\\ \times
\int d^2 \zt
\Big[
G(\xt-\zt)G(\xt-\zt)
+ G(\yt-\zt)G(\yt-\zt) 
\\
-2G(\xt-\zt)G(\yt-\zt)
\Big] 
F_1(\zt,\Bt)),
\end{multline}
where $F_1$ is the same function as defined before in Eq.~\eqref{eq:F1}. This can be written more compactly by defining
\begin{multline}
\Omega(\xt,\yt,\zt,\vt) = G(\xt-\zt)G(\xt-\vt)
\\ 
+G(\yt-\zt)G(\yt-\vt) -2G(\xt-\zt)G(\yt-\vt),
\end{multline}
and writing
\begin{multline}
\left\langle \left\langle \frac{d\sigma  ^{\text{p}}_{\text{dip}}}{d^2\bt}(\bt,\rt) \right\rangle \right\rangle \approx
\frac{g^2(N_c^2-1)N_q}{2N_c}
\int d^2 \zt
\Omega(\xt,\yt,\zt,\zt)
\\ 
\times F_1(\zt,\Bt),
\end{multline}
which can be written as
\begin{multline}
\left\langle \left\langle
\frac{d \sigma  ^{\text{p}}_{\text{dip}}}{d^2 \bt}(\bt, \rt)
\right\rangle \right\rangle
=
\frac{g^2 (N_c^2 - 1) N_q}{2 N_c} 
\\  \times
\int d^2 \zt
\Omega \left( \bt + \frac{\rt}{2}, \bt - \frac{\rt}{2}, \zt, \zt \right) F_1(\zt , \Bt).
\end{multline}

\section{Average of the dipole cross section squared} \label{app:dip2point}

Now we start from the definition of the dipole cross section squared. It is easy get from the definition of the cross section and it reads
\begin{equation}
\begin{split}
&
\frac{d\sigma  ^{\text{p}}_{\text{dip}}}{d^2\bt}(\bt,\rt)\frac{d\sigma  ^{\text{p}}_{\text{dip}}}{d^2\btbar}(\btbar,\rtbar) 
\\ &
= 2\left( 1-\frac{1}{N_c}\Tr\left[ V(\xt)V^{\dagger}(\yt) \right] \right)2\left( 1-\frac{1}{N_c}\Tr\left[ V(\xtbar)V^{\dagger}(\ytbar) \right] \right).
\end{split}
\end{equation}

We again expand the Wilson lines to the lowest nontrivial order in the proton color charge. After this, we take the contractions of the color charges as we did before. Now when we take hot spot averages, we take averages of two hot spots instead of just one. Thus we get a term proportional to $N_q$, which corresponds to measuring two gluons originating from the same hot spot, and then we have a term proportional to $N_q(N_q-1)$ that corresponds to measuring the gluons from different hot spots. After taking both of these averages, we get
\begin{equation}
\begin{split}
&
\left\langle \left\langle
\frac{d \sigma ^{\text{p}}_{\text{dip}}}{d^2 \bt}(\bt, \rt)
\frac{d \sigma ^{\text{p}}_{\text{dip}}}{d^2 \bar{\bt}}(\bar{\bt}, \bar{\rt})
\right\rangle \right\rangle
\\ &
=
\frac{g^4}{4 N_c^2} \int d^2 \zt d^2 \vt 
\Big \{ 
(N_c^2-1)^2 \Omega(\xt,\yt,\zt,\zt) \Omega(\bar{\xt},\bar{\yt},\vt,\vt)
\\ & +
(N_c^2-1) \Omega(\xt,\yt,\zt,\vt) \Omega(\bar{\xt},\bar{\yt},\zt,\vt) 
\\ &
+ (N_c^2-1) \Omega(\xt,\yt,\zt,\vt) \Omega(\bar{\xt},\bar{\yt},\vt,\zt)
\Big \}
\\ & \times
\Big \{ 
N_q F_2(\zt,\vt,\Bt) + N_q(N_q-1)F_3(\zt,\vt,\Bt)
\Big \},
\end{split}
\end{equation}
which can be written, in the impact parameter and dipole size coordinates, as
\begin{equation} \label{eq:2PointDipoleOmegas}
\begin{split}
&
\left\langle \left\langle
\frac{d \sigma ^{\text{p}}_{\text{dip}}}{d^2 \bt}(\bt, \rt)
\frac{d \sigma ^{\text{p}}_{\text{dip}}}{d^2 \bar{\bt}}(\bar{\bt}, \bar{\rt})
\right\rangle \right\rangle
\\ &
=
\frac{g^4}{4 N_c^2} \int d^2 \zt d^2 \vt 
\\ & \times
\Big \{ 
(N_c^2-1)^2 \Omega \left(\bt + \frac{\rt}{2},\bt - \frac{\rt}{2},\zt,\zt \right) \Omega \left( \bar{\bt} + \frac{\bar{\rt}}{2},\bar{\bt} - \frac{\bar{\rt}}{2},\vt,\vt \right)
\\ & +
(N_c^2-1) \Omega \left( \bt + \frac{\rt}{2},\bt - \frac{\rt}{2},\zt,\vt \right) \Omega \left( \bar{\bt} + \frac{\bar{\rt}}{2},\bar{\bt} - \frac{\bar{\rt}}{2},\zt,\vt \right)
\\ &
+ (N_c^2-1) \Omega \left( \bt + \frac{\rt}{2},\bt - \frac{\rt}{2},\zt,\vt \right) \Omega \left( \bar{\bt} + \frac{\bar{\rt}}{2},\bar{\bt} - \frac{\bar{\rt}}{2},\vt,\zt \right)
\Big \}
\\ & \times
\Big \{ 
N_q F_2(\zt,\vt,\Bt) + N_q(N_q-1)F_3(\zt,\vt,\Bt)
\Big \}.
\end{split}
\end{equation}
Here we have defined
\begin{multline} \label{eq:F2}
F_2(\zt, \vt, \Bt) \equiv \langle \mu^2(\zt-\bt_i)\mu^2(\vt-\bt_i) \rangle_{\text{Hot spot}}
\\  
=
\left( \frac{\mu^2_0}{2 \pi r_H^2} \right) ^2 \left( \frac{1}{1+2\left(\frac{N_q-1}{N_q}\right)\frac{R^2}{r_H^2}} \right)
\\  \times
\exp \left\{ 
- \frac{(\zt+\vt-2\Bt)^2}{4r_H^2(1+2\left(\frac{N_q-1}{N_q}\right)\frac{R^2}{r_H^2})} 
-\frac{(\zt-\vt)^2}{4r_H^2} 
\right\}
\end{multline}
and
\begin{multline} \label{eq:F3}
F_3(\zt, \vt, \Bt) \equiv \langle \mu^2(\zt-\bt_i)\mu^2(\vt-\bt_j) \rangle_{\text{Hot spot}}
\\  
=
\left( \frac{\mu^4_0}{(2 \pi)^2 (R^2+r_H^2)} \right) \left( \frac{1}{r_H^2+\left(\frac{N_q-2}{N_q}\right)R^2} \right)
\\  \times
\exp \left\{ 
- \frac{(\zt+\vt-2\Bt)^2}{4(r_H^2+\left(\frac{N_q-2}{N_q}\right)R^2)} 
-\frac{(\zt-\vt)^2}{4(R^2+r_H^2)} 
\right\}
\end{multline}
which we first computed in \cite{Demirci:2021kya}.

The integrals in the first term, the term proportional to $(N_c^2-1)^2$, decouple and the result is proportional to the coherent amplitude squared. The second two terms however are different. For the Fourier transform of the latter terms we use the identity
\begin{equation} \label{eq:ColorFlucIdentity}
\begin{split}
&
\int d^2 \bt d^2 \bar{\bt} d^2 \vt d^2 \wt e^{-i \bt \cdot \Delta + i \bar{\bt} \cdot \Delta}
\\ & \times
G(\bt + \Rt_1 - \vt)G(\bt + \Rt_2 - \wt)
\\ & \times
G(\bar{\bt} + \bar{\Rt}_1 - \vt)G(\bar{\bt} + \bar{\Rt}_2 - \wt)e^{-B(\vt + \wt)^2 - C(\vt - \wt)^2}
\\ & =
\frac{1}{16(2\pi)^2}\frac{1}{BC}\int d^2 \kt d^2 \bar{\kt}
\\ & \times
\exp\Big\{ 
-\frac{1}{4C} \left( \kt + \bar{\kt} \right)^2
\\ &
+i\kt \cdot \left( \Rt_2 - \Rt_1 \right) +i\bar{\kt}\cdot \left( \bar{\Rt}_2-\bar{\Rt}_1 \right)
\\ &
+\frac{i\Delta}{2} \cdot \left( \Rt_1 + \Rt_2 - \bar{\Rt}_1 - \bar{\Rt}_2 \right)
\Big\}
\\ & \times
\frac{1}{( \left( \kt+\frac{\Delta}{2} \right)^2+m^2)( \left( \kt-\frac{\Delta}{2})^2 + m^2 \right) }
\\ & \times
\frac{1}{( \left( \bar{\kt}+\frac{\Delta}{2} \right)^2+m^2)( \left( \bar{\kt}-\frac{\Delta}{2})^2+m^2 \right) }.
\end{split}
\end{equation}
Here $B$ and $C$ are the coefficients of the exponentials \eqref{eq:F2} or \eqref{eq:F3} depending on which term we are calculating.

Now focusing on the latter two terms of \eqref{eq:2PointDipoleOmegas}, using the identity \eqref{eq:ColorFlucIdentity} and simplifying, we may write
\begin{equation}
\begin{split}
&
\int d^2 \bt d^2 \bar{\bt} e^{-i \bt \cdot \Delta + i \bar{\bt} \cdot \Delta}
\\ & \times
\int d^2 \zt d^2 \vt
\exp
\Big \{ 
-B(\zt+\vt)^2-C(\zt-\vt)^2
\Big \}
\\ & \times
\Big \{
\Omega \left( \bt + \frac{\rt}{2},\bt - \frac{\rt}{2},\zt,\vt \right) \Omega \left( \bar{\bt} + \frac{\bar{\rt}}{2},\bar{\bt} - \frac{\bar{\rt}}{2},\zt,\vt \right)
\\ &
+ \Omega \left( \bt + \frac{\rt}{2},\bt - \frac{\rt}{2},\zt,\vt \right) \Omega \left( \bar{\bt} + \frac{\bar{\rt}}{2},\bar{\bt} - \frac{\bar{\rt}}{2},\vt,\zt \right)
\Big \}
\\ &
=
\frac{1}{16(2\pi)^2BC}
\int d^2 \kt d^2 \ktbar
\\ & \times
\frac{\exp \big\{ -\frac{1}{4C} \left( \kt+\ktbar \right)^2 \big\} }{( \left( \kt+\frac{\Delta}{2} \right)^2+m^2)( \left( \kt-\frac{\Delta}{2} \right)^2+m^2)}
\\ & \times
\frac{1}{( \left( \ktbar+\frac{\Delta}{2} \right)^2+m^2)( \left( \ktbar-\frac{\Delta}{2} \right)^2+m^2)}
\\ & \times
\Big[
8\cos \left( \frac{\Delta}{2}\cdot \rt \right) \cos \left( \frac{\Delta}{2}\cdot \rtbar \right)
\\ &
-8\cos \left( \frac{\Delta}{2}\cdot \rt \right) \exp  \left( i\ktbar \cdot \rtbar \right)
-8\cos \left( \frac{\Delta}{2}\cdot \rtbar \right) \exp \left( i\kt \cdot \rt \right)
\\ &
+4\exp \left( i\kt \cdot \rt + i\ktbar \cdot \rtbar \right)
+4\exp \left( i\kt \cdot \rt - i\ktbar \cdot \rtbar \right)
\Big].
\end{split}
\end{equation}

Using this and the definitions \eqref{eq:F2} and \eqref{eq:F3} we can substitute the appropriate values for $B$ and $C$.  This is enough to find the color connected contributions to the Fourier transform of \eqref{eq:2PointDipoleOmegas} i.e. the average of the Fourier transformed dipole cross section squared.

Now we can write the Fourier transform of the different parts of the dipole cross section squared as
\begin{equation}
\begin{split}
&
\left[
\int d^2 \bt d^2 \bar{\bt} e^{-i \bt \cdot \Delta + i \bar{\bt} \cdot \Delta}
\left\langle \left\langle
\frac{d \sigma ^{\text{p}}_{\text{dip}}}{d^2 \bt}(\bt, \rt)
\frac{d \sigma ^{\text{p}}_{\text{dip}}}{d^2 \bar{\bt}}(\bar{\bt}, \bar{\rt})
\right\rangle \right\rangle \right]_{1,DC}
\\ &
=
\frac{g^4\mu_0^4(N_c^2-1)^2N_q}{(2\pi)^2N_c^2}\exp \left( -r^2_H\Delta^2 \right)
\\ & \times
\Big\{ \Psi(\rt)\Psi(\bar{\rt}) +\cos \left( \frac{1}{2}\Delta\cdot\rt \right) \cos \left( \frac{1}{2}\Delta\cdot\bar{\rt} \right) \Psi(0)\Psi(0)
\\ &
- \cos \left( \frac{1}{2}\Delta\cdot\bar{\rt} \right)\Psi(\rt)\Psi(0) - \cos \left( \frac{1}{2}\Delta\cdot\rt \right) \Psi(\bar{\rt})\Psi(0)
 \Big\}
\end{split}
\end{equation}
\begin{equation}
\begin{split}
&
\left[
\int d^2 \bt d^2 \bar{\bt} e^{-i \bt \cdot \Delta + i \bar{\bt} \cdot \Delta}
\left\langle \left\langle
\frac{d \sigma ^{\text{p}}_{\text{dip}}}{d^2 \bt}(\bt, \rt)
\frac{d \sigma ^{\text{p}}_{\text{dip}}}{d^2 \bar{\bt}}(\bar{\bt}, \bar{\rt})
\right\rangle \right\rangle \right]_{2,DC}
\\ &
=
\frac{g^4\mu_0^4(N_c^2-1)^2N_q(N_q-1)}{N_c^2(2\pi)^2}\exp \left(- \left(R^2+r^2_H \right)\Delta^2 \right)
\\ & \times
\Big\{ \Psi(\rt)\Psi(\bar{\rt}) +\cos \left( \frac{1}{2}\Delta\cdot\rt \right) \cos \left( \frac{1}{2}\Delta\cdot\bar{\rt} \right) \Psi(0)\Psi(0)
\\ &
- \cos \left( \frac{1}{2}\Delta\cdot\bar{\rt} \right)\Psi(\rt)\Psi(0) - \cos \left( \frac{1}{2}\Delta\cdot\rt \right) \Psi(\bar{\rt})\Psi(0)
 \Big\}
\end{split}
\end{equation}
\begin{equation}
\begin{split}
&
\left[
\int d^2 \bt d^2 \bar{\bt} e^{-i \bt \cdot \Delta + i \bar{\bt} \cdot \Delta}
\left\langle \left\langle
\frac{d \sigma ^{\text{p}}_{\text{dip}}}{d^2 \bt}(\bt, \rt)
\frac{d \sigma ^{\text{p}}_{\text{dip}}}{d^2 \bar{\bt}}(\bar{\bt}, \bar{\rt})
\right\rangle \right\rangle \right]_{1,C}
\\ &
=
\frac{g^4\mu_0^4(N_c^2-1)}{64\pi^4N_c^2}N_q
\\ & \times
\int d^2\kt d^2 \bar{\kt} 
\frac{\exp \left( -r^2_H \left( \kt+\bar{\kt} \right)^2 \right)}{( \left(\kt+\frac{\Delta}{2} \right)^2+m^2)( \left(\kt-\frac{\Delta}{2} \right)^2+m^2)}
\\ & \times
\frac{1}{( \left(\bar{\kt}+\frac{\Delta}{2} \right)^2+m^2)( \left(\bar{\kt}-\frac{\Delta}{2} \right)^2+m^2)}
\\ & \times 
\Bigg\{
8\cos \left(\frac{\Delta}{2}\cdot\rt \right)\cos \left(\frac{\Delta}{2}\cdot\bar{\rt} \right)
\\ &
-8\cos \left(\frac{\Delta}{2} \cdot \rt \right)\exp \left(i\bar{\kt} \cdot \bar{\rt} \right)
-8\cos \left(\frac{\Delta}{2} \cdot \bar{\rt} \right)\exp \left(i\kt \cdot \rt \right)
\\ &
+4\exp \left(i\kt \cdot \rt +i \bar{\kt} \cdot \bar{\rt} \right)
+4\exp \left( i\kt\cdot\rt - i\bar{\kt}\cdot\bar{\rt} \right)
\Bigg\}
\end{split}
\end{equation}
and
\begin{equation}
\begin{split}
&
\left[
\int d^2 \bt d^2 \bar{\bt} e^{-i \bt \cdot \Delta + i \bar{\bt} \cdot \Delta}
\left\langle \left\langle
\frac{d \sigma ^{\text{p}}_{\text{dip}}}{d^2 \bt}(\bt, \rt)
\frac{d \sigma ^{\text{p}}_{\text{dip}}}{d^2 \bar{\bt}}(\bar{\bt}, \bar{\rt})
\right\rangle \right\rangle \right]_{2,C}
\\ &
=
\frac{g^4\mu_0^4(N_c^2-1)}{64\pi^4N_c^2}N_q(N_q-1)
\\ & \times
\int d^2\kt d^2 \bar{\kt} 
\frac{\exp \left(- \left(R^2+r^2_H \right) \left(\kt+\bar{\kt} \right)^2 \right)}{( \left(\kt+\frac{\Delta}{2} \right)^2+m^2)( \left(\kt-\frac{\Delta}{2} \right)^2+m^2)}
\\ & \times
\frac{1}{( \left(\bar{\kt}+\frac{\Delta}{2} \right)^2+m^2)( \left(\bar{\kt}-\frac{\Delta}{2} \right)^2+m^2)}
\\ & \times 
\Bigg\{
8\cos \left( \frac{\Delta}{2}\cdot\rt \right)\cos \left( \frac{\Delta}{2}\cdot\bar{\rt} \right)
\\ &
-8\cos \left( \frac{\Delta}{2} \cdot \rt \right) \exp \left(i\bar{\kt} \cdot \bar{\rt} \right)
-8\cos \left(\frac{\Delta}{2} \cdot \bar{\rt} \right)\exp \left(i\kt \cdot \rt \right)
\\ &
+4\exp \left(i\kt \cdot \rt +i \bar{\kt} \cdot \bar{\rt} \right)
+4\exp \left(i\kt\cdot\rt - i\bar{\kt}\cdot\bar{\rt} \right)
\Bigg\}
\end{split}
\end{equation}

The latter two color connected terms can be integrated over the angles of the dipole size $\rt,\rtbar$ as the non-relativistic wave functions remove the Fourier exponent depending on the dipole orientation. This yields
\begin{equation}
\begin{split}
&
\Bigg[
\int d\theta_{\rt} d\theta_{\rtbar}
\int d^2 \bt d^2 \bar{\bt} e^{-i \bt \cdot \Delta + i \bar{\bt} \cdot \Delta}
\\ & \times
\left\langle \left\langle
\frac{d \sigma ^{\text{p}}_{\text{dip}}}{d^2 \bt}(\bt, \rt)
\frac{d \sigma ^{\text{p}}_{\text{dip}}}{d^2 \bar{\bt}}(\bar{\bt}, \bar{\rt})
\right\rangle \right\rangle \Bigg]_{1,C}
\\ &
=
\frac{g^4\mu_0^4(N_c^2-1)}{2\pi^2 N_c^2}N_q
\\ & \times
\int d^2\kt d^2 \bar{\kt} 
\frac{\exp \left(-r^2_H \left(\kt+\bar{\kt} \right)^2 \right)}{( \left(\kt+\frac{\Delta}{2} \right)^2+m^2)( \left(\kt-\frac{\Delta}{2} \right)^2+m^2)}
\\ & \times
\frac{1}{( \left(\bar{\kt}+\frac{\Delta}{2} \right)^2+m^2)( \left(\bar{\kt}-\frac{\Delta}{2} \right)^2+m^2)}
\\ & \times 
\Bigg\{
J_0 \left(\frac{|\Delta||\rt|}{2} \right)J_0 \left(\frac{|\Delta||\rtbar|}{2} \right)
-J_0 \left(\frac{|\Delta||\rt|}{2} \right)J_0 \left(|\ktbar||\rtbar| \right)
\\ &
-J_0 \left(\frac{|\Delta||\rtbar|}{2} \right)J_0 \left(|\kt||\rt| \right)
+J_0 \left(|\kt||\rt| \right)J_0 \left(|\ktbar||\rtbar| \right)
\Bigg\}
\end{split}
\end{equation}
and
\begin{equation}
\begin{split}
&
\Bigg[
\int d\theta_{\rt} d\theta_{\rtbar} d^2 \bt d^2 \bar{\bt} e^{-i \bt \cdot \Delta + i \bar{\bt} \cdot \Delta}
\\ & \times
\left\langle \left\langle
\frac{d \sigma ^{\text{p}}_{\text{dip}}}{d^2 \bt}(\bt, \rt)
\frac{d \sigma ^{\text{p}}_{\text{dip}}}{d^2 \bar{\bt}}(\bar{\bt}, \bar{\rt})
\right\rangle \right\rangle \Bigg]_{2,C}
\\ &
=
\frac{g^4\mu_0^4(N_c^2-1)}{2\pi^2 N_c^2}N_q(N_q-1)
\\ & \times
\int d^2\kt d^2 \bar{\kt} 
\frac{\exp \left(- \left(R^2+r^2_H \right) \left(\kt+\bar{\kt} \right)^2 \right)}{( \left(\kt+\frac{\Delta}{2} \right)^2+m^2)( \left(\kt-\frac{\Delta}{2} \right)^2+m^2)}
\\ & \times
\frac{1}{( \left(\bar{\kt}+\frac{\Delta}{2} \right)^2+m^2)( \left(\bar{\kt}-\frac{\Delta}{2} \right)^2+m^2)}
\\ & \times 
\Bigg\{
J_0 \left(\frac{|\Delta||\rt|}{2} \right)J_0 \left(\frac{|\Delta||\rtbar|}{2} \right)
-J_0 \left(\frac{|\Delta||\rt|}{2} \right)J_0 \left(|\ktbar||\rtbar| \right)
\\ &
-J_0 \left(\frac{|\Delta||\rtbar|}{2} \right)J_0 \left(|\kt||\rt| \right)
+J_0 \left(|\kt||\rt| \right)J_0 \left(|\ktbar||\rtbar| \right)
\Bigg\}
\end{split}
\end{equation}

These are now the forms of the parts of the cross section that we can convolute with the vector meson-photon overlaps and integrate over the dipole sizes. This is also the form which we can easily expand in the small-$\rt$ limit   as we do in Appendix~\ref{app:smallR}. 

\section{$\Delta=0$ expansion of the coherent cross section} \label{app:Delta0Coh}

The $Z$-and $K$ functions have a nontrivial dependence on $m$ and $\Delta$. To proceed further we must look at specific limits in terms of these variables. In particular we are interested in the behavior in the limits of small and large $\Delta$.

Using the definitions \eqref{eq:CohIntegral},\eqref{eq:PsiFunction} and \eqref{eq:PsiFunctionR0}, the coherent cross section \eqref{eq:CohAmplitudeSq} at the limit of $\Delta=0$, or $t=0$, can be written as
\begin{multline}
\left. \frac{d \sigma_{T,L}^{\gamma^*p\rightarrow Vp}}{dt}\right|_{t=0}
 = \frac{1}{16\pi}
\left[ C^2_T +  C^2_L \right]
\frac{g^4 \mu_0^4 (N_c^2-1)^2 N_q^2}{(2\pi N_c)^2} 
\\  \times
\left[ 
\int d^2\rt
K_0( \varepsilon' |\rt|)
\left \{ 
\frac{1}{2m^2}
-\frac{|\rt|K_1(m|\rt|)}{2m}
\right \}
\right]^2 .
\end{multline}
The $\rt$-integral yields
\begin{multline}
\left.\frac{d \sigma_{T,L}^{\gamma^*p\rightarrow Vp}}{dt}\right|_{t=0}
= \frac{1}{16\pi}
\left[ C^2_T +  C^2_L \right]
\frac{g^4 \mu_0^4 (N_c^2-1)^2 N_q^2}{(2\pi N_c)^2} 
\\  \times
\left[ 
\frac{\pi}{m^2}
\left(
\frac{1}{\varepsilon'^2}
-
\frac{\varepsilon'^2 - m^2 + 2m^2 \log(\frac{m}{\varepsilon'})}{(\varepsilon'^2-m^2)^2}
\right)
\right]^2 .
\end{multline}
Expanding this at small $m$ to the lowest nontrivial order yields
\begin{multline}\label{eq:coherentzerot}
\left.
\frac{d \sigma_{T,L}^{\gamma^*p\rightarrow Vp}}{dt}\right|_{t=0}
\approx \frac{1}{16\pi}
\left[ C^2_T +  C^2_L \right]
\frac{g^4 \mu_0^4 (N_c^2-1)^2 N_q^2}{(2\pi N_c)^2} 
\\  \times
\left[ 
-\frac{\pi(1+2\log(\frac{m}{\varepsilon'}))}{\varepsilon'^4}
\right]^2 + \mathcal{O}(m^2).
\end{multline}
Firstly this limit serves as a convenient check of the numerical evaluation of the coherent cross section. Moreover, it demonstrates the important effect of the IR regulator $m$ on the magnitude of the coherent cross section. From this expression, using $\varepsilon' \sim m_Q$ for $m_Q^2 \gg Q^2$ and the definition of the constants $C_{T,L}$ from \eqref{eq:CT} and \eqref{eq:CL} one can deduce the power law dependence of the cross section in the heavy quark limit $d \sigma/dt \sim \Gamma_V/m_Q^5$.

\section{The small dipole size expansion of the cross sections} \label{app:smallR}

\subsection{The small dipole size approximation} \label{app:smallRCrossSections}

In this appendix we study the small dipole size expansion of our results. Here the argument is that for a heavy quark, we would expect the photon wave function to set $r\sim 1/m_Q$, so that both the meson wave function and the dipole amplitude can be expanded to lowest nontrivial order in $r$. In this limit several of the integrals that we were left with in the full case can be calculated analytically. We expand the dipole cross section Fourier transforms to the lowest nontrivial order in the dipole size $\rt$. We keep the photon wave function contribution intact as it kills off the large dipole size contributions at a scale dependent on the quark masses and the photon virtuality.

The Fourier-transform of the coherent amplitude \eqref{eq:1PointDipFourier} has a small $\rt$ expansion which reads
\begin{multline}
\int d^2 \bt e^{-i \bt \cdot \Delta}
\left\langle \left\langle
\frac{d \sigma ^{\text{p}}_{\text{dip}}}{d^2 \bt}(\bt, \rt)
\right\rangle \right\rangle
\\  
=
\frac{g^2 \mu_0^2 (N_c^2-1) N_q}{2 \pi N_c} \exp \left \{ -\frac{1}{2} \left( r^2_H + \left( \frac{N_q-1}{N_q}  \right) R^2 \right) \Delta^2 \right \}
\\  \times
\Bigg[ 
\frac{(\Delta \cdot \rt)^2}{4 \Delta^2} \left\{ -1 
+ \frac{4m^2}{\Delta \sqrt{\Delta^2 + 4m^2}}\arctanh \left( \frac{\Delta}{\sqrt{\Delta^2 + 4m^2}} \right) \right\}
\\ 
+ \rt^2 \Bigg\{ -\frac{1}{4}\log(m\rt) + \frac{1}{8}\log(4e^{3-2\gamma}) 
\\ 
- \frac{1}{4} \sqrt{1+4\frac{m^2}{\Delta^2}}\arccoth \left( \sqrt{1+4\frac{m^2}{\Delta^2}} \right) \Bigg\}
\Bigg].
\end{multline}
Here one should note that the term depending on the orientation of the dipole with respect to the momentum transfer (the $(\Delta \cdot \rt)^2$-term) does not vanish in the lowest order in the dipole size. This feature has an interpretation in terms of the Wigner distribution of the proton, and corresponds to an angular correlation between the jet axis and momentum transfer in exclusive dijet production~\cite{Hatta:2016dxp,Altinoluk:2015dpi,Mantysaari:2019csc}.
However, this term vanishes in the limit of large $m/\Delta$. 
  This effect has an interesting interpretation in terms of  the shape of the proton. When $m$ is large, the gluon field generated by a pointlike color charge is more or less pointlike, which translates to the gluon field of the proton being Gaussian because the color charges are distributed as a Gaussian. This reduces the size of the boundary, which is where the orientation of the small dipole matters, and this makes the angle dependence vanish at the lowest order of the dipole size. On the other hand when the $m$ is not large, the shape of the gluon field of the proton is decreasing more as a power law (in momentum space, a modified Bessel function in coordinate space), which makes the boundary more relevant than in the Gaussian shape and results in  the angle dependent term having a significant contribution.

To get the small-$r$ expressions for the cross sections, we expand the dipole cross section to the lowest nontrivial order in the dipole size $r$ before integrating it with the wave function overlap. The small-$r$ expressions for the $Z-$ and $K$-functions defined in \eqref{eq:CohIntegral} and \eqref{eq:IncohIntegral} then read
\begin{equation} \label{eq:CohIntegralSmallr}
\begin{split}
&
Z_{\rt\approx0}(\Delta,m^2,\varepsilon')
\equiv
\\ &
\frac{2\pi}{\varepsilon'^4}
\left[
-\frac{2m^2+\Delta^2}{\Delta \sqrt{4m^2+\Delta^2}} \arctanh \left( \frac{\Delta}{\sqrt{4m^2+\Delta^2}} \right) 
+ \log \left(\frac{\varepsilon'}{m} \right)
\right]
\end{split}
\end{equation}
and
\begin{equation} \label{eq:IncohIntegralSmallr}
\begin{split}
&
K_{\rt\approx0}(A,\Delta,m^2,\varepsilon') \equiv 
\\ &
\frac{1}{\varepsilon'^8}
\int d^2\kt d^2 \bar{\kt} 
\frac{\exp \left(-A \left(\kt+\bar{\kt} \right)^2 \right)}{( \left(\kt+\frac{\Delta}{2} \right)^2+m^2)( \left(\kt-\frac{\Delta}{2} \right)^2+m^2)}
\\ & \times
\frac{1}{( \left(\bar{\kt}+\frac{\Delta}{2} \right)^2+m^2)( \left(\bar{\kt}-\frac{\Delta}{2} \right)^2+m^2)}
\\ & \times
\left[ 
\frac{\Delta^4}{16} - \frac{\Delta^2 \ktbar^2}{4} - \frac{\Delta^2 \kt^2}{4} + \kt^2 \ktbar^2
\right].
\end{split}
\end{equation}
We see that the $Z$-function, and consequently the coherent and hot spot fluctuation parts of the cross section, retain a logarithmic dependence on $ \varepsilon'$ and thus the quark mass even in the small dipole limit. Only in the color fluctuation part, i.e. the function $K$, the dependence on the quark mass becomes a pure power law.

In the small $\rt/\rtbar$ limit of the dipole cross section, the coherent cross section becomes
\begin{equation} \label{eq:smallRCoh}
\begin{split}
&
\left. \frac{d \sigma_{T,L}^{\gamma^*p\rightarrow Vp}}{dt}
\right|_{\rt\approx0}
=
\frac{1}{16\pi} C^2_{T,L}\frac{g^4 \mu_0^4 (N_c^2-1)^2 N_q^2}{(2\pi N_c)^2} 
\\ & \times
\exp \left( - \left[r^2_H + \left(\frac{N_q-1}{N_q} \right)R^2 \right]\Delta^2 \right)
Z_{\rt\approx0}(\Delta,m^2,\varepsilon')^2
.
\end{split}
\end{equation}
Now by also expanding \eqref{eq:IncohIntegral}, the parts of the incoherent cross section, in the small-$\rt$ limit, read
\begin{equation} \label{eq:smallRIncoh1DC}
\begin{split}
&
\left. \frac{d \sigma_{T,L}^{\gamma^*p\rightarrow Vp^*}}{dt} \right|_{1,DC,\rt\approx0}
\\ &
=
\frac{1}{16\pi}C^2_{T,L} \frac{g^4\mu_0^4(N_c^2-1)^2}{(2\pi N_c)^2}
N_q 
\\ & \times
\left[
\exp \left(-\Delta^2r_H^2 \right)
- \exp\left( - \left[r^2_H + \left(\frac{N_q-1}{N_q} \right)R^2 \right]\Delta^2 \right)
\right]
\\ & \times
Z_{\rt\approx0}(\Delta,m^2,\varepsilon')^2 ,
\end{split}
\end{equation}

\begin{equation} \label{eq:smallRIncoh2DC}
\begin{split}
&
\left. \frac{d \sigma_{T,L}^{\gamma^*p\rightarrow Vp^*}}{dt} \right|_{2,DC,\rt\approx0}
\\ &
=
\frac{1}{16\pi}C^2_{T,L} \frac{g^4\mu_0^4(N_c^2-1)^2}{(2\pi N_c)^2}
N_q(N_q-1) 
\\ & \times
\Bigg[
\exp \left(-\Delta^2 \left[ R^2+r_H^2 \right] \right)
\\ &
- \exp\left( - \left[r^2_H + \left(\frac{N_q-1}{N_q} \right)R^2 \right]\Delta^2 \right)
\Bigg]
\\ & \times
Z_{\rt\approx0}(\Delta,m^2,\varepsilon')^2 ,
\end{split}
\end{equation}

\begin{equation} \label{eq:smallRIncoh1C}
\begin{split}
&
\left. \frac{d \sigma_{T,L}^{\gamma^*p\rightarrow Vp^*}}{dt} \right|_{1,C,\rt\approx0}
\\ &
=
\frac{1}{16\pi}C^2_{T,L} \frac{g^4 \mu_0^4 (N_c^2-1)}{2\pi^2 N_c^2} N_q
K_{\rt\approx0}(r_H^2,\Delta,m^2,\varepsilon') ,
\end{split}
\end{equation}

\begin{equation} \label{eq:smallRIncoh2C}
\begin{split}
&
\left. \frac{d \sigma_{T,L}^{\gamma^*p\rightarrow Vp^*}}{dt} \right|_{2,C,\rt\approx0}
\\ &
=
\frac{1}{16\pi}C^2_{T,L} \frac{g^4 \mu_0^4 (N_c^2-1)}{2\pi^2 N_c^2} N_q(N_q-1)
\\ & \times
K_{\rt\approx0}(R^2+r_H^2,\Delta,m^2,\varepsilon') ,
\end{split}
\end{equation}
where the subscripts are the same as before, referring to the number of hot spots and whether the color structure is connected or not.

\begin{figure*} 
\centering
\includegraphics[scale=0.12]{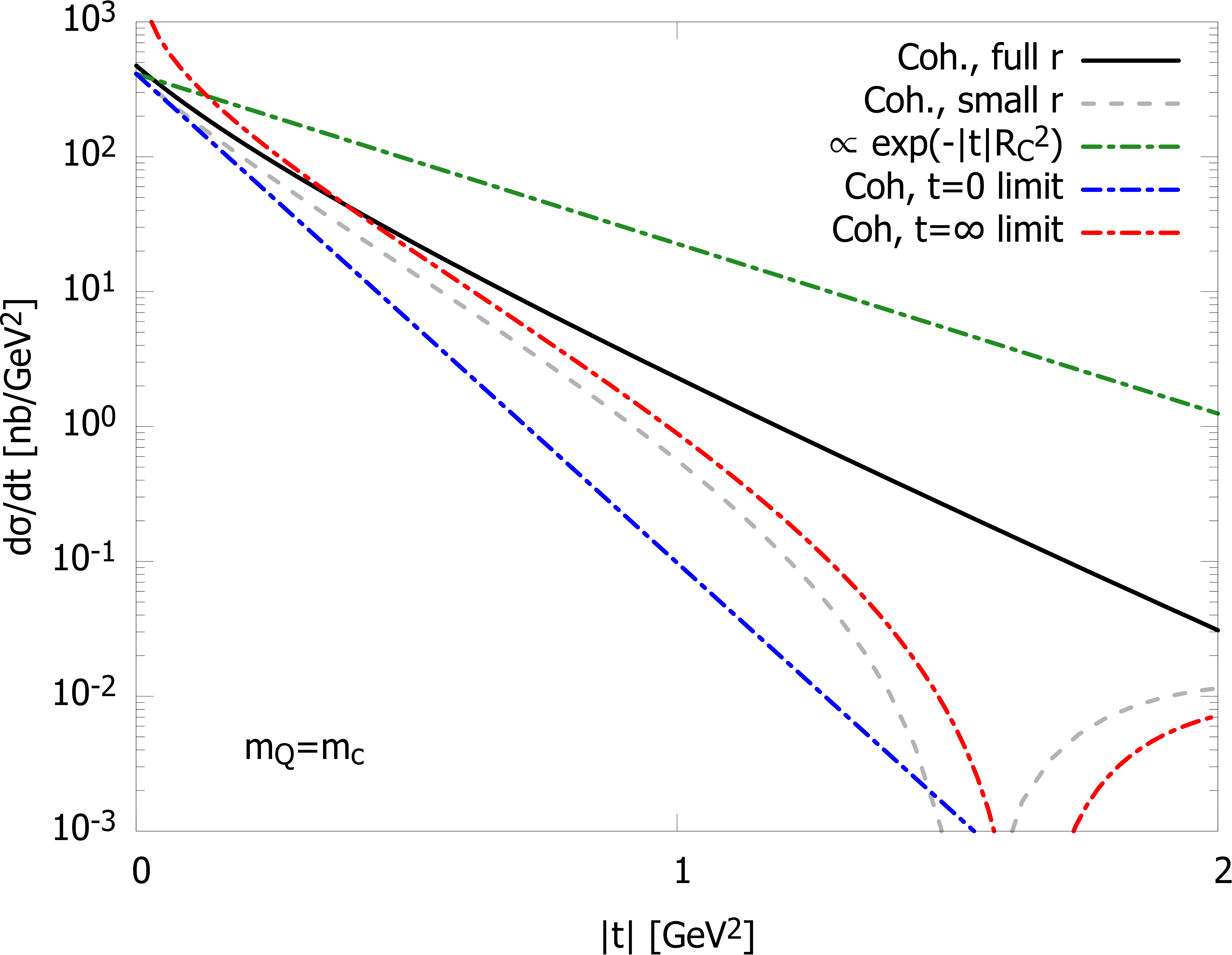}
\includegraphics[scale=0.12]{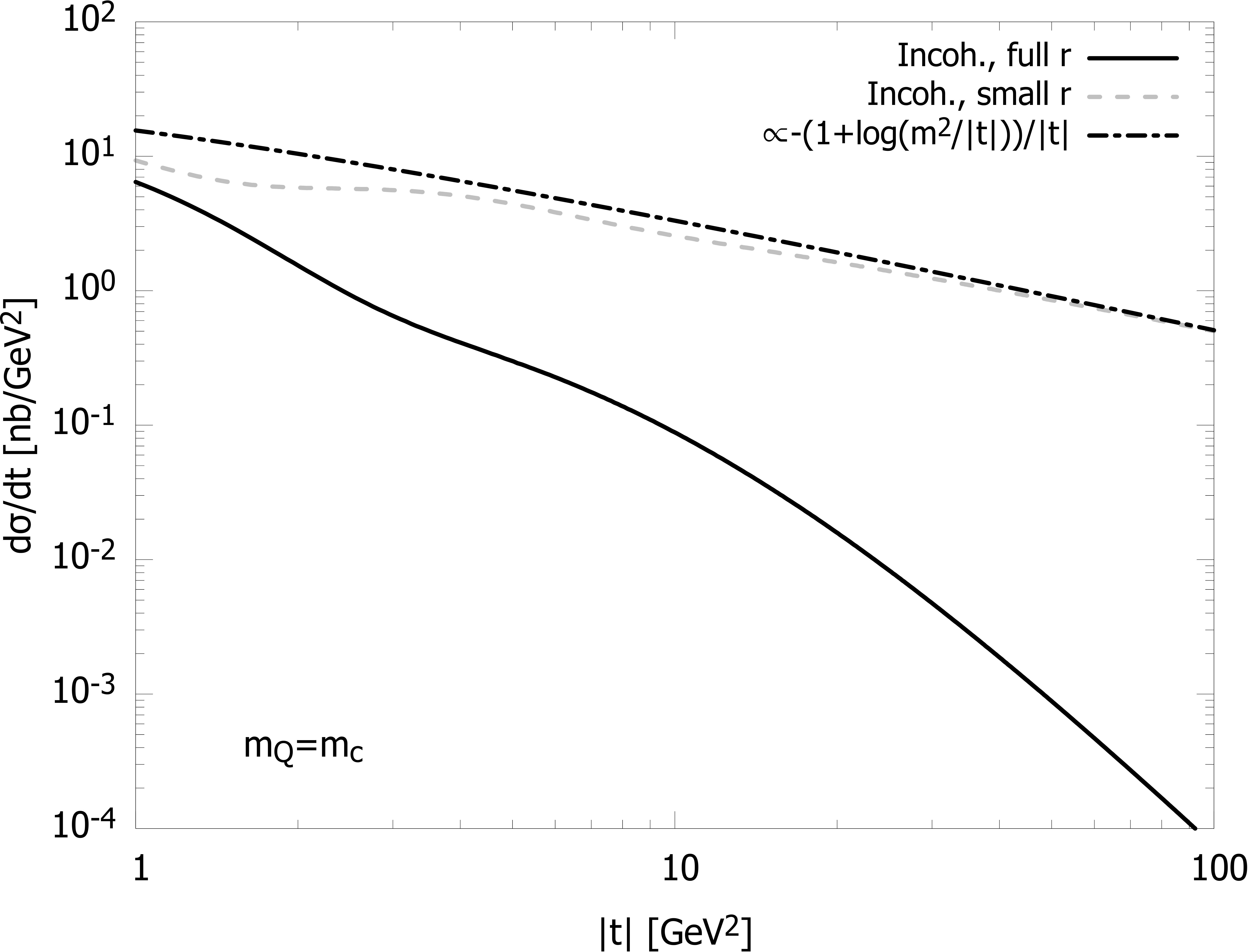} \\
\includegraphics[scale=0.12]{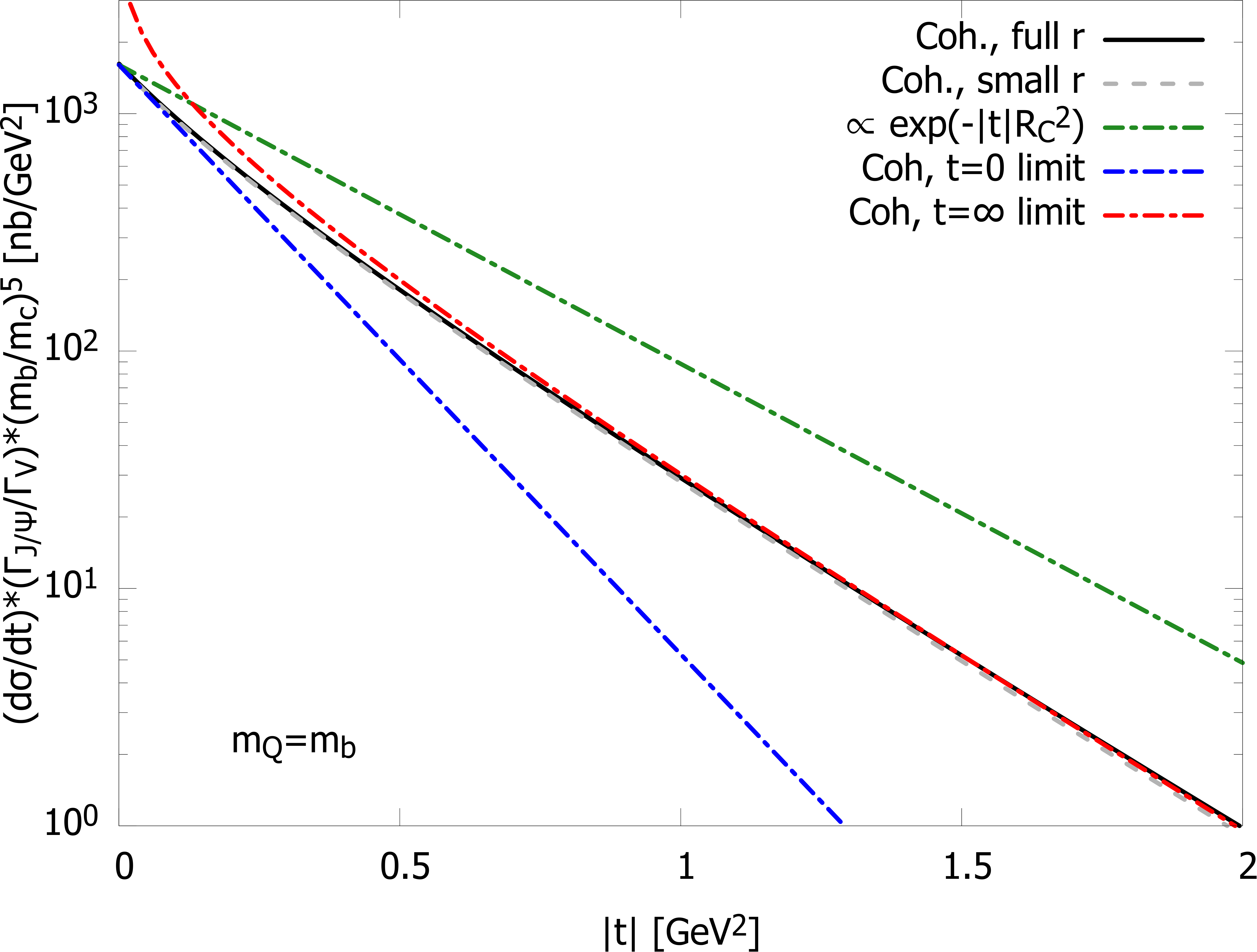}
\includegraphics[scale=0.12]{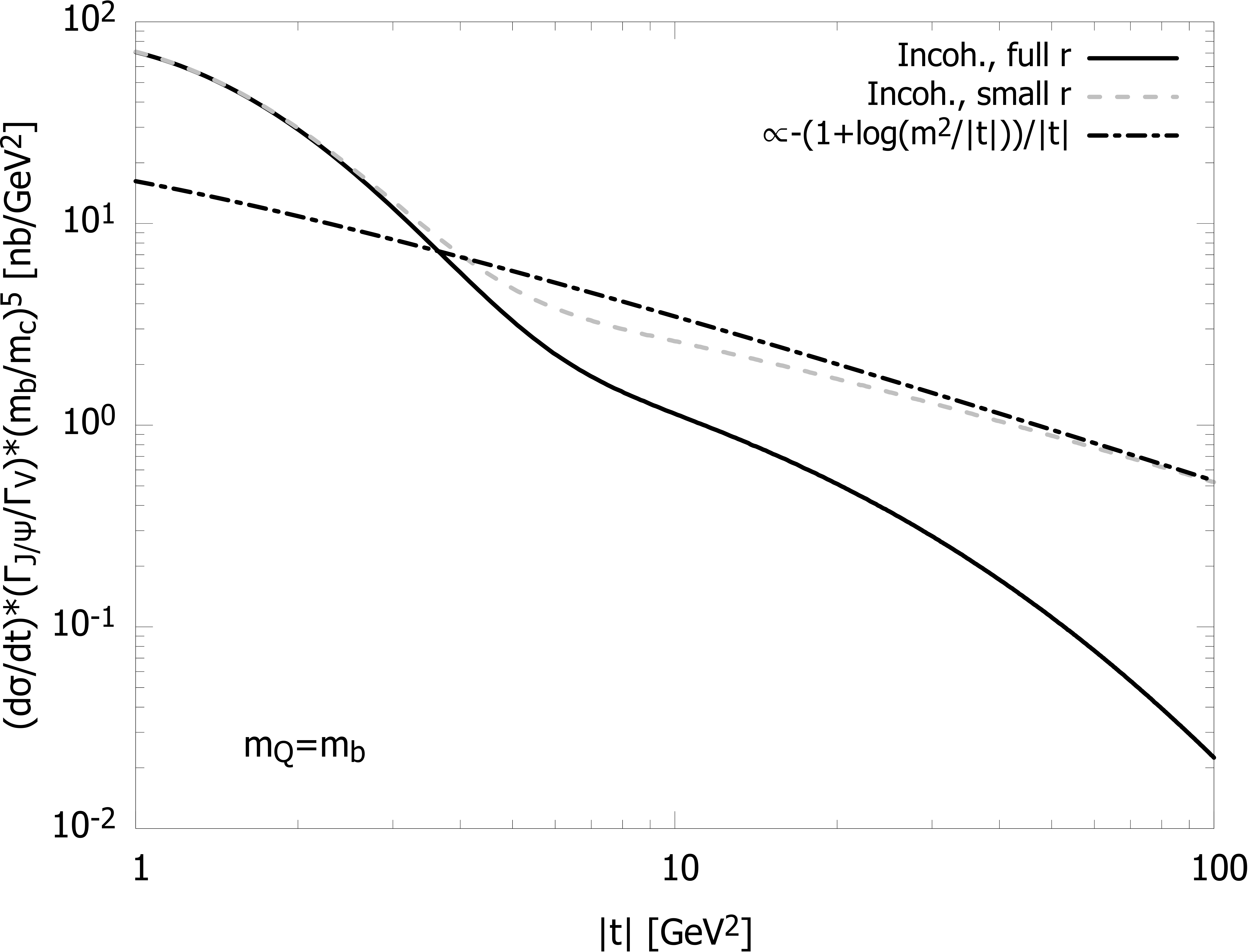} \\
\includegraphics[scale=0.12]{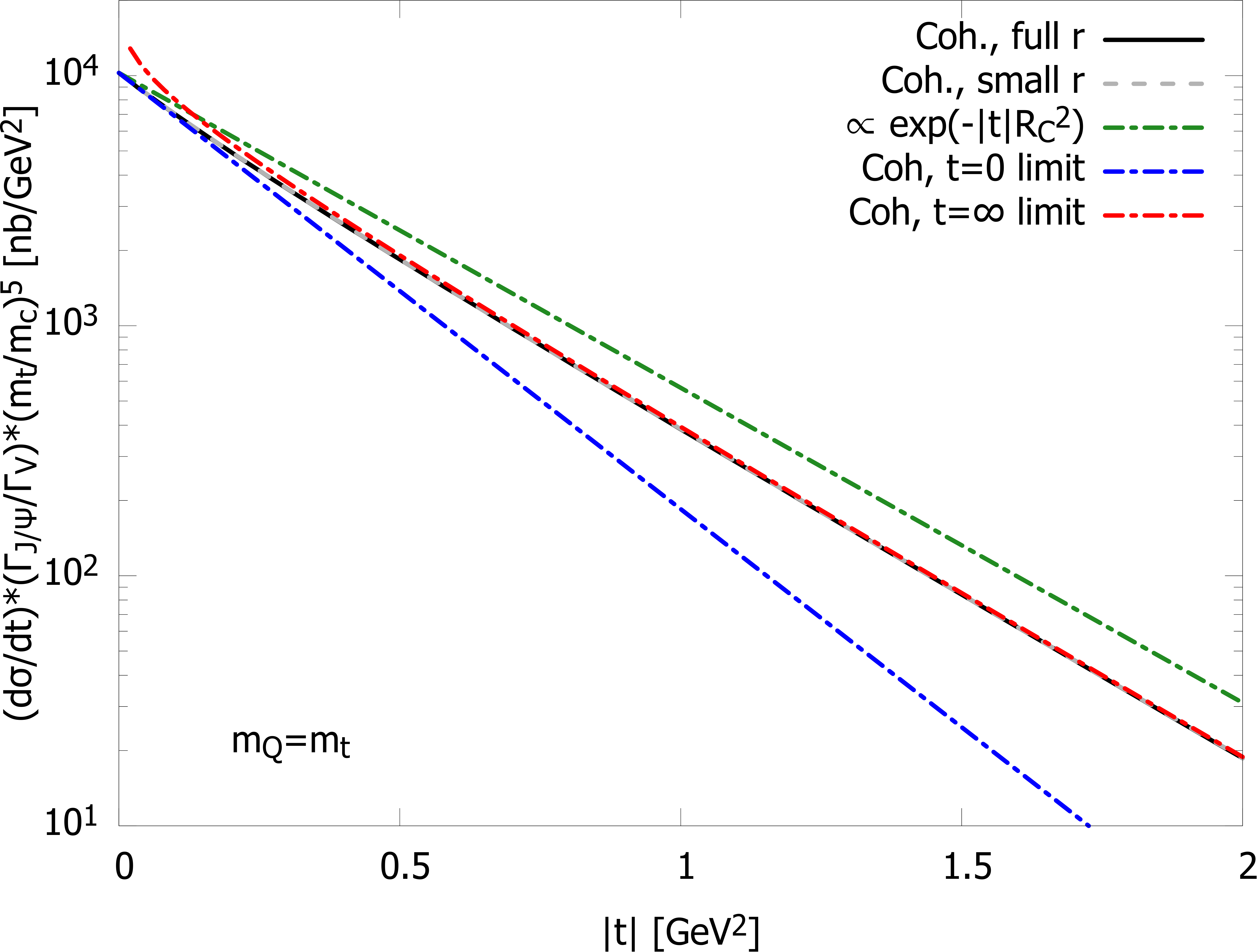}
\includegraphics[scale=0.12]{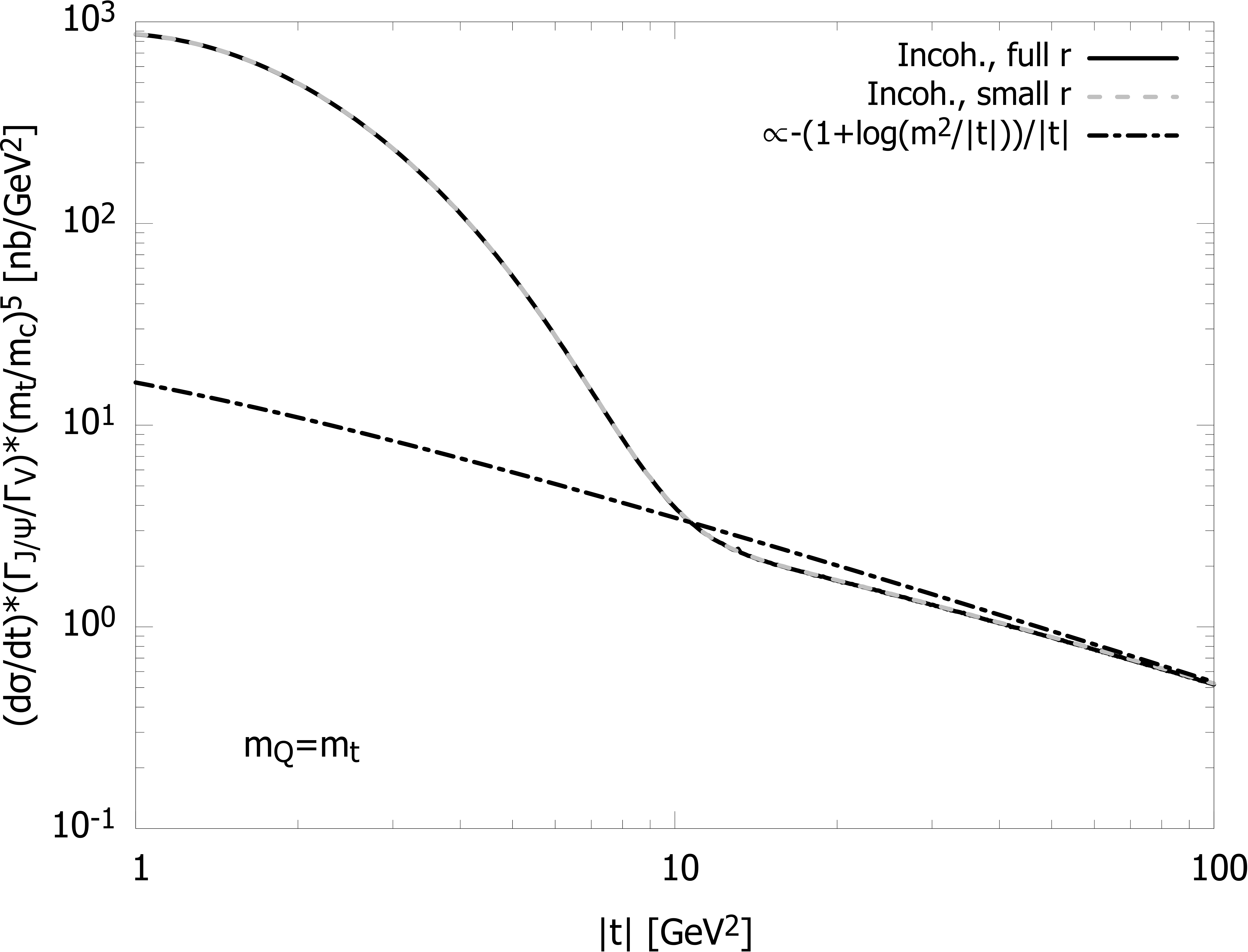}
\caption{Comparison of the analytic small-$r$ limits of the coherent \eqref{eq:smallRCoh} and incoherent cross sections \eqref{eq:smallRIncoh1DC}, \eqref{eq:smallRIncoh2DC}, \eqref{eq:smallRIncoh1C}, \eqref{eq:smallRIncoh2C} to the full numerical result. Also shown are analytical results obtained in the small or large $t$ limits, \eqs\eqref{eq:Coht0}, \eqref{eq:CohSmallRLargeTSmallM}, \eqref{eq:largetIncoh} and an exponential with a value at $t=0$, which can be easily seen from Eq.~\eqref{eq:CohSmallTExpansion} by setting $t=0$, and a slope given by the coherent radius. As usual, the top row shows the result for the $c$ mass, the middle row for $b$ and the bottom row for $t$.
}
\label{fig:CSAsymptotics}
\end{figure*}

\subsection{Small-$\rt$ coherent cross section at $\Delta=\infty$ and $\Delta=0$} \label{app:DeltaCohApproximations0AndInfty}

The small-$\rt$ limit of the coherent cross section reads
\begin{equation} \label{eq:CohCSSmallRApp}
\begin{split}
&
\frac{d \sigma_{T,L}^{\gamma^*p\rightarrow Vp}}{dt}
=
\frac{1}{16\pi}
C^2_{T,L}\frac{g^4 \mu_0^4 (N_c^2-1)^2 N_q^2}{N_c^2 \varepsilon'^8} 
\\ & \times
\exp \left( - \left[r^2_H + \left(\frac{N_q-1}{N_q} \right)R^2 \right]\Delta^2 \right)
\\ & \times
\Bigg[
-\frac{2m^2+\Delta^2}{\Delta \sqrt{4m^2+\Delta^2}} \arctanh \left( \frac{\Delta}{\sqrt{4m^2+\Delta^2}} \right) 
+ \log \left(\frac{\varepsilon'}{m} \right)
\Bigg]^2
.
\end{split}
\end{equation}
Now to approximate this at large $t$, we expand the function in the square brackets in large $-t=\Delta^2$. This yields
\begin{multline} \label{eq:CohSmallRLargeTSmallM}
\left. \frac{d \sigma_{T,L}^{\gamma^*p\rightarrow Vp}}{dt} \right|_{-t\to \infty}
=
\frac{1}{16\pi}
C^2_{T,L}\frac{g^4 \mu_0^4 (N_c^2-1)^2 N_q^2}{N_c^2 \varepsilon'^8} 
\\  \times
\exp \left( \left[r^2_H + \left(\frac{N_q-1}{N_q} \right)R^2 \right]t \right)
\frac{1}{4}\log \left(\frac{\varepsilon'^2}{-t} \right)^2
.
\end{multline}
This result also coincides with the small-$m$ expansion of \eqref{eq:CohCSSmallRApp}.

On the other hand, we can approximate the slope at $-t=\Delta^2=0$ by first expanding the square bracket at small $t$. Doing this we get
\begin{equation} \label{eq:CohSmallTExpansion}
\begin{split}
&
\left. \frac{d \sigma_{T,L}^{\gamma^*p\rightarrow Vp}}{dt}
\right|_{t\approx 0}
=
\frac{1}{16\pi}
C^2_{T,L}\frac{g^4 \mu_0^4 (N_c^2-1)^2 N_q^2}{N_c^2 \varepsilon'^8} 
\\ & \times
\exp \left( \left[r^2_H + \left(\frac{N_q-1}{N_q} \right)R^2 \right]t \right)
\\ & \times
\left(
\log \left(\frac{\varepsilon'}{m} \right)-\frac{1}{2}
\right)^2
\\ & \times
\left[
1
-\frac{-t}{3m^2\left( \log \left(\frac{\varepsilon'}{m} \right) - \frac{1}{2} \right)}
\right]
.
\end{split}
\end{equation}
Near $t=0$ we can use this to get an expression that is an exponential in $t$, but with a slope that takes into account the Coulomb tails regulated by $m$.
\begin{multline} \label{eq:Coht0}
\left. \frac{d \sigma_{T,L}^{\gamma^*p\rightarrow Vp}}{dt} \right|_{t\approx 0}
\\
=
\frac{1}{16\pi}
C^2_{T,L}\frac{g^4 \mu_0^4 (N_c^2-1)^2 N_q^2}{N_c^2 \varepsilon'^8} 
\left(
\log \left(\frac{\varepsilon'}{m} \right)-\frac{1}{2}
\right)^2
\\  \times
\exp \Bigg( \bigg[r^2_H + \left(\frac{N_q-1}{N_q} \right)R^2  
+\frac{1}{3m^2\left( \log \left(\frac{\varepsilon'}{m} \right) - \frac{1}{2} \right)}
\bigg]t
\Bigg).
\end{multline}
For a finite IR regulator $m$, the correction to the slope $\sim 1/[m^2 \log (m_Q/m)]$ does indeed vanish in the limit $m_Q\to \infty,$ but this happens only extremely slowly. 
Note also that this re-exponentiation that gives an interpretation as a correction to the coherent slope is based in an expansion in $t/m^2$, and is thus valid for only very small values of $t$.

\subsection{Large-$t$ behavior of the small-$r$ incoherent cross section} \label{app:LargeT}

In the large $t$ limit the incoherent cross section is dominated by the color fluctuation contributions. For this we need to look at the large $t$ behavior of the the integral
\begin{equation}
\begin{split}
&
\int d^2\kt d^2 \bar{\kt} \frac{\exp \left(-A \left(\kt+\bar{\kt} \right)^2 \right)}{( \left(\kt+\frac{\Delta}{2} \right)^2+m^2)( \left(\kt-\frac{\Delta}{2} \right)^2+m^2)}
\\ & \times
\frac{1}{( \left(\bar{\kt}+\frac{\Delta}{2} \right)^2+m^2)( \left(\bar{\kt}-\frac{\Delta}{2} \right)^2+m^2)}
\\ & \times
\left[ 
\frac{\Delta^4}{16} - \frac{\Delta^2 \ktbar^2}{4} - \frac{\Delta^2 \kt^2}{4} + \kt^2 \ktbar^2
\right] ,
\end{split}
\end{equation}
with $A=R^2+r_H^2$ for the two hot spot case and $A=r_H^2$ for the one hot spot case. Here  the former has a much smaller effect for the large-$t$ tail, as we shall see.

Firstly let us change the coordinates to $\qt = \kt + \ktbar$ and $\qtbar = \kt - \ktbar$. Doing this, and by simplifying a bit, we get
\begin{equation}
\begin{split}
&
=
4
\int d^2\qt d^2 \qtbar \frac{\exp \left(-A\qt^2 \right)}{( \left(\qt+\qtbar+\Delta \right)^2+4m^2)( \left(\qt+\qtbar-\Delta \right)^2+4m^2)}
\\ & \times
\frac{1}{( \left(\qt-\qtbar+\Delta \right)^2+4m^2)( \left(\qt-\qtbar-\Delta \right)^2+4m^2)}
\\ & \times
\left[ 
\Delta^4 - \Delta^2 (\qt-\qtbar)^2 - \Delta^2 (\qt+\qtbar)^2 + (\qt+\qtbar)^2(\qt-\qtbar)^2
\right] .
\end{split}
\end{equation}
Now by assuming that the Gaussian restricts $\qt$ to small values, we can set $\qt=0$ everywhere else. Performing the Gaussian integration, we get
\begin{equation}
\begin{split}
&
=
\frac{4\pi}{A}
\int d^2 \qtbar \frac{\Delta^4 - 2\Delta^2 \qtbar^2 + \qtbar^4}{( \left(\qtbar+\Delta \right)^2+4m^2)^2( \left(\qtbar-\Delta \right)^2+4m^2)^2}
.
\end{split}
\end{equation}
The angle of $\Delta$ does not matter in our calculation so we can choose $\Delta$ to be fully in the $x$ direction. This integration can be performed in the Cartesian coordinates, and it yields
\begin{equation}
\begin{split}
&
=
-\frac{\pi^2}{2A\Delta^3} \Bigg[ 
\frac{2 \left(4m^4\Delta+2m^2\Delta^3+\Delta^5 \right)}{ \left(4m^2+\Delta^2 \right)^2}
\\ &
-\frac{4 \left(8m^6+16m^4\Delta^2+6m^2\Delta^4+\Delta^6 \right)\arctanh \left( \frac{\Delta}{\sqrt{4m^2+\Delta^2}} \right) }{ \left( 4m^2+\Delta^2 \right)^{\frac{5}{2}}}
\Bigg]
.
\end{split}
\end{equation}
Now expanding this to the leading order at  large $\Delta$, we get
\begin{equation} \label{eq:largetIncoh}
\begin{split}
&
\approx
-\frac{\pi^2 \left(1+\log \left(\frac{m^2}{\Delta^2} \right) \right)}{A\Delta^2} + \mathcal{O} \left( \frac{1}{\Delta^4} \right)
.
\end{split}
\end{equation}
We can also write this as
\begin{equation}
\label{eq:larget}
\begin{split}
&
\approx
\frac{\pi^2 \left( 1+\log \left(-\frac{m^2}{t} \right) \right)}{At} + \mathcal{O} \left(\frac{1}{t^2} \right)
\end{split}
\end{equation}
Now it is clear that the smaller $A$ term with $A=r_H^2$ dominates over the $A=R^2+r_H^2$ one. Thus setting 
 $A=r_H^2$ in \eq\nr{eq:larget} gives the large-$t$ limit for the incoherent cross section.

\subsection{Comparing the full result with the analytical limits} \label{app:FullVsLimits}

In Fig.~\ref{fig:CSAsymptotics} we have plotted the cross sections for both the full and small-$r$ expressions together with their expected behavior in the different regimes. For the coherent cross section we have plotted the full result \eqref{eq:CohCS}, its small-$\rt$ expansion \eqref{eq:smallRCoh}, an exponential $\propto \exp\left(tR_C^2\right)$ with the coherent radius \eqref{eq:CohRadius} and the two limits where we approximate the $t$ behavior with exponentials obtained from the slope of the small-$\rt$ expansion at $\Delta=0$ \eqref{eq:Coht0} and $\Delta=\infty$ \eqref{eq:CohSmallRLargeTSmallM}. In these plots we use the actual normalizations of the limits as they are seen in the equations. 

Firstly looking at the charm quark coherent cross sections, we see that both the full and the small-$r$ cross sections stray from both the expected exponential and from each other   already at small $|t|$. 
We also see a dip in the small-$\rt$ limit of the coherent cross section. This dip appears because the $Z_{\rt\approx0}$-function changes sign. It also exists for the heavier quarks but its location is at a higher value of $t$ the higher the quark mass $m_Q$, or photon virtuality $Q^2$, is. 
We see that in the range of $t$ we study here, the bottom quark mass is sufficient to make the full and small-$\rt$ results of the coherent cross section coincide quite well. Also the $t=\infty$ limit expansion of the coherent cross section works quite well except for the very low $t$ values. For the top quark case, we see that both the full and the small-$\rt$ coherent cross sections almost have the expected exponential decay and thus measure the overall size of the proton much better than the $c$ and $b$ quarks. The $t=0$ limit expansion of the coherent cross section does not work very well for any of the quark masses except at very low values of $t$.

Next let us look at the incoherent cross section. In Fig.~\ref{fig:CSAsymptotics} we have also plotted the full incoherent cross section \eqref{eq:OneHSDC},\eqref{eq:TwoHSDC},\eqref{eq:OneHSConn},\eqref{eq:TwoHSConn}, the small-$\rt$ expansion of the incoherent cross section \eqref{eq:smallRIncoh1DC}, \eqref{eq:smallRIncoh2DC}, \eqref{eq:smallRIncoh1C}, \eqref{eq:smallRIncoh2C} and the small-$\rt$ and large-$t$ expansion of the incoherent cross section (\eqref{eq:smallRIncoh1C} with the integral substituted with \eqref{eq:largetIncoh}). Also here we use also  the normalization resulting from the calculation in the plot.

Firstly we notice that for the charm quark, the small-$\rt$ incoherent cross section goes down much too slowly in comparison to the full-$\rt$ result. The large-$t$ expansion does capture the behavior of the small-$\rt$ limit of the incoherent cross section,  however. We seem to have to have extremely heavy quarks to make the dipole size to be small enough to warrant for the small-$\rt$ expansion. This can be seen in the bottom quark plot where the mass of this quark is not high enough to really make the expansion work. Only in the top quark case do we see that our small-$\rt$ limit is the correct one for the asymptotically heavy quark limit.

\bibliographystyle{JHEP-2modlong}
\bibliography{draft}

\end{document}